%% file: acs.tex
\documentclass[journal=jpccck,manuscript=article,layout=traditional]{achemso}

\setkeys{acs}{keywords = true}

\usepackage[version=3]{mhchem} 
\usepackage{siunitx}
\usepackage{amsmath}
\usepackage[bookmarks=false]{hyperref}

\usepackage{cleveref}

\usepackage{multirow}
\usepackage{subfigure}
\usepackage{verbatim}
\usepackage{xr}
\usepackage{booktabs}
\usepackage{caption}
\usepackage{array}

\usepackage{subcaption} 
\usepackage{tikz}

\usepackage[T1]{fontenc}
\usepackage{float}
\usepackage{xspace} 

\DeclareSIUnit{\volpercent}{vol\%}
\DeclareSIUnit{\kcal}{kcal}
\DeclareSIUnit{\Bohr}{Bohr}

\usepackage{mathtools}

\SectionNumbersOn

\crefname{figure}{fig.}{Fig.}
\crefname{section}{Section}{Sections}

\newcolumntype{I}[1]{>{\hspace{1em}}p{#1}}
\newcolumntype{C}[1]{>{\centering\arraybackslash}p{#1}}

\makeatletter
\newcommand*{\addFileDependency}[1]{
\typeout{(#1)}
\@addtofilelist{#1}
\IfFileExists{#1}{}{\typeout{No file #1.}}
}\makeatother

\newcommand{\HXX}{HXX\xspace}
\newcommand{\optHXX}{optHXX\xspace}
\newcommand{\srpbe}{LC-srPBEx25\xspace}
\newcommand{\optVo}{optRPA24\xspace}
\newcommand{\optVt}{optRPA26\xspace}
\newcommand{\Ecut}{\ensuremath{E_{\text{cutoff}}}\xspace}

\newcommand{\EHXX}{\ensuremath{E_{\text{HXX}}}\xspace}
\newcommand{\EoptHXX}{\ensuremath{E_{\text{optHXX}}}\xspace}
\newcommand{\Ecrpa}{\ensuremath{E_{\text{c,RPA}}}\xspace}
\newcommand{\dEcut}{\ensuremath{\Delta \text{cutoff}}\xspace}

\newcommand{\Eads}{\ensuremath{E_\text{ads}}\xspace}
\newcommand{\Nads}{\ensuremath{27}\xspace}
\newcommand{\embed}{\ensuremath{\textit{PBE+D3/M06}}\xspace}

\author{Neung-Kyung Yu}%
 \affiliation{ 
School of Chemical \& Biomolecular Engineering, Georgia Institute of Technology, Atlanta, Georgia 30332, USA
}%

\author{Johannes Voss}%
 \email{vossj@slac.stanford.edu}
 \affiliation{ 
SUNCAT Center for Interface Science and Catalysis, SLAC National Accelerator Laboratory, Menlo Park, California 94025, USA
}%

\author{Andrew J. Medford}%
 \email{ajm@gatech.edu}
 \affiliation{ 
School of Chemical \& Biomolecular Engineering, Georgia Institute of Technology, Atlanta, Georgia 30332, USA
}%

\title[ ]
    {
    Optimization of random phase approximation calculations for improved energies of molecules, solids, and surfaces
    }

\abbreviations{}

\mciteErrorOnUnknownfalse

\begin{document}


\begin{tocentry}
    
    \label{TOC Graphic}
    \includegraphics[width=2.75 in]{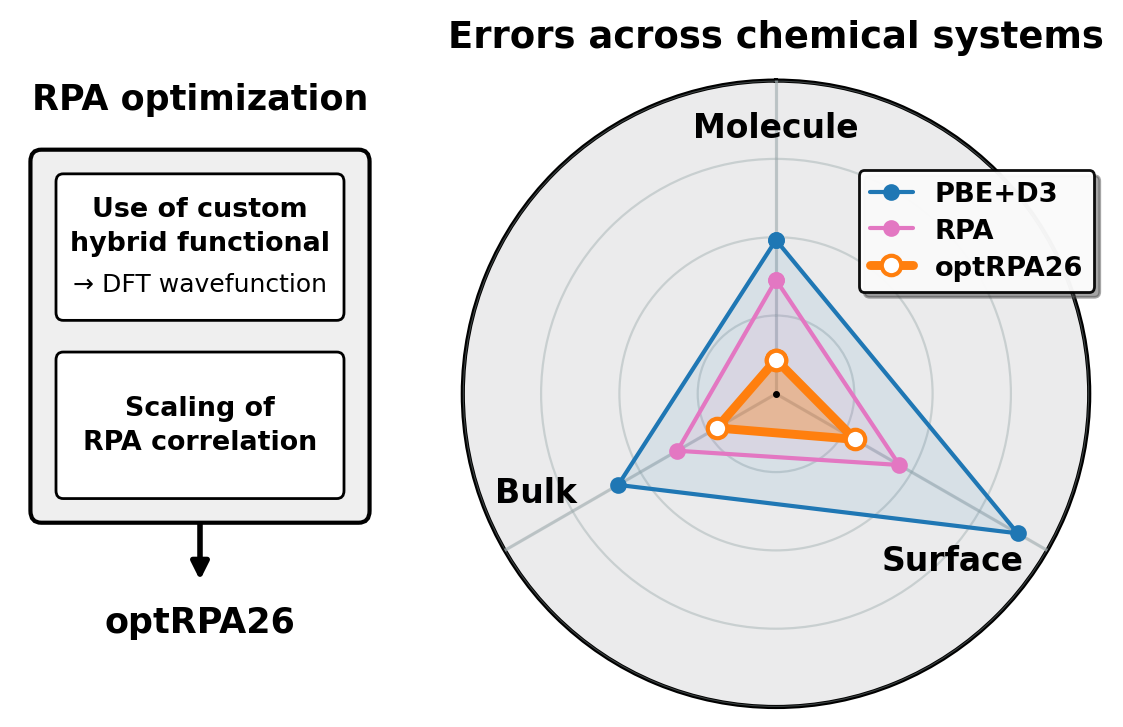}

\end{tocentry}

\begin{abstract}

We present an optimized random phase approximation method (\optVt) that significantly improves upon conventional RPA. 
The method employs an empirically constructed hybrid functional to generate DFT orbitals to evaluate the RPA correlation energy, which is then scaled by a constant. Comprehensive benchmarks across molecules, bulk solids, and surface systems demonstrate that \optVt consistently achieves high accuracy, with mean absolute errors of 0.05 eV for W4-11 reaction energies, 0.07 eV for cohesive energies, 0.09 eV for metal oxide formation energies, 0.11--0.12 eV for adsorption of small molecules on metals, and 0.06 eV for adsorption on oxides. In addition, \optVt{} correctly captures phase stability in metal oxides and magnetic metals. The \optVt{} approach can be run using standard RPA implementations, highlighting its potential as a general-purpose reference method that can accurately capture covalent, ionic, and metallic, and van der Waals bonding in molecules, solids, and interfaces.

\end{abstract}


\section{\label{sec:intro}Introduction}

Density functional theory (DFT) is the most widely used approach for quantum mechanical (QM) simulations in physics, materials science, and chemistry \cite{dumaz2021citeDFT}. 
Many problems in applied sciences require computing properties of systems with $\sim$100 atoms or more, often with periodic or mixed boundary conditions. 
DFT is capable of performing these simulations within a practical time frame due to its relatively low computational cost and the wide availability of efficient implementations. 
However, the lack of a universally accurate exchange-correlation (xc) functional leads to errors that can greatly exceed the ``chemical accuracy'' (\SI{1}{\kcal\per\mol} or \SI{0.043}{eV}) needed for quantitative predictions of reaction rates \cite{medvedev2017DFTaccuracy}. 
For transition metal systems, the uncertainty is even greater from both experimental and theoretical points of view, so the chemical precision of the transition metals (\SI{3}{\kcal\per\mol} or \SI{0.130}{eV}) has been defined more loosely \cite{deyonker2007TMaccuracy,jiang2012TMaccuracy,shee2019TMaccuracy}.
Although numerous specialized functionals exist, the lack of systematic improvement in functionals makes it difficult to determine a priori which approximation will be most accurate for a particular chemical system.
The development of an xc functional that achieves chemical accuracy across all types of systems---including atoms, molecules, liquids, ionic solids, metals, and interfaces---is a grand challenge in the field of computational chemistry \cite{cohen2012DFTbroad,peverati2014DFTbroad}. The problem is particularly acute in surface science in catalysis, where there is currently no method capable of achieving chemical accuracy for adsorption at metal surfaces \cite{ye2024ccsdt_oxide}.

Wavefunction methods offer a more accurate alternative to DFT. However, their high computational cost and limited availability of periodic implementations make them impractical for many problems in materials science and solid-state physics. Nevertheless, two wavefunction approaches have become increasingly efficient and have been applied to solid-state problems with periodic boundary conditions. 
The first is coupled cluster with singles, doubles, and partial triples (CCSD(T)), widely regarded as the gold standard for chemical calculations, achieving chemical accuracy in most systems where it can be applied \cite{feller2007ccsdt,peterson2012ccsdt}.
Although CCSD(T) is computationally very expensive, with formal $\mathcal{O}(N^7)$ scaling \cite{feller2007ccsdt,peterson2012ccsdt}, more efficient approaches have been developed \cite{riplinger2016ccsdt_scaling}.
Specifically, recent progress in local approximations to CCSD(T) led to the development of variants such as PNO-CCSD(T) \cite{neese2009ccsdt_PNO}, and DLPNO-CCSD(T) \cite{riplinger2013ccsdt_DLPNO}, and LNO-CCSD(T) \cite{rolik2013ccsdt_LNO,nagy2017ccsdt_LNO,nagy2019ccsdt_LNO}.
In practice, CCSD(T) is primarily applied for gas molecules since most implementations do not support periodic boundary conditions. However, periodic CCSD(T) implementations have recently become available \cite{ye2024ccsdt_periodic}.  
Several groups have successfully applied local CCSD(T) methods (both periodic and non-periodic) to calculate highly accurate adsorption energies on oxide surfaces \cite{kubas2016ccsdt_oxide,shi2023ccsdt_oxide, ye2024ccsdt_oxide,ye2024ccsdt_oxide2,shi2025autoSKZCAM}.
Despite this progress, the implementation of partial triples (T) remains problematic for metallic systems due to energy divergence, also known as the infrared catastrophe \cite{alavi2024ccsdt_periodic,ye2024ccsdt_periodic}. Only very recently, a modified approximation to the triples (cT) has been proposed to address this limitation. \cite{masios2023CCSDcT,carbone2024CCSDcT,schafer2025CCSDcT}.

In addition, quantum Monte Carlo (QMC) methods, such as diffusion Monte Carlo (DMC), have been applied even more widely to problems in solid-state physics and surface science \cite{needs2010qmc_casino,kim2018qmcpack,dd2017qmc_ads}. 
In principle, DMC is an exact method, but its accuracy is typically limited by practical factors such as the number of moves for statistical sampling, time step size, specialized pseudopotentials, and slow convergence of Brillouin zone sampling via supercell extrapolation, as well as the fixed-node approximation\cite{needs2010qmc_casino,kim2018qmcpack,kent2020qmcpack,krogel2017qmc_psp}. 
Although generally much more accurate than DFT, statistical error bars on practical quantities such as formation energies and adsorption energies often exceed chemical precision \cite{saritas2017dmc_Eform,dd2017qmc_ads,hsing2019qmc_ads}.
Due to the lack of highly accurate and practical computational methods for surfaces, there are only a few well-established reference datasets, such as the CE39 \cite{wellendorff2015ce39} or ADS41 dataset \cite{mall2019ads41} of experimental and DFT adsorption energies on simple metal surfaces.

Another strategy to improve upon DFT methods is the use of embedding schemes that combine low-level and high-level QM methods. The high-level method is typically applied to small clusters, whereas the low-level method is used to describe larger systems, such as periodic models. Embedding approaches have been applied to describe adsorption on oxide surfaces using MP2 or CCSD(T) as the high-level method \cite{boese2013embed_ccsdt_ox,kubas2016ccsdt_oxide,sauer2019embed_ccsdt_ox,Maristella2019embed_mp2-ccsdt_ox,shi2023ccsdt_oxide,shi2025autoSKZCAM} and on metal surfaces using hybrid functionals, RPA, or wavefunction theories as the high-level method \cite{huang2006embed_pc_metal,ren2009rpa-hybrid,araujo2022ads38,wei2023embed_rpa_metal,sheldon2024embed_rpa_metal}. 
In particular, the \embed approach is an embedding scheme that combines cluster (at hybrid M06 level) and periodic models (at PBE$+$D3 level). This approach has been reported as one of the most accurate methods for adsorption energies on metals and is therefore used as a comparison in this benchmarking study. Despite the potential for high accuracy in embedding methods, they generally require specialized software and expertise, and are not widely used in the computational materials science community.

Returning to standard DFT, the concept of Jacob's ladder of density functional approximations describes the physical hierarchy of density-based quantum chemical methods into five levels \cite{perdew2005jacob}.
Different levels of DFT methods are positioned on each rung, starting with the local density approximation (LDA) on the first rung, followed by the generalized gradient approximation (GGA), meta-GGA, and hybrid functionals.
Unlike those on the lower rungs, the methods on the fifth rung consider unoccupied orbitals for \textit{ab initio} calculation of correlation energy, with the random phase approximation (RPA) located at this level.
RPA correlation energies are computed with help of the adiabatic connection fluctuation-dissipation theorem (ACFDT), where the interacting density response function is approximated at the RPA level. The ACFDT was established over 40 years ago \cite{harl2008rpa_30y}, but ACFDT-RPA correlation energies have not become a standard choice. While RPA is generally applicable to solids, it remains computationally expensive and exhibits limited accuracy, particularly for molecular systems, despite its physical soundness \cite{furche2005rpa_atom, olsen2013rpa_gpaw}. 

There have been continued efforts to improve the accuracy of RPA. 
One strategy involves correcting the relatively less accurate short-range RPA correlation using the short-range LDA or GGA correlation. This concept underlies the RPA$+$ \cite{yan2000rpa_poorSR} and the generalized RPA$+$ \cite{gould2019gRPAp}, which have been applied to atomic or molecular systems. 
An alternative approach introduces a second-order screened exchange (SOSEX) term to the RPA energy, leading to the RPA$+$SOSEX method \cite{gruneis2009sosex}. Similarly, singles excitation (SE) and renormalized singles excitation (rSE) contributions have been proposed, resulting in RPA$+$SE, RPA$+$rSE, and RPA$+$SOSEX$+$rSE \cite{paier2012rSE,ren2013rSE}. Although these corrections improve the accuracy of RPA, their overall performance remains limited and often unbalanced \cite{paier2012rSE,ruzsinszky2015sosex_rse}.  Other beyond-RPA methods, based on time-dependent DFT, include the renormalized adiabatic local density approximation (rALDA) and the renormalized adiabatic Perdew–Burke–Ernzerhof (rAPBE) xc kernels \cite{olsen2014rA}. These approaches show even higher accuracy across different classes of systems \cite{olsen2019rA_benchmark}. 
Despite the promise of these approaches, most are not widely available in standard electronic structure packages, and thus require specialized expertise or software, limiting their application in practice.

In contrast, the approach introduced in this work focuses on optimizing the standard RPA calculation, often referred to as direct RPA \cite{ruzsinszky2010dRPA}, to achieve substantially higher accuracy at the cost of generating hybrid-level wavefunctions, while leveraging widely implemented RPA algorithms. 
In standard RPA, the total RPA energy is typically evaluated on PBE orbitals (referred to as RPA$@$PBE), but RPA calculations using a hybrid PBE0 functional (RPA$@$PBE0) have also been reported  \cite{furche2005rpa-hybrid,ren2009rpa-hybrid}.
In a prior study, we used a global PBE hybrid with 50\% exact exchange (EXX) to incorporate non-local interactions and then scaled the RPA correlation energy \cite{kim2024nitrogen}, which we refer to here as \optVo. 
This approach yielded accurate results for molecular atomization and formation energies, but suffered from energy convergence issues in the adsorption energies on metallic surfaces. In addition, the hybrid functional used for the wavefunctions did not yield the correct magnetic properties for bulk magnetic metals.
Here, we use a long-range corrected (LC) custom hybrid functional that includes 25$\%$ EXX only in the long range to construct DFT orbitals.
The RPA correlation energy is then scaled by a constant, which is fitted using the W4-11 dataset of highly accurate molecular atomization energies. 
Finally, short-range PBE correlation is applied when evaluating the non-RPA correlation part of total energy, which reduces the optimal scaling constant to $\sim$1 and improves energy convergence. 
The approach of scaling the RPA correlation energy has been reported in LC double-hybrid frameworks, where the total energy includes long-range EXX and long-range RPA correlation \cite{janesko2009LC-RPA,irelan2011LC-RPA}. 
Our optimized RPA approach, which we refer to as \optVt, can be run using a standard VASP installation (compiled with the Libxc library \cite{lehtola2018libxc}) without modification to the source code, making it a practical alternative to more specialized approaches. 
Here, we present the details of the formulation and benchmark it against a diverse dataset of molecules, bulk solids, and surfaces, demonstrating that \optVt can serve as a highly accurate reference method for general systems in computational chemistry and catalysis.

\section{\label{sec:methods}Methods}

Details of the benchmark datasets and electronic structure calculations are summarized here. Additional details are provided in the SI.

\subsection{Benchmark Datasets}

The following benchmark datasets were used as ground-truth reference values in this work. 
Additional details of data curation and computational protocols are provided in the \nameref{sec:SI_datasets} section of the SI.

\begin{itemize}

    \item Molecular atomization energies (W4-11) \cite{karton2011w411}:
    \begin{itemize}
        \item 140 atomization energies of small molecules and radicals, calculated using highly accurate W4 theory \cite{karton2006w4}.
        \item We considered 124 molecules with less multireference character (TAE\_nonMR124 subset \cite{karton2011w411}), which should be more relevant for catalysis. A list of excluded molecules can be found in Ref. \citenum{W411_nonMR}. 
    \end{itemize}

    \item Molecular reaction energies (W4-11-RE) \cite{margraf2017w411re}:
    \begin{itemize}
        \item 11,247 reaction energies derived from the W4-11 dataset.
        \item We selected only reactions involving molecules in TAE\_nonMR124, resulting in 8,868 reactions.
    \end{itemize}

    \item Molecular barrier heights (BH76) \cite{zhao2005BH76a,zhao2005BH76b,lars2010gmtkn24}:
    \begin{itemize}
        \item 76 forward and reverse barrier heights for hydrogen transfer, heavy-atom transfer, nucleophilic substitution, unimolecular, and association reactions.
    \end{itemize}

    \item Molecular reaction energies (BH76RC) \cite{lars2010gmtkn24}:
    \begin{itemize}
        \item 30 reaction energies derived from the BH76 dataset.
    \end{itemize}
    
    \item Reactions of organometallic transition metal (TM) complexes (MOBH29) \cite{iron2019mobh35,dohm2020mobh29,semidalas2022mobh35rev,grotjahn2023mobh28}:
    \begin{itemize}
        \item A revised version of MOBH35 \cite{iron2019mobh35}.
        \item 29 forward and 29 backward barrier heights, as well as 29 reaction energies.
    \end{itemize}

    \item Non-covalent interaction energies (S19):
    \begin{itemize}
        \item We selected 19 interaction energies from the S66 dataset \cite{rezac2011s66}.
    \end{itemize}

    \item Bulk solid datasets:
    \begin{itemize}
        \item 24 lattice constants, 24 bulk moduli, and 24 atomization energies of non-oxides \cite{harl2010Ecorr,zhang2018bulk_prop}.
        \item 23 oxide formation energies \cite{yan2013rpa_oxde,voss2022oxide}.
        \item Magnetic moments of magnetic solids (Fe, Co, and Ni) \cite{wijn1997bulk_mag_expt}.
        \item Relative stability of bulk phases of Co (hcp, fcc, and bcc phases), \ce{MoO3} ($\alpha$ and $\beta$ phases), and \ce{TiO2} (rutile and anatase phases).
    \end{itemize}

    \item Surface datasets:
    \begin{itemize}
        \item \Nads experimental adsorption energies on transition metal surfaces derived from ADS41 \cite{wellendorff2015ce39,duanmu2017ce39reref,mall2019ads41}.
        \item 7 surface reaction energies on oxide surfaces (5 CCSD(T)-level calculation values \cite{ye2024ccsdt_oxide,ye2024ccsdt_oxide2} and 2 experimental values \cite{shi2025autoSKZCAM}).
        \item Surface energies of four transition metal surfaces with (111) facet \cite{vitos1998Esurf_expt}.
    \end{itemize}

\end{itemize}

\subsection{Computational Details}

All periodic calculations were performed using the Vienna \latin{Ab initio} Simulation Package (VASP) version 6.3.2, employing projector-augmented wave (PAW) pseudopotentials \cite{kresse1993vasp, kresse1996vasp1, kresse1996vasp2, kresse1999vasp_paw}.
GW-type PAW pseudopotentials (version 54) were used throughout. Specifically, the pseudopotentials recommended by VASP \cite{paw_vasp} were employed, except for C, N, O, and F, for which ``\texttt{GW\_new}'' pseudopotentials were used.
Another exception was for metal slab calculations, where GW pseudopotentials with the fewest valence electrons were selected. 
The GW pseudopotentials used for the metal slabs do not include semi-core s and p states in Ni, Cu, Rh, Pd, and Pt, while those for Ru and Ir include them.

Spin polarization was considered only for magnetic materials, including Fe, Ni, and Co–containing systems and alkali-metal superoxides (\ce{NaO2}, \ce{KO2}, \ce{RbO2}, and \ce{CsO2}).
The fast Fourier transform (FFT) grids for GGA (\texttt{PREC}) and exact exchange (\texttt{PRECFOCK}) were set to ``Accurate'' and ``Normal'', respectively, with the exception of calculations for non-oxide bulk systems (see Section~S\ref{sec:SI_methods_bulk} of the SI).

For DFT calculations, the following dispersion-corrected DFT functionals were used: PBE$+$D3 (GGA) \cite{perdew1996pbe,grimme2011d3bj}, r2SCAN$+$rVV10 (meta-GGA) \cite{sabatini2013rvv10,galimberti2021r2scan,ning2022r2scan_rvv10}, and HSE06$+$D3 (hybrid) \cite{krukau2006hse06,grimme2011d3bj}.  
All other calculation parameters were identical to those used for RPA calculations, unless otherwise specified. Structural relaxations were performed using PBE$+$D3, and the same structures were used for single-point calculations with all functionals.

For RPA calculations, a cubic-scaling algorithm was used \cite{kaltak2014rpa_cubic} with a frequency integration grid containing 24 points. We used an empirically constructed custom hybrid functional, \srpbe, to generate DFT orbitals for evaluating the RPA correlation energy. This functional uses a long-range corrected hybrid scheme with a range separation parameter ($\mu$) of \SI{3.0}{\per\angstrom} and 25\% exact exchange. 
The GGA xc component was described using the short-range PBE (srPBE) functional \cite{toulouse2005srpbe,goll2005srpbe,goll2006srpbe} with the default range separation parameter of \SI{0.94}{\per\angstrom} (\SI{0.5}{\per\Bohr}) for both exchange ($\mu$) and correlation ($\mu_c$) via the Libxc library (version 7.0.0) \cite{lehtola2018libxc}. 
The \srpbe functional includes 25\% long-range (lr) exact exchange, 75\% srPBE exchange, and 500\% srPBE correlation (see Section~\ref{sec:discuss_orbital} in the Discussion), as defined below. 
This empirically constructed functional is used solely to generate Kohn--Sham orbitals, which have been found to serve as suitable inputs for RPA energy evaluation. Its exchange-correlation energy is calculated as:
\begin{equation}
    \label{eqn:Exc_srpbe}
    E_{\text{xc,\,\srpbe}} = (0.25 \cdot E_{\text{x,EXX}}^{\text{lr},\mu=3.0\text{\AA}^{-1}} + 0.75 \cdot E_{\text{x,srPBE}}^{\text{sr},\mu=0.94\text{\AA}^{-1}}) + 5.00 \cdot E_{\text{c,srPBE}}^{\text{sr},\mu_c=0.94\text{\AA}^{-1}}
\end{equation}
where $E_{\text{x,EXX}}^{\text{lr},\mu=3.0\text{\AA}^{-1}}$ is long-range exact exchange, $E_{\text{x,srPBE}}^{\text{sr},\mu=0.94\text{\AA}^{-1}}$ is short-range PBE exchange, and $E_{\text{c,srPBE}}^{\text{sr},\mu_c=0.94\text{\AA}^{-1}}$ is short-range PBE correlation.

The total energy in the standard RPA calculation is defined as
\begin{equation}
    E_{\text{total}}^{\text{RPA}} = \EHXX + \Ecrpa
\end{equation}
where \EHXX is HF energy (HF$@$PBE) and \Ecrpa is the RPA correlation energy (RPA$_{\text{c}}@$PBE), both evaluated using PBE orbitals.
In \optVo \cite{kim2024nitrogen}, the previous version of \optVt, the total energy is defined as
\begin{equation}
    E_{\text{total}}^{\,\text{\optVo}} = \EHXX + r_c \cdot \Ecrpa
\end{equation}
where \EHXX (HF$@$PBEx50) and \Ecrpa (RPA$_{\text{c}}@$PBEx50) are evaluated using PBEx50 (a PBE global hybrid with 50\% EXX) orbitals, and $r_c$ (=1.180) is a scaling constant for \Ecrpa. 
In \optVt, the total energy is defined as
\begin{equation}
    \label{eqn:Etotal}
    E_{\text{total}}^{\,\text{\optVt}} = \EoptHXX + r_c \cdot \Ecrpa
\end{equation}
where \EoptHXX is the modified \HXX energy that includes a non-RPA correlation contribution of the total energy. Both \EoptHXX (\text{\optHXX}$@$\srpbe) and \Ecrpa (RPA$_{\text{c}}@$\srpbe) are evaluated using the same \srpbe orbitals. 
Each term in Eq. \eqref{eqn:Etotal} is obtained from a separate single-point calculation.
The xc part of \optHXX consists of 100\% full-range EXX ($E_{\text{x,EXX}}$) and 20\% srPBE correlation ($E_{\text{c,srPBE}}^{\text{sr},\mu_c=1.89\text{\AA}^{-1}}$) with a range separation parameter for DFT correlation of \SI{1.89}{\per\angstrom} (=\SI{1.0}{\per\Bohr}):
\begin{equation}
    \label{eqn:optHXX}
    E_{\text{xc,\optHXX}} = E_{\text{x,EXX}} + 0.20 \cdot E_{\text{c,srPBE}}^{\mu_c=1.89\text{\AA}^{-1}}
\end{equation}
The srPBE correlation was introduced to reduce $r_c$ for \optVt (from 1.114 to 1.02), which improves energy convergence (see Section~\ref{sec:discuss_etc} in the Discussion). 
We did not include a correction to the non-RPA correlation part for partial occupancies \cite{harl2010Ecorr}, as it did not help energy convergence (\Cref{fig:conv_bulk_surf}).


For RPA calculations, we applied a plane-wave energy cutoff (\Ecut) correction term \dEcut for all calculations, except for molecular datasets and non-oxide bulk solids, for which a sufficiently large \Ecut of 500--600 eV can be used without this correction. The correction is defined as:
\begin{equation}
    \dEcut = \Delta E_{\text{low-k, high-E}} - \Delta E_{\text{low-k, low-E}}
    \label{eqn:dEcut}
\end{equation}
The \dEcut is the difference in the calculated energy change $\Delta E$ (e.g., reaction energy) obtained using low and high \Ecut (low-E and high-E, respectively) with a coarse k-point grid (low-k). The correction is added to $\Delta E_{\text{high-k, low-E}}$ to obtain an accurate energy $\Delta E_{\text{high-k, high-E}}$ at a dense k-point grid (high-k),
\begin{align}
    \label{eqn:Ecut}
    \Delta E_{\text{high-k, high-E}} &\approx \Delta E_{\text{high-k, low-E}} + \dEcut
\end{align}
The \dEcut has been reported to be largely independent of k-point meshes and can recover $\Delta E_{\text{high-k, high-E}}$ almost perfectly \cite{jauho2015rapbe_oxide, schmidt2018benchmarkRPA,sheldon2021rpa_ads, oudot2024rpa_ads} (\Cref{fig:conv_ecut}).

For RPA calculations of molecules involved in bulk solid and surface datasets, we extrapolated \EoptHXX and \Ecrpa with respect to cell volumes ($V$) to the limit of infinitely dilute gas, using the form $E = E_{\infty} + \alpha V^{-1} + \beta V^{-2}$, where $E_{\infty}$ is the energy of a molecule in an infinitely large cell \cite{schafer2021fs_mol} (\Cref{fig:fs_atom,fig:fs_mol}).
%
$E_{\infty}$ is correlated with the total number of electrons ($N$) and the number of valence electrons ($N_\mathrm{v}$) in gas species (\Cref{fig:fs_mol_fit}). 
Therefore, the volume dependence of molecular energies should be considered for gas species with a large number of electrons, such as I and \ce{CH3I}, to obtain accurate adsorption energies.
For DFT calculations, a large cubic cell (\SI{20}{\angstrom}) was used without applying the extrapolation scheme. The energy difference between HSE06 results using \SI{15}{\angstrom} and \SI{20}{\angstrom} cells was less than 0.001 eV. The Coulomb divergence of exact exchange was corrected using the probe-charge method \cite{gygi1986hfrcut0,massidda1993hfrcut0} ($\texttt{HFRCUT}$=0 in VASP, which is the default), except for bulk calculations used for surface energy evaluations (see Section~S\ref{sec:SI_methods_surf} of the SI).

More computational details can be found in Sections~S\ref{sec:SI_methods} and S\ref{sec:SI_note_RPA} of the SI.

\subsection{\label{sec:aa}Adsorption energy definitions}

Evaluating the thermochemistry of adsorbates is non-trivial, since both gas-phase and surface properties are involved, and it is not always straightforward to establish whether error occurs due to the gas-phase reference state or the adsorbed state. Several definitions of adsorption energies exist in the literature \cite{kreitz2025thermo}, and here we adopt the following three definitions:

\begin{equation}
    \label{eqn:Eads}
    E_{\mathrm{ads}} = E_{\mathrm{slab+ads}} - E_{\mathrm{slab}} - E_{\mathrm{ads(g)}}
\end{equation}
\begin{equation}
\label{eqn:Ediss}
E_{\mathrm{diss}} = \sum_i \nu_i (E_{\mathrm{slab+ads},i} - E_{\mathrm{slab}}) - E_{\mathrm{mol}}
\end{equation}
\begin{equation}
\label{eqn:EadsLS}
    E_{\mathrm{ads,LS}} = E_{\mathrm{slab+ads}} - E_{\mathrm{slab}} - \sum_k n_k \mu_{\mathrm{LS}, k}
\end{equation}
where $E_{slab+ads}$ is the combined slab and adsorbate system, $E_{slab}$ is the bare slab, $E_{ads(g)}$ is the adsorbate directly in the gas-phase, 
$E_{mol}$ is a gas-phase molecule that dissociates into $E_{slab+ads,i}$, $\nu_i$ denotes the stoichiometric coefficient of adsorbate $i$ in dissociative adsorption,
$n_k$ is the number of elemental species $k$ in the adsorbate, and $\mu_{LS,k}$ is a chemical potential of an elemental species that is obtained through least-squares regression \cite{kreitz2025thermo}. 

The first definition is the most straightforward, as it represents non-dissociative adsorption. 
However, significant errors can be introduced because the gas-phase analogues of the adsorbates may be unstable open-shell radicals with complex electronic structure not well captured by DFT (e.g. bare O atoms, CH fragments, etc.). 
We obtained the reference values for the first definition by applying the same approach to ADS41 as in Ref.~\citenum{duanmu2017ce39reref}, but using different molecular energies obtained from the AtCT database\cite{atct2025} and zero point energies (ZPE) from literature\cite{irikura2007zpe_diatom,shin201zpe_C,khiri2018soc_zpe_I,cccbdb}. For adsorption of CH and \ce{CH3}, ZPE-corrected experimental values were used \cite{karp2013CH3ads,wolcott2014CHads}. 
 
The second definition minimizes the use of open-shell reference states by treating the adsorption event as a dissociation of a stable gas-phase species, and normalizing the energy to be per mole of adsorbate ($E_{diss}(H)=\Delta E$/2, considering \ce{H2}$\to$2H). 
One advantage of referencing \Eads to molecules is that spin–orbit coupling (SOC) effects are diminished. For example, the I atom has a spin–orbit stabilization energy of -0.31 eV, whereas the value for the \ce{I2} molecule is -0.09 eV \cite{isaacs2022soc_I2}. Given that SOC effects are not accounted for in the methods investigated here, we applied SOC correction terms to all calculated energies involving I-containing species in both adsorption definitions. 
Still, this approach does not guarantee reduced gas-phase error because some molecules are difficult to describe accurately using DFT, such as \ce{O2} (whose triplet state is the ground state).     
In cases where the adsorbate does not dissociate (e.g., CO, NO, \ce{CH4}, etc.) the two definitions are identical. For other species, conversion between the definitions requires knowledge of the gas-phase molecular energies. For the \embed method, only the first definition ($E_{ads}$) was reported in the original paper, and gas-phase energies were not reported. Therefore, we calculated molecular energies at the same level of theory, using ORCA version 5.0.4 \cite{neese2012orca} and the def2-QZVPP basis set to convert between the two \cite{weigend2005qzv}: 
\begin{equation}
    E_\text{gas}^\text{\embed}=E_\text{gas}^\text{PBE+D3,\,periodic} \ 
    +(E_\text{gas}^\text{M06,\,cluster}-E_\text{gas}^\text{PBE+D3,\,cluster})  \nonumber
\end{equation}
where $E_\text{gas}$ includes both $E_\text{ads(g)}$ and $E_\text{mol}$, $E_\text{gas}^\text{PBE+D3,\,periodic}$ is molecular energies at PBE+D3 level under periodic boundary conditions (calculated using VASP), and $E_\text{gas}^\text{M06,\,cluster}$ and $E_\text{gas}^\text{PBE+D3,\,cluster}$ are the corresponding energies computed in a cluster model at M06 and PBE+D3 levels (calculated using ORCA).

The third definition of adsorption energies, $E_{\mathrm{ads, LS}}$, avoids issues of reference state error by obtaining reference chemical potentials using least-squares (LS) regression. 
The $\mu_{\mathrm{LS},i}$ are determined by minimizing the squared magnitude of the adsorption energy $E_i$ \cite{kreitz2025thermo}:

\begin{equation}
\label{eqn:mu_ls}
\boldsymbol{\mu}_{\mathrm{LS}}
=
\underset{\boldsymbol{\mu}}{\mathrm{arg\,min}}\, 
\sum_{i=1}
\left(E_i - \sum_{k=1} n_{i,k} \mu_k
\right)^2
\end{equation}
%
where $\boldsymbol{\mu}$ is a ($k \times 1$) vector of elemental chemical potentials $\mu_k$, $E_i$ is QM energy of species $i$, and $n_{i,k}$ is the number of element $k$ in species $i$ (and includes only elements in the adsorbates).
In this case, $E_i$ corresponds to $(E_{\mathrm{slab+ads}} - E_{\mathrm{slab}})$ for computed adsorption energies and ($E_{\mathrm{ads}} + \Delta E_{\mathrm{f,\,ads(g)}}$) for experimental adsorption energies, where $\Delta E_{\mathrm{f,\,ads(g)}}$ is the formation energy of gas-phase adsorbate (i.e., the formation enthalpy at 0 K with ZPE contributions removed) obtained from AtCT.
This minimization can be efficiently solved using linear algebra:
\begin{equation}
\boldsymbol{\mu}_{\mathrm{LS}}
=
\left(\mathbf{N}^{\mathrm{T}}\mathbf{N}\right)^{-1}
\mathbf{N}^{\mathrm{T}}\mathbf{E}
\label{eqn:mu_pinv}
\end{equation}
\begin{align}
\boldsymbol{\mu}_{\mathrm{LS}} =
\begin{pmatrix}
\mu_{\mathrm{LS,1}} \\ \mu_{\mathrm{LS,2}} \\ \vdots \\ \mu_{\mathrm{LS,k}}
\end{pmatrix},\,\,
\mathbf{N} =
\begin{pmatrix}
n_{1,1} & n_{1,2} & \cdots & n_{1,k} \\
n_{2,1} & n_{2,2} & \cdots & n_{2,k} \\
\vdots & \vdots & \ddots & \vdots \\
n_{i,1} & n_{i,2} & \cdots & n_{i,k}
\end{pmatrix},\,\,
\mathbf{E} =
\begin{pmatrix}
E_1 \\ E_2 \\ \vdots \\ E_{i}
\end{pmatrix}
\end{align}
where $\mathbf{N}$ is a stoichiometry matrix ($ i\times k$) with elements $n_{i,k}$ and $\boldsymbol{E}$ is an ($i \times 1$) vector of QM energies $E_i$ of each species.
This approach eliminates the need for any gas-phase reference, and can be shown to minimize the squared errors between energies from different sources \cite{kreitz2025thermo}. However, it introduces one degree of freedom for each element involved and can thus underestimate the error for datasets with few examples of an element (e.g. iodine). It also removes all systematic error by construction, so that systematic over- or under-binding of molecules will not show up in this definition since the error will be pushed into the reference states. Finally, the specific values of $\mu_{\mathrm{LS},i}$ are also dependent on the dataset used, and it is thus not a physically meaningful quantity in general. 
Nonetheless, it provides a useful perspective on the adsorption error since it directly probes the error that arises from the surface and adsorbate and establishes a lower bound on a practical adsorption energy error.

\section{\label{sec:results}Results}

The performance of the proposed \optVt method was benchmarked against the various datasets introduced earlier. For comparison, we also evaluated representative DFT functionals (PBE$+$D3, r2SCAN$+$rVV10, and HSE06$+$D3) and included data from higher-level methods from the literature where available.
We report energy errors using mean absolute deviation (MAD), root mean square deviation (RMSD), and maximum absolute deviation. For lattice constants and bulk moduli, mean absolute percentage deviation (MAPD), root mean square percentage deviation (RMSPD), and maximum absolute percentage deviation are reported.

\subsection{\label{sec:}Molecular Datasets}

\begin{figure} [H]
    \centering
    \includegraphics[width=0.97\textwidth]{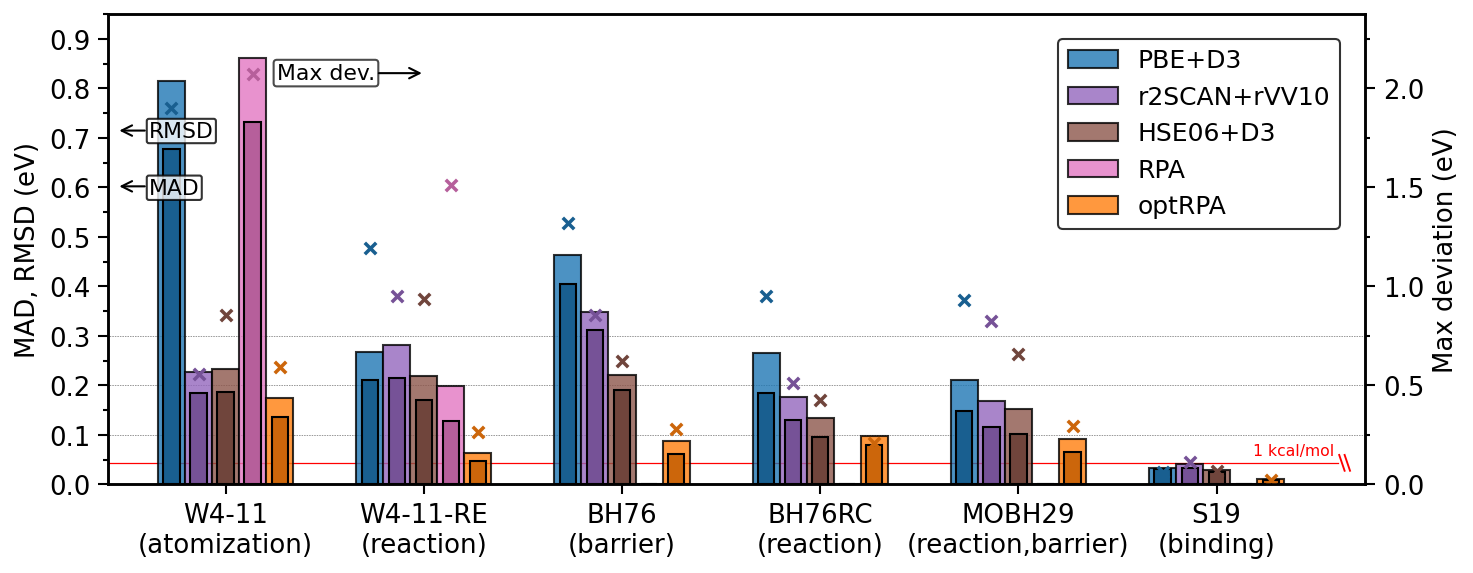}
    \caption{
    Absolute deviations from reference values for each molecular dataset. Datasets include W4-11 (n=124$^*$)\cite{karton2011w411}, W4-11-RE (n=8,868$^*$)\cite{margraf2017w411re}, BH76 (n=76)\cite{zhao2005BH76a,zhao2005BH76b,lars2010gmtkn24}, BH76RC (n=30)\cite{lars2010gmtkn24}, MOBH29 (n=87) \cite{iron2019mobh35,dohm2020mobh29,semidalas2022mobh35rev,grotjahn2023mobh28}, and S19 (n=19)\cite{rezac2011s66}.
    Bars represent MAD (dark color) and RMSD (light color), and \texttt{×} symbols indicate maximum deviations. A horizontal red line denotes the chemical accuracy of \SI{1}{kcal\per\mol}.
    $^*$W4-11 and W4-11-RE contain only the molecules listed in TAE\_nonMR124.
    }
    \label{fig:mol}
\end{figure}

Benchmarking results for different molecular datasets are presented in \Cref{fig:mol}. For W4-11 (atomization energy), PBE$+$D3 showed a significantly higher RMSD (0.81 eV) than other DFT functionals, but performed comparably for W4-11-RE (reaction energy) with an RMSD of 0.27 eV, suggesting substantial error cancellation. 
Similar behavior was observed for RPA, with RMSDs of 0.86 eV (W4-11) and 0.20 eV (W4-11-RE). The largest errors of RPA in W4-11-RE were associated with the HS molecule, and the comparable performance of RPA and PBE$+$D3 on gas-phase properties is consistent with prior benchmarking for ethane dehydrogenation \cite{szaro2023rpa_ads_c2}.
For BH76 (barrier height), the accuracy of DFT functionals followed their hierarchy on Jacob's ladder, with RMSDs of 0.46 eV (PBE$+$D3), 0.35 eV (r2SCAN$+$rVV10), and 0.22 eV (HSE06$+$D3). Similar trends were observed for BH76RC (reaction energy) and MOBH29 (barrier height and reaction energy of TM complex), albeit with smaller differences among the functionals. 
DFT methods showed mutually similar accuracy for S19 (non-covalent binding energy), with RMSDs of 0.03--0.04 eV.
The \optVt method consistently outperformed all tested methods, yielding 
RMSDs of 0.17 eV (W4-11), 0.06 eV (W4-11-RE), 0.09 eV (BH76), 0.10 eV (BH76RC), 0.09 eV (MOBH29), and 0.01 eV (S19). 
Notably, the maximum error for \optVt on W4-11-RE is 0.27 eV, which is substantially lower than all other methods ($\sim 1$ eV), indicating that the approach will be generally reliable for reaction energies, with similar trends observed for barrier heights. 




%
%
%
%
%

\subsection{\label{sec:}Bulk Solid Datasets}

\subsubsection{\label{sec:}Lattice constant and bulk modulus}

\begin{figure} [H]
    \centering
    \includegraphics[width=0.6\textwidth]{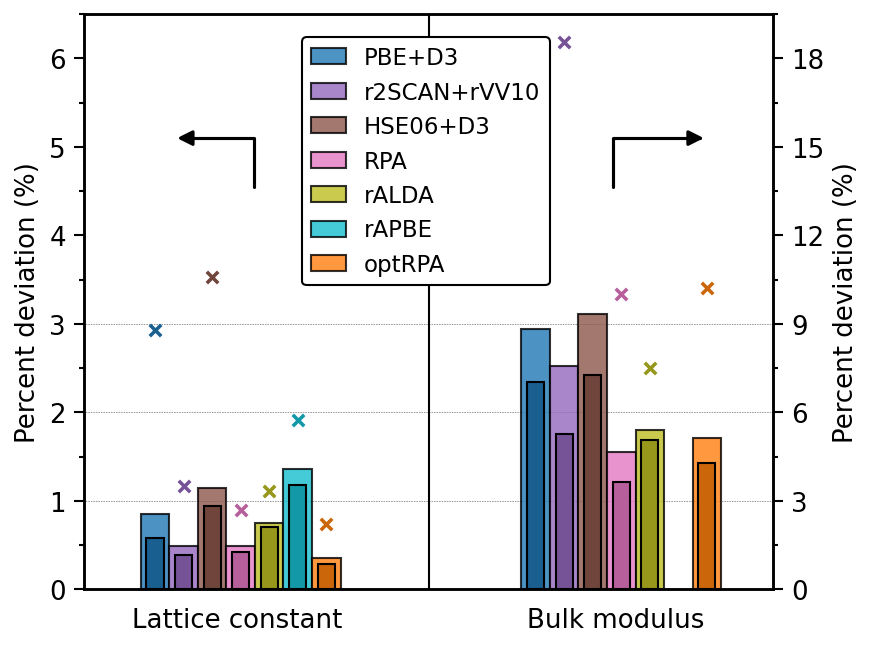}
    \caption{
    Absolute percentage deviations from ZPE-corrected experimental values for lattice constants \cite{harl2010Ecorr} and bulk moduli \cite{harl2010Ecorr,zhang2018bulk_prop} for 24 (non-oxide) bulk solids. 
    RPA (n=24) \cite{harl2010Ecorr}, rALDA (n=10) \cite{patrick2015ralda_bulk}, and rAPBE (n=4) \cite{olsen2014rA} values are taken from the literature.
    Bars represent MAPD (dark color) and RMSPD (light color), and \texttt{×} symbols indicate maximum percentage deviations.
    }
    \label{fig:bulk}
\end{figure}

The calculated bulk lattice constants and bulk moduli for non-oxide solids are presented in \Cref{fig:bulk}.
In both lattice constants and bulk moduli, \optVt showed good accuracy with RMSPD of 0.37\% and 5.16\%, respectively, comparable to the best-performing methods, RPA (0.48\% and 4.64\%) and rALDA (0.75\% and 5.39\%). 
The accuracy of rAPBE is difficult to assess because only four data points are available (for lattice constants only).
PBE$+$D3 delivered good performance for lattice constants (RMSPD of 0.85\%) and bulk moduli (8.83\%), but the D3 correction was necessary because, without it, RMSPD increased to 1.64\% and 14.51\%, respectively.
Among DFT methods, r2SCAN+rVV10 produced the most accurate results, outperforming HSE06+D3.

\subsubsection{\label{sec:}Atomization energy of non-oxide and oxide formation energy}

\begin{figure} [H]
    \centering
    \includegraphics[width=0.6\textwidth]{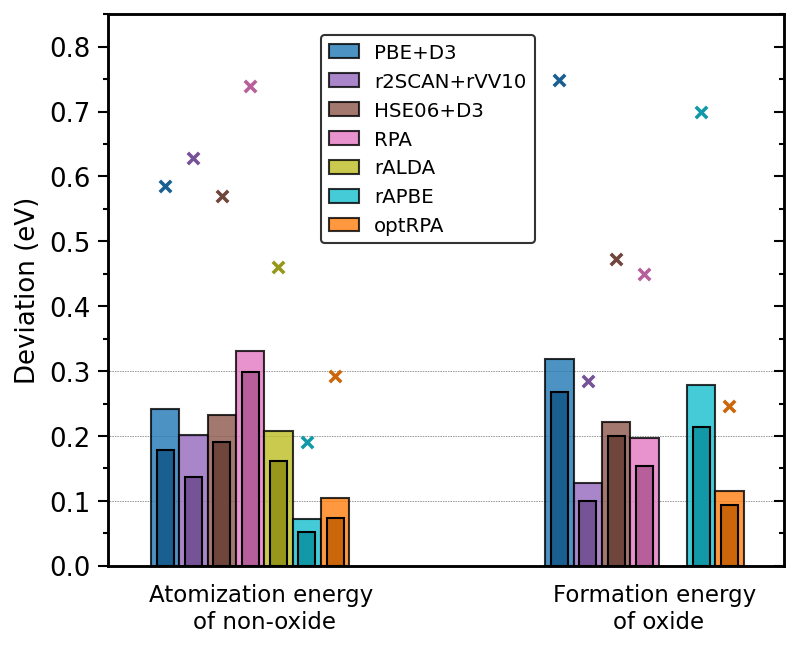}
    \caption{
    Absolute deviations from (ZPE-corrected) experimental atomization energies (eV per formula unit) of 24 non-oxides \cite{harl2010Ecorr} and (non-ZPE-corrected) formation energies (eV per oxygen) of 23 oxides \cite{yan2013rpa_oxde,voss2022oxide}. 
    For atomization energies, RPA values (n=24) are from Ref.~\citenum{harl2010Ecorr}, and rALDA (n=19) and rAPBE (n=19) are from Ref.~\citenum{olsen2014rA}.
    For oxide formation energies, RPA values (n=22) are from Ref.~\citenum{yan2013rpa_oxde}, and rAPBE values (n=21) are from Ref.~\citenum{jauho2015rapbe_oxide}.
    Bars represent MAD (dark color) and RMSD (light color), and \texttt{×} symbols indicate maximum deviations. 
    }
    \label{fig:oxide}
\end{figure}

The calculated atomization energies of non-oxides and oxide formation energies are presented in \Cref{fig:oxide}.
\optVt performed well for both datasets with RMSDs of 0.10 eV (atomization energy) and 0.12 eV (oxide formation energy), comparable to the best-performing methods for each dataset, which are rAPBE (0.07 eV) for atomization energy and r2SCAN+rVV10 (0.13 eV) for oxide formation energy, respectively. 
As observed earlier for lattice constants and bulk moduli, r2SCAN+rVV10 again showed the highest accuracy for the bulk properties among DFT methods. The reference values for oxide formation are not ZPE-corrected, but this effect is reported to change the MAD of the current dataset by, at most, 0.01 eV \cite{yan2013rpa_oxde}.

\subsubsection{\label{sec:}Magnetic moments of metals}

We evaluated the magnetic moments of elemental ferromagnets (Fe, Co, and Ni) to assess the performance of different functionals in capturing the correct spin polarization, as shown in \Cref{tab:bulk_mag}. 
PBE values show good agreement with experimental values, while HSE06 (and hybrid functionals in general) tends to significantly overestimate the spin magnetic moments, with a similar trend observed for the meta-GGA r2SCAN. 
In contrast, \srpbe reproduces the correct experimental magnetic moments, similar to PBE. We note that this property depends only on the \srpbe method, not \optVt itself, but it confirms that the orbitals used in \optVt for spin-polarized metals have qualitatively accurate magnetic properties.

\input{ACS/tables/bulk_mag.tex}

\subsubsection{\label{sec:}Relative stability of bulk solids}

We evaluated the relative stability of bulk polymorphs for Co, \ce{MoO3}, and \ce{TiO2}, where standard DFT often fails to predict the correct ground-state phase \cite{jang2012Co_dft,labat2007tio2_dft,zhang2019MoO3_dft}. 
\optVt correctly predicts the most stable phases for all systems considered, albeit by very small energy differences, as shown in \Cref{tab:bulk_rel}. We note that it is possible that the experimentally observed ground-state phase may not have the lowest energy at \SI{0}{K}, as thermal effects may play a role \cite{luo2016tio2_dmc}.

\input{ACS/tables/bulk_rel.tex}

\subsection{\label{sec:}Surface Datasets}

\subsubsection{\label{sec:}Surface reactions on metals}

The performance of \optVt was benchmarked for adsorption energies on metallic surfaces using experimental values\cite{duanmu2017ce39reref,mall2019ads41}, as shown in \Cref{fig:ads_metal}.
A total of \Nads surface reactions were investigated on non-magnetic Cu(100), Cu(111), Ru(0001), Rh(100), Rh(111), Pd(100), Pd(111), Ir(111), and Pt(111) surfaces, as well as on the magnetic Ni(111) surface.

In \Cref{fig:ads_metal}(a), the reaction energies from ADS41 were recalculated to represent non-dissociative adsorption energies, as defined in Eq. \eqref{eqn:Eads}. We compare the results to other density functionals as in prior sections, as well as \embed.
The \embed method has been reported as one of the most accurate approaches for adsorption energy calculations, showing the lowest RMSD of 0.13 eV for the current set. 
The \optVt approach achieved a comparable accuracy with an RMSD of 0.16 eV, followed by RPA (0.25 eV).
Despite substantially different computational costs, various other DFT methods showed similar accuracies with RMSDs of 0.49 eV (PBE+D3), 0.48 eV (r2SCAN+rVV10), and 0.53 eV (HSE06+D3), consistent with results from ADS41 \cite{mall2019ads41}. 
RPA performed significantly better than DFT methods, confirming its high accuracy for adsorption systems, as previously reported \cite{schmidt2018benchmarkRPA,sauer2024rpa_review}, and further supported here using a larger dataset.
Despite the good performance of RPA for adsorption energies, the prior benchmarking study of ethane dehydrogenation on Pt(111) showed that RPA was not superior in predicting macroscopic observables through microkinetic modeling, compared to other DFT functionals \cite{szaro2023rpa_ads_c2}. The reduced accuracy of these macroscopic observables is likely related to the lower accuracy of RPA for molecular energies and, in part, to the neglect of the volume dependence of molecular energies (\Cref{fig:fs_mol}).

In \Cref{fig:ads_metal}(b), the same adsorption energy results are reported using dissociative surface reaction energies, as defined in Eq. \eqref{eqn:Ediss}. 
The results did not change significantly, and post-DFT methods showed RMSDs of 0.13 eV (\embed), 0.13 eV (\optVt), and 0.19 eV (RPA), while standard DFT methods exhibited RMSDs of 0.44 eV (PBE+D3), 0.46 eV (r2SCAN+rVV10), and 0.54 eV (HSE06+D3).
PBE+D3 and RPA values differed the most ($\sim$0.05 eV) between the two adsorption energy definitions, reflecting their large errors in molecular atomization energies, revealing a significant conflation of gas-phase and adsorbed-state errors in adsorption energies. 

To deconvolute these contributions, the least-squares adsorption energies (Eq. \eqref{eqn:EadsLS}) are shown in \Cref{fig:ads_metal}(c).
These adsorption energies avoid any explicit gas-phase reference by determining elemental chemical potentials by minimizing the squared magnitude of the relative reaction energies ($E_{\mathrm{slab+ads}} - E_{\mathrm{slab}}$) in the dataset. The resulting elemental potentials will implicitly depend on the compositions of all adsorbed species, but does not rely on any computed gas-phase energies and is thus purely a dependent on the surface-adsorbate interactions  \cite{kreitz2025thermo}. 
The least-squares adsorption energies minimize bias, giving a zero mean signed error (MSD), whereas in the former two definitions, DFT methods give MSDs that are similar in magnitude to the MAD values (\Cref{fig:ads_metal1_noD}), indicating systematic error in the surface--adsorbate interactions or gas-phase energies. 
The strong overbinding trend is reduced when excluding the dispersion correction, leading to lower RMSDs for all DFT functionals (\Cref{fig:ads_metal1_noD}). 
The least-squares approach yields RMSDs of 0.12 eV (\embed), 0.11 eV (\optVt), and 0.17 eV (RPA), with corresponding DFT errors of 0.18 eV (PBE+D3), 0.17 eV (r2SCAN+rVV10), and 0.28 eV (HSE06).
The result is different if the least-squares approach is used to correct gas-phase energies only, in which case the gas-phase least-squares RMSDs are always lower than the RMSDs for adsorption energies, indicating that slab-adsorbate errors dominate for all methods (\Cref{tab:ls_rmsd}).
Excluding the dispersion correction for DFT functionals leads to very similar RMSDs, confirming that the least-square approach reduces the systematic bias (\Cref{tab:ls_rmsd}).
Interestingly, HSE06+D3 is consistently outperformed by other DFT methods across the different referencing schemes, likely due to issues of hybrid functionals in treating metallic systems \cite{spencer2008singularity,sund2013singularity}.
The strong performance of \embed and \optVt across all referencing schemes reflects their low gas-phase errors, and suggests that methods which maximize error cancellation between gas-phase and adsorbed species are promising for reducing adsorption energy errors \cite{kreitz2023error_cancel, porter2026error_cancel}. 
However, for calculations that intrinsically involve both gas-phase and surface-mediated interactions, such as transition states of dissociative adsorption, methods like \optVt or \embed with inherently high gas-phase accuracy will be required.

We note a substantial difference in the adsorption energies of O and CH between our PBE+D3 values and the PBE+D3 values from Ref.\citenum{araujo2022ads38}.
The energy offsets for O adsorption were consistent at 0.27–0.29 eV, with the values from Ref.\citenum{araujo2022ads38} indicating stronger adsorption, which coincides with half of the \ce{O2} energy difference between the doublet and triplet states (0.33 eV). 
When converting adsorption energies from Ref.\citenum{araujo2022ads38} to the second definition, the PBE+D3 values for O adsorptions in Ref.\citenum{araujo2022ads38} also show a similar energy shift relative to the PBE values in ADS41\cite{mall2019ads41} (beyond D3 effects) and to the PBE+D3 values for O/Pt(111) from the literature \cite{gautier2015Oads,wexler2023Oads}. 
A similar discrepancy was found for CH adsorption, with a larger energy difference of 0.59 eV. 
It is possible that these exaggerated adsorption energies may have propagated into the \embed values, and using the corrected energies could increase the error in the adsorption energy of the method by $\sim$0.3 eV for these adsorbates. 
However, these issues should not affect the least-squares results in Fig. \ref{fig:ads_metal}c, where the effects of gas-phase reference states have been removed. The similar RMSDs for E$_{\mathrm{ads,LS}}$ values at PBE+D3 level support this statement, with the values being 0.135 eV (Ref.\citenum{araujo2022ads38}) and 0.132 eV (our calculations).
We also note a structural aspect of \ce{CH3I} adsorption on Pt(111). In previous benchmarking studies of adsorption on metal surfaces \cite{wellendorff2015ce39,duanmu2017ce39reref,mall2019ads41,araujo2022ads38}, \ce{CH3I} was modeled as adsorbing perpendicular to Pt(111). However, experimental studies have shown that \ce{CH3I} adsorbs in a tilted configuration \cite{zaera1991CH3I_ads,fan1994CH3I_ads,french1995CH3I_ads}. Our DFT calculations using the tilted geometry predict that this configuration adsorbs more strongly by 0.35 eV at PBE+D3 level, compared to the perpendicular configuration. The tilted geometry is more stable with all functionals, including \optVt, but this stronger adsorption increases error compared to experiment for functionals that already overbind \ce{CH3I}, while it reduces the error for \optVt.  

\input{ACS/figures/ads/ads_metal.tex}

\begin{figure} [h!]
    \centering
    \includegraphics[width=0.55\textwidth]{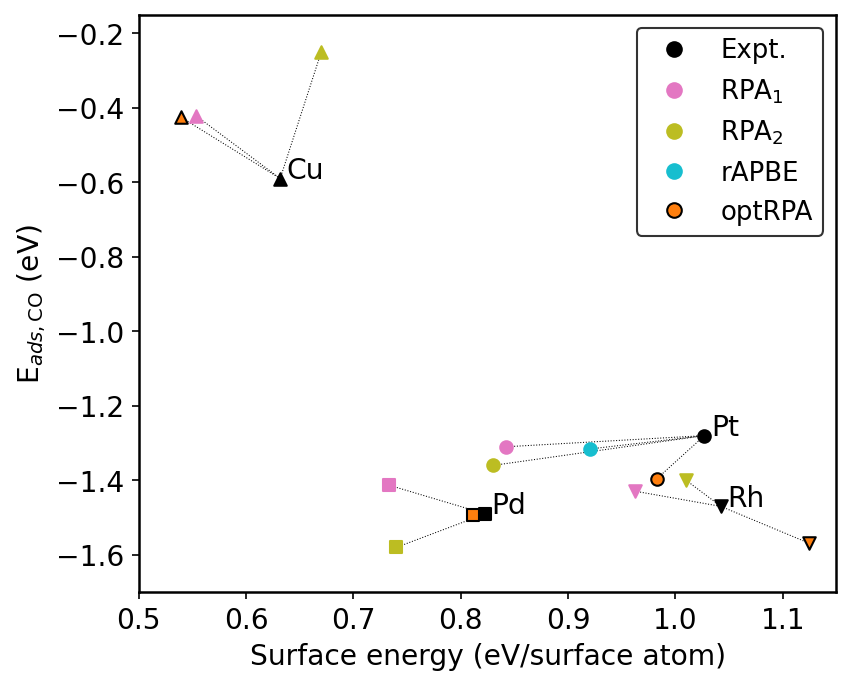}
    \caption{
    Surface energies and CO adsorption energies on four TM surfaces. The \optVt values for \Eads correspond to those in \Cref{fig:ads_metal}. The values of RPA$_\text{1}$ \cite{schimka2010rpa_surf}, RPA$_\text{2}$ \cite{schmidt2018benchmarkRPA}, rAPBE\cite{olsen2014rA}, and experiments\cite{vitos1998Esurf_expt} are taken from the literature.
    }
    \label{fig:Esurf}
\end{figure}

Finally, there is a well-known tradeoff between the accuracy of surface energies and adsorption energies for TM surfaces treated with standard density functionals, and RPA methods are known to perform well on both quantities \cite{schimka2010rpa_surf, olsen2019rA_benchmark}. To evaluate whether \optVt is able to simultaneously capture surface and adsorption energies, we evaluated the surface energies of four TM (Cu, Rh, Pd, and Pt) surfaces with the (111) facet and plotted them against the CO adsorption energy (used in \Cref{fig:ads_metal}a-b), as shown in \Cref{fig:Esurf}. 
The post-DFT methods generally showed good agreement with experimental values of both surface and adsorption energies of CO, and \optVt achieved an accuracy on surface energies comparable to or better than other post-DFT methods.

\subsubsection{\label{sec:}Surface reactions on oxide}

\begin{figure} [h!]
    \centering
    \includegraphics[width=0.55\textwidth]{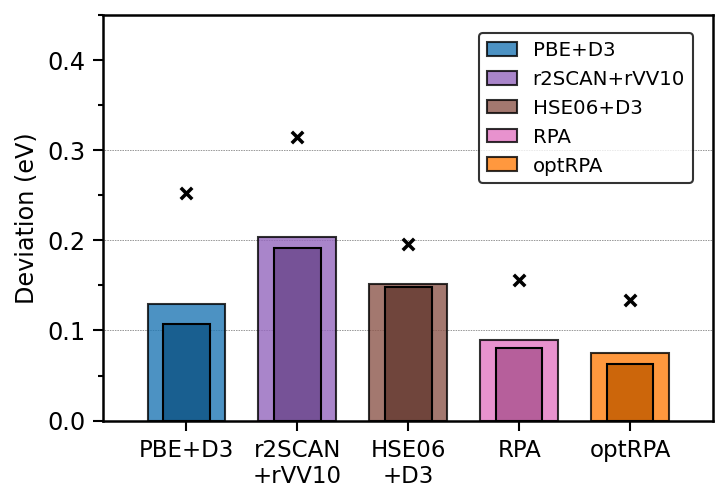}
    \caption{
    Absolute deviations from CCSD(T) values (n=5) \cite{ye2024ccsdt_oxide,ye2024ccsdt_oxide2} or ZPE-corrected experimental values (n=2) (for \ce{H2O} adsorption on MgO and \ce{TiO2})\cite{shi2025autoSKZCAM} for surface reaction energies on oxides (\ce{MgO}, \ce{Al2O3}, and \ce{TiO2}) (n=7).
    Bars represent MAD (dark color) and RMSD (light color), and \texttt{x} symbols indicate maximum deviations.
    }
    \label{fig:ads_oxide}
\end{figure}

We evaluated the performance of \optVt for surface reactions on oxide surfaces using reference values from local CCSD(T) \cite{ye2024ccsdt_oxide,ye2024ccsdt_oxide2} or experiments when available (for non-dissociative \ce{H2O} adsorption on MgO and \ce{TiO2})\cite{shi2025autoSKZCAM}, as presented in \Cref{fig:ads_oxide}. The dataset includes CO adsorption energy on MgO, and \ce{H2O} adsorption and dissociation energies and barrier heights on \ce{Al2O3} and \ce{TiO2} surfaces. The systems were selected because both high-level energies and corresponding structures are available.
A recent study of the autoSKZCAM framework \cite{shi2025autoSKZCAM} (an embedding method combining local CCSD(T)) provides high-quality structures and energies, that could serve as a basis for further benchmarking of \optVt in future work.

Within the set of DFT methods examined, PBE+D3 showed the lowest RMSD (0.13 eV), followed by HSE06+D3 (0.15 eV) and r2SCAN+rVV10 (0.20 eV).
DFT methods significantly underestimated barrier heights for \ce{H2O} dissociation on both the \ce{Al2O3} surface ($E_{a}^{\text{CCSD(T)}}$=0.38 eV), with DFT values of 0.13 eV (PBE+D3), 0.07 eV (r2SCAN+rVV10), and 0.19 eV (HSE06+D3), and the \ce{TiO2} surface ($E_{a}^{\text{CCSD(T)}}$=0.32 eV) with DFT values of 0.26 eV, 0.13 eV, and 0.17 eV, respectively.
The \optVt method showed the lowest error metrics (MAD of 0.06 eV and RMSD of 0.08 eV), but RPA performed comparably (0.08 and 0.09 eV). These RPA-based methods yield barrier heights closer to the CCSD(T) values, namely 0.34 eV (RPA) and 0.32 eV (\optVt) on the \ce{Al2O3} surface, and 0.46 eV (RPA) and 0.37 eV (\optVt) on the \ce{TiO2} surface.

\section{\label{sec:discussion}Discussion}

\subsection{\label{sec:discuss_orbital}Optimization of DFT orbitals}

In our previous study \cite{kim2024nitrogen}, we introduced an optimized RPA method ($\text{\optVo}$), which used PBEx50 (a global PBE hybrid with 50\% exact exchange), to calculate DFT orbitals. Although this approach worked well for gas-phase atomization and reaction energies and for non-oxide bulk properties, further testing revealed a formation energy error exceeding \SI{0.5}{eV} for the TM oxide \ce{RuO2} (despite an RMSD of around \SI{0.1}{eV} for formation energies of non-TM oxides). Moreover, the convergence of the PBEx50 wavefunctions was slow with respect to k-points for the adsorption energies in metals, creating a significant practical barrier to the large-scale application of \optVo. 
The convergence issue is likely related to multiple local minima observed in EXX calculations of metal surfaces \cite{mihm2021mb_conv,carbone2024CCSDcT}, even with tight energy convergence thresholds (\(10^{-8}\,\)eV).
In addition, the Coulomb singularity \cite{ashcroft1976physics} worsens the energy convergence of calculations involving EXX on metals.
Although singularity correction schemes have been successfully applied to approximate the $\mathbf{G}$=0 contribution in DFT hybrid calculations \cite{spencer2008singularity,sund2013singularity}, RPA is far more sensitive to changes in wavefunctions, making convergence particularly challenging. Finally, we found that incorporation of EXX led to quantitative and qualitative failures in the magnetic moments of magnetic metals. For example, PBE gives a magnetic moment of 2.12 $\mu_\mathrm{B}$ for bulk Fe, which is close to the experimental value of 2.13 $\mu_\mathrm{B}$ \cite{wijn1997bulk_mag_expt}, while hybrid functionals such as PBE0 and HSE06 predict $\sim$2.8 $\mu_\mathrm{B}$, consistent with previous reports \cite{tran2020mag_bulk,jana2020mag_bulk,voss2024Fe_mag}, and PBEx50 predicts the same value. Also, similar to other hybrid functionals, PBEx50 incorrectly predicts the ground-state crystal structure of Co to be BCC instead of HCP, raising further questions about its suitability for treating the electronic structure of metallic systems.

Using a lower EXX fraction can alleviate some of these issues, but does not fully resolve them.
Instead, we explored a long-range corrected hybrid scheme (LC) with 25\% EXX to generate orbitals for RPA evaluation.
Applying 25\% long-range EXX did not exaggerate the spin magnetic moment of Fe, unlike PBE0 and HSE06, which use the same EXX fraction.
This behavior is consistent with the results of RSHXLDA (a long-range corrected hybrid functional with 100\% EXX), where the magnetic moments of magnetic oxides were lower than PBE0 values \cite{gerber2007rshxlda}.
In addition, using long-range EXX can destroy the Fermi surface and make a metallic system non-metallic, as observed in RSHXLDA, generating a vanishing density of states at the Fermi level in bulk Na \cite{gerber2007rshxlda}. 
This behavior can be moderated by applying a smaller fraction of EXX. For example, RSHXLDA with 25\% EXX retains states at the Fermi level in bulk Na, similar to \srpbe (\Cref{fig:dos_Na}).  
Nevertheless, the LC-hybrid based on the standard PBE functional required a very small range separation parameter for EXX ($\mu$<\SI{0.5}{\per\angstrom}) to reproduce the correct Fe magnetic moment. 
A low $\mu$ limits the EXX contribution to only the very long-range region, substantially degrading the accuracy of RPA compared to non-LC approaches.
The need for a low $\mu$ to reproduce the correct Fe magnetic moment may stem from the incompatibility between EXX and PBE correlation \cite{chai2008hf_incompatible}. 
To avoid relying on such a small $\mu$ value, we replaced PBE with the srPBE functional \cite{toulouse2005srpbe,goll2005srpbe,goll2006srpbe}, specifically designed to handle short-range exchange and correlation contributions. 
This replacement enabled the reproduction of the correct Fe magnetic moment at a larger $\mu$ value of \SI{3.0}{\per\angstrom}, while largely preserving the RPA accuracy.
However, the use of srPBE led to incorrect orbital orderings in halogen anions, most notably in F anion, where the degeneracy of the outermost $p$ orbitals was broken. 
We attribute this qualitatively incorrect orbital description to two factors.
The first is insufficient exchange contribution, as only 25\% EXX (only in mid- to long range) is included to prevent energy convergence failure in metallic slabs.
Second, range separation is handled improperly in VASP version 6.3.2 when using Libxc functionals. In a correct scheme, the sum of GGA exchange and EXX contributions should be 100\%. 
However, in \srpbe, GGA exchange contributes only 75\% at short range ($\mu$=\SI{0}{\per\angstrom}) (instead of the expected 100\%), where EXX is absent (\Cref{fig:RS}(a)). 
We note that this issue has been partially resolved in newer versions of VASP, where an arbitrary fraction of GGA exchange can be applied, but without proper range separation for Libxc functionals (i.e. 100\% GGA exchange at $\mu$=\SI{0}{\per\angstrom}, but GGA exchange and EXX do not sum to 100\% at $\mu$$\neq$\SI{0}{\per\angstrom}).

We addressed this issue by increasing the srPBE correlation contribution to 500\% (\Cref{fig:RS}(b)), which largely compensates for the underestimation of srPBE exchange (\Cref{fig:Exc}). Moreover, this increase in correlation energy improved both the orbital ordering and Fe magnetic moment.
The xc energy and the expectation value of xc potential of \srpbe (75\% x and 500\% c) are much larger in magnitude than those of srPBE (100\% x and 100\% c) or a modified version of\srpbe with 100\% srPBE correlation, and are instead closer to PBE and HSE06 values (\Cref{fig:Exc}(a)). 
Although this approach of exaggerating the correlation is highly empirical and not physically justified, it fully leverages the existing implementation in VASP, one of the most widely used DFT codes, without requiring any modifications to its source code, ensuring that \optVt is practically accessible to any practitioner with access to VASP version 6.3.2 or greater.
Despite these seemingly large modifications, the resulting charge densities from \srpbe remained largely consistent with those obtained from other functionals (\Cref{tab:chg}).
In addition, \ce{H2O} formation energies evaluated non-self-consistently with different functionals on \srpbe densities and orbitals were very similar to those obtained using their own self-consistent densities and orbitals (\Cref{tab:dE_H2O}). The total energies showed similarly small differences, suggesting that the functional is physically reasonable. Although \srpbe is highly empirical and largely ad hoc, it may help guide the development of more rigorous xc functionals with sophisticated range-separation strategies for generating RPA wavefunctions. We also re-emphasize that only the wavefunctions from the \srpbe functional are used in \optVt, and all energies reported ultimately use a combination of standard exact exchange energy, RPA correlation, and short-range PBE correlation as described in the Methods section.

\subsection{\label{sec:opt_xc}Optimization of exchange and correlation parts of \optVt}

The scaling constant for the RPA correlation in \optVt was optimized by fitting to W4-11\cite{karton2011w411}. For this procedure, we used the W4-11 values provided in GMTKN55\cite{goerigk2017gmtkn55}, which are zero-point exclusive, clamped-nuclei, non-relativistic (i.e., no corrections for scalar relativistic and spin--orbit coupling effects), core--valence-corrected, designed for benchmarking of DFT functionals.
However, PAW pseudopotentials in VASP are scalar relativistic and use the frozen-core approximation \cite{kresse1993vasp}. To make our results consistent with the reference values, we subtracted the scalar relativistic correction and added core--valence correction to our calculated values, leaving the original reference values unchanged. The same procedure was applied to W4-11-RE\cite{margraf2017w411re}. 
To improve numerical stability and magnetic properties in metallic surfaces, the non-RPA correlation part of the total energy is defined using \optHXX, which combines 100\% EXX and 20\% srPBE correlation (with $\mu_c$=\SI{1.89}{\per\angstrom}=\SI{1.0}{\per\Bohr}).
This construction is motivated by the observation that energy convergence with respect to k-points is more easily achieved when evaluating the total energy directly, rather than treating the non-RPA correlation term and RPA correlation term separately. 
In metallic slabs, the non-RPA correlation and RPA correlation energies tend to be highly correlated. As a result, scaling the RPA correlation energy alone---especially with large $r_c$---worsens the energy convergence (\Cref{fig:Etot_conv_kpoint}). 
Moreover, using a large $r_c$ can result in physically unreasonable behavior in magnetic systems, where the spin-polarized state has a higher total energy (i.e., is more unstable) than the spin-paired state.
The range separation parameter $\mu_c$ and srPBE correlation fraction were tuned to minimize $r_c$ while preserving most of the accuracy on the W4-11 and W4-11-RE datasets. Different choices of these two parameters did not lead to a significant change in accuracy as long as a similar $r_c$ was obtained. 
Defining the non-RPA correlation part through \optHXX (obtained by applying 20\% srPBE correlation with $\mu_c$=$\SI{1.0}{Bohr}$ in the \HXX calculation) reduces the optimized scaling constant from 1.114 to 1.02, indicating that the underestimation of \Ecrpa can be largely compensated by incorporating a fraction of GGA correlation. Alternatively, adding a fraction of srPBE or PBE correlation energy ($E_{\text{c,(sr)PBE}}$) (or subtracting -$V_{\text{c,(sr)PBE}}$ + $E_{\text{c,(sr)PBE}}$, which is printed in the VASP output) to \HXX yields similar results. We note that inclusion of short-range PBE correlation is conceptually similar to the RPA+ approach \cite{yan2000rpa_poorSR,gould2019gRPAp}, although our results were similar even if full range PBE correlation is used.

\subsection{\label{sec:discuss_etc}Numerical convergence, physical accuracy, and computational cost}

In standard DFT approaches it is straightforward to perform convergence testing to control numerical error associated with k-point sampling and planewave cutoff. However, these issues demand more attention when using RPA-based approaches.
Despite the significantly improved k-point convergence of adsorption energies on metals with \optVt, residual RMSDs (relative to k60 values) of less than \SI{0.1}{eV} were still observed for k35--k55 meshes (\Cref{fig:conv_kpoint}). 
The total energy convergence is similar to or slightly worse than that of the adsorption energies (\Cref{fig:Etot_conv_kpoint}), and bulk systems tend to be more stable (\Cref{fig:conv_bulk_surf}).
Magnetic slabs and 3$\times$3 slabs exhibit poorer total energy convergence behavior, but the adsorption energy starts to converge around k40--k45.
As noted earlier, this likely originates from difficulties in k-point sampling for EXX calculations involving metals with highly non‑analytic Brillouin‑zone integrals, where a more accurate description requires statistical sampling of k-points, for example, through twist averaging\cite{lin2001twist_average,mihm2021mb_conv}, which is computationally much more demanding.
A more approximate k-point sampling (i.e., averaging adsorption energies over three consecutive k-mesh scales) also helps achieve energy convergence using only coarse k-point grids (\Cref{fig:conv_kpoint_mean}). When computing RPA-based adsorption energies on metals it is critical to carefully check k-point convergence and increase k-point density beyond what is typically needed for semi-local calculations; all results in the main text use a k50 mesh, which is much higher than the k20-k30 meshes typical with semi-local functionals.
Conversely, oxide surfaces show significantly faster energy convergence, making it straightforward to include additional system-size correction terms, such as surface thickness and supercell size.

We also note that, when comparing to experimental values, there are other sources of uncertainty beyond the electronic structure. Uncertainty in experimental measurements for adsorption energies is typically on the order of 0.05 eV, with an average reported uncertainty of 0.07 eV \cite{araujo2022ads38}.
Furthermore, physical approximations such as the precise atomic geometry of the active site and adsorbate, the number of layers in the slab, the size of the supercell and vacuum spacing, and the slab approximation itself can introduce errors that are likely also on the order of 0.05 -- 0.1 eV. Finally, numerical approximations such as the pseudopotentials used \cite{lej2016psp,borlido2020psp} and the residual errors from k-point sampling, plane wave cutoff, smearing width, and FFT grid can also introduce errors of 0.01 -- 0.05 eV, or higher. 
Taken together, it becomes difficult to differentiate signal from noise when evaluating the slow convergence with respect to k-points, and it is likely unreasonable to expect errors below $\sim$0.05 -- 0.1 eV when comparing to experimental adsorption energies on metals. The case of oxide surface reaction energies is more straightforward, since comparisons to other computational results eliminate the uncertainty around the structural model used, and the insulating nature of oxides makes k-point convergence much faster. Indeed, the errors of \optVt for oxide surface reaction energies are somewhat lower ($\sim$0.08 eV), which supports that at least some of the errors in metallic adsorption energies may arise from these relatively small but uncontrolled sources of uncertainty.

It is also worth considering the computational cost of \optVt as compared to other commonly used methods. In general, \optVt calculations consist of three parts: (1) optimization of occupied bands in \srpbe wavefuctions (via a self--consistent field (SCF) cycle), (2) one-step determination of the unoccupied bands by exact diagonalization of the KS Hamiltonian using the previous wavefunction, and (3) evaluation of the ACFDT-RPA correlation energy.
The computational cost of \optVt is highest for metal adsorption systems.
For \srpbe calculations of a typical metal slab, one SCF iteration within an SCF cycle takes 20--100 cpu hours at k50 mesh with $\Ecut$=300 eV, whereas the cost is negligible for PBE calculations (0.1--0.2 cpu hours). The high cost of \srpbe is driven by the exact exchange calculation.
HSE06 has an even higher computational cost of 50--300 cpu hours per SCF iteration because a larger \Ecut of 600 eV was used, compared to 300 eV for \srpbe.
Naturally, exact timings and performance will vary depending on computing resources, but in general the step that requires the most cpu time in \optVt is the SCF cycle to obtain the \srpbe wavefunctions, while the step that requires the most memory is evaluating the RPA correlation energy.

\section{\label{sec:conclusion}Conclusion}

We have improved the previously reported $\text{\optVo}$ method to make it applicable to general systems, including magnetic metal surfaces. This improvement was achieved by fine-tuning the underlying xc functional to produce orbitals with correct non-energy properties, such as magnetic moments and orbital orderings, while also ensuring energy convergence in metallic systems. In benchmarking across diverse systems, \optVt showed high and well-balanced accuracy, with RMSDs of 0.06--0.16 eV. The accuracy for chemisorption at metal surfaces is particularly noteworthy, since the approach is competitive with the most accurate PBE+D3/M06 embedding method reported. This high accuracy and broad applicability for chemisorption on metals and other divers properties demonstrates the capability of \optVo to be used as as a reference method, especially when accurate experimental data is not available and CCSD(T)-level calculations are not feasible. This study also highlights the potential of optimized RPA and related double-hybrid approaches for achieving higher accuracy than standard RPA, and suggests that further work in the development of specialized functionals for creating wavefunctions for RPA correlation calculations could enable methods that are more accurate and robust.

\begin{acknowledgement}

A.J.M. and N-.K.Y. acknowledge funding from the U.S. Department of Energy, Office of Science, Office of Basic Energy Sciences, Chemical Sciences, Geosciences, and Biosciences Division under Award Number DE-SC0016486 and A.J.M. acknowledges further support under Grant No. DE-SC0023445. J.V.\ acknowledges support by the U.S.\ Department of Energy, Office of Science, Office of Basic Energy Sciences, Chemical Sciences, Geosciences, and Biosciences Division, Catalysis Science Program to the SUNCAT Center for Interface Science and Catalysis.

\end{acknowledgement}


\bibliography{ACS/acs}

\clearpage

\begin{suppinfo}

\renewcommand{\thefigure}{S\arabic{figure}}
\renewcommand{\thetable}{S\arabic{table}}
\renewcommand{\theequation}{S\arabic{equation}}
\setcounter{figure}{0}
\setcounter{table}{0} 
\setcounter{equation}{0}
\setcounter{section}{0}

\section{Supporting Information}

\subsection{\label{sec:SI_datasets}Details of Benchmark Datasets}

\subsubsection{\label{sec:}Molecular datasets}
The datasets for molecules include subsets of W4-11 (molecular atomization energy) \cite{karton2011w411} and W4-11-RE (reaction energies derived from W4-11) \cite{margraf2017w411re}, BH76 (barrier heights of molecular reactions) \cite{zhao2005BH76a,zhao2005BH76b,lars2010gmtkn24}, BH76RC (reaction energies derived from BH76) \cite{lars2010gmtkn24}, MOBH29 (barrier heights and reaction energies of transition metal (TM) complexes) \cite{iron2019mobh35,dohm2020mobh29,semidalas2022mobh35rev,grotjahn2023mobh28}, and the S19 subset of S66 (non-covalent interaction energies of molecular complexes) \cite{rezac2011s66}. 
W4-11, BH76, BH76RC, and S66 are subsets of the GMTKN55 benchmark database \cite{goerigk2017gmtkn55}, whose values were used as reference values in this study. 
Charge and multiplicity values reported in the datasets were used.


The \textbf{W4-11} dataset contains 140 atomization energies of small first- and second-row molecules and radicals \cite{karton2011w411}. The reference energies were calculated using highly accurate W4 theory, which includes post-CCSD(T) contributions \cite{karton2006w4}. 
Its superior accuracy, along with the fact that error cancellation is less likely in atomization energies than in reaction energies, makes it well suited for optimizing the scaling constant for RPA correlation energy.
In W4-11, we only considered 124 molecules with less multireference character (TAE\_nonMR124 subset of W4-11\cite{karton2011w411}), which should be more relevant for catalysis. The excluded molecules include \ce{C2}, FOOF, and \ce{S4}, and a complete list can be found in Ref. \citenum{W411_nonMR}.  
%
The \textbf{W4-11-RE} dataset \cite{margraf2017w411re}, derived from W4-11, contains 11,247 reaction energies. Similarly to W4-11, we selected only reactions involving molecules in TAE\_nonMR124, resulting in 8,868 reactions.
The \textbf{BH76} dataset contains 76 barrier heights of hydrogen transfer, heavy atom transfer, nucleophilic substitution, unimolecular and association reactions \cite{zhao2005BH76a,zhao2005BH76b,lars2010gmtkn24}.
The \textbf{BH76RC} dataset, derived from BH76, consists of 30 reaction energies \cite{lars2010gmtkn24}.
The \textbf{MOBH29} dataset, which is a revised version of MOBH35 \cite{iron2019mobh35}, contains 29 forward and 29 backward barrier heights, as well as 29 reaction energies of organometallic TM complexes \cite{iron2019mobh35,dohm2020mobh29,semidalas2022mobh35rev,grotjahn2023mobh28}. MOBH29 does not contain reactions 17--20 and 24--25 from the original dataset, and reference values were obtained from Ref.~\citenum{semidalas2022mobh35rev}.
The S66 dataset contains 66 interaction energies of biomolecules containing H, C, N, and O elements \cite{rezac2011s66}. 
We considered the \textbf{S19} subset, introduced in this work and derived from the original S66 set \cite{rezac2011s66}, which includes 19 interaction energies. 
Of these, 14 data points were selected based on the 10 worst-performing cases for each DFT functional considered in this study.

\subsubsection{\label{sec:}Bulk solid datasets}
For bulk solids, we considered datasets containing 24 lattice constants, 24 bulk moduli, and 24 atomization energies of non-oxides \cite{harl2010Ecorr,olsen2013ralda_bulk,patrick2015ralda_bulk,olsen2014rA}, as well as a dataset of 23 oxide formation energies \cite{yan2013rpa_oxde,jauho2015rapbe_oxide,voss2022oxide}. For reference values of bulk moduli, zero-point energy (ZPE)-corrected values from a previous study were used \cite{zhang2018bulk_prop}. 
The oxide formation dataset includes all data points from Ref.~\citenum{yan2013rpa_oxde}, except for \ce{LiO2}, whose most stable form was not available in the Materials Project \cite{jain2013MP}, and it includes one additional TM oxide, \ce{MoO3}\cite{voss2022oxide}.

We also tested the magnetic moments of magnetic solids (Fe, Co, and Ni) and examined the relative stability of bulk phases of Co (hcp, fcc, and bcc phases), \ce{MoO3} ($\alpha$ and $\beta$ phases), and \ce{TiO2} (rutile and anatase phases), as DFT (or specific functionals) fails to predict the correct ground-state phase \cite{jang2012Co_dft,labat2007tio2_dft,zhang2019MoO3_dft}.

\subsubsection{\label{sec:}Surface datasets}

We considered surface reactions involving metal and oxide surfaces, as well as surface energies of selected metals. For surface reactions on metals, we used a subset of the ADS41 dataset \cite{mall2019ads41}, which includes CE39 \cite{wellendorff2015ce39} and contains 41 experimental adsorption energies, and selected \Nads data points for evaluation.

For surface reactions on oxide surfaces, seven reactions on MgO, \ce{Al2O3}, and \ce{TiO2} surfaces were examined using CCSD(T)-level reference values \cite{ye2024ccsdt_oxide,ye2024ccsdt_oxide2} and two experimental values (for non-dissociative \ce{H2O} adsorption on MgO and \ce{TiO2})\cite{shi2025autoSKZCAM}.
Here, adsorption energies were evaluated using the second definition, with the energies referenced to a gas-phase reactant (CO or \ce{H2O}).

Surface energies were evaluated for four transition metal (Cu, Rh, Pd, and Pt) surfaces with (111) orientation.
Data from experiment \cite{vitos1998Esurf_expt} and other RPA-based methods \cite{schimka2010rpa_surf,schmidt2018benchmarkRPA,olsen2019rA_benchmark} were taken directly from literature sources.

\clearpage

\subsection{\label{sec:SI_methods}Dataset-Specific Computational Details}

\subsubsection{\label{sec:SI_methods_mol}Settings for molecular datasets }

For the RPA calculations of molecules, a single $\Gamma$-point and \Ecut of \SI{600}{eV} were used, except for MOBH29. All DFT calculations used the same \Ecut of \SI{600}{eV}.
Gaussian smearing was applied with a smearing width of \SI{0.001}{eV}.
The molecular structures from the original datasets were used. Each molecule was placed in a 10 Å cubic cell, except for MOBH29 and S66. 
Due to the large molecular sizes in MOBH29, the molecules were placed in a larger cubic cell (14.5 Å) and rotated to ensure that the minimum distance between any two atoms in adjacent periodic images was larger than 5.5 Å, with \Ecut set to 500 eV for RPA calculations only. Similarly, for S66, a larger cubic cell (12 Å) was used to maintain a minimum inter-image atom--atom distance of 6 Å.

\subsubsection{\label{sec:SI_methods_bulk} Settings for bulk solid datasets}
Gaussian smearing was applied with a smearing width of \SI{0.01}{eV}.
The $\Gamma$-point-centered k-point grids were generated using a density-based scheme equivalent to \texttt{KSPACING} in VASP, defined as $\max(1, \operatorname{ceil}(2\pi  \cdot |\mathbf{b}_i| / d_k) ) $, where $\mathbf{b}_i$ is the reciprocal lattice vector, $d_k$ is the target k-point effective length scale (in \si{\per\angstrom}), and $\operatorname{ceil}(x)$ returns the smallest integer greater than or equal to $x$. 
The k-point density is defined as 1/$d_k$ (in Å).

For non-oxide solids, we used \Ecut of \SI{600}{eV} and k-point densities of \SI[parse-numbers=false]{3.2}{\angstrom} for insulators and \SI[parse-numbers=false]{4.5}{\angstrom} for metals to calculate lattice constants, bulk moduli, and atomization energies in both DFT and RPA calculations.
When evaluating \EoptHXX, the FFT grid for the exact exchange was set to ``Accurate'' to ensure more reliable energy derivatives for computing the bulk modulus.
The reported lattice constants and bulk moduli were obtained by fitting the Birch–Murnaghan equation of state to energies computed at seven volumes centered around the experimental values (spanning 90–110\%). The atomization energies were evaluated using the structures corresponding to the fitted lattice constants.

Oxide formation energies were calculated using PBE$+$D3-optimized structures for all systems involved.
The \dEcut was applied for the calculation of oxide formation energy. 
The ``low-k'' and ``high-k'' in Eqs. (\ref{eqn:dEcut}--\ref{eqn:Ecut}) correspond to k-point densities of \SI{2.0}{\per\angstrom} and \SI{4.0}{\per\angstrom} for oxides, and \SI{3.0}{\per\angstrom} and \SI{6.0}{\per\angstrom} for metals, respectively. The ``low-E'' value corresponds to $E_\mathrm{cutoff}$ of 400 eV, and ``high-E'' values are 900 eV (for exchange) and 600 eV (for correlation).
DFT calculations were performed using the ``high-k'' and ``high-E'' (for exchange).

\subsubsection{\label{sec:SI_methods_surf}Settings for surface datasets}

Unlike in bulk calculations, the k-point grids here are defined using a k-mesh scale, based on real-space supercell dimensions. 
For example, the k-mesh scale ``k35'' corresponds to a k35 mesh or a $\Gamma$-point-centered k-point grid of $[\operatorname{round}(35/A), \operatorname{round}(35/B), 1]$, where $A$ and $B$ are the x and y dimensions of the supercell and $\operatorname{round}$ denotes rounding to the nearest integer. 

For most surface reactions on metals, the calculations employed 2$\times$2, four-layer slabs with a vacuum spacing of \SI{13.5}{\angstrom}. 
The reactions involving O/Pt(111) and \ce{C3H8}/Pt(111) were calculated using 3$\times$3 slabs.
The surface structures were optimized using PBE$+$D3, with the two bottom layers fixed.
Gaussian smearing was applied with smearing widths of \SI{0.1}{eV} (for DFT) and \SI{0.01}{eV} (for RPA).
The \dEcut was applied for RPA calculations. 
The ``low-k'' and ``high-k'' correspond to k35 (or k25 for 3$\times$3 slabs) and k50 meshes.
The k50 mesh corresponds to k-point densities of 6--7.5 Å.
The ``low-E'' and ``high-E'' correspond to $E_\mathrm{cutoff}$ values of \SI{300}{eV} and \SI{500}{eV}. 
Results of the convergence tests can be found in \Cref{fig:conv_ecut} (\Ecut), \Cref{fig:conv_sigma} (smearing width), and \Cref{fig:conv_kpoint} (k-point).
DFT calculations were conducted using k50 mesh and \Ecut of \SI{600}{eV}.

For oxide surfaces, we used structures from the literature from which the CCSD(T) reference values were obtained \cite{ye2024ccsdt_oxide,ye2024ccsdt_oxide2} and applied a vacuum spacing of 13 Å. 
The \dEcut was applied for RPA calculations. 
The ``low-k'' corresponds to k1 mesh, while ``high-k'' corresponds to k20 (for \ce{TiO2}) and k25 meshes (for MgO and \ce{Al2O3}).
The ``low-E'' is 400 eV, and ``high-E'' values are 900 eV (for exchange) and 600 eV (for correlation). 
DFT calculations were performed using the ``high-k'' and ``high-E'' (for exchange). 
More details about the structures, as well as an additional correction term for slab thickness ($\Delta layer < \SI{0.05}{eV}$) (applied to MgO and \ce{TiO2}) and for supercell size ($\Delta supercell \sim \SI{0.01}{eV}$) (applied to MgO) are provided in the \nameref{sec:SI_ads_ox} section of the SI. The results of energy convergence can be found in \Cref{fig:conv_ox,fig:conv_MgO}.

For metal surface energies, PBE$+$D3-optimized bulk and surface structures were used. The surface models consisted of 1$\times$1 six-layer slabs, with the two middle layers fixed during structural relaxation.
The \dEcut was applied for both bulk and surface systems.
For bulk calculations, the ``low-k'' and ``high-k'' correspond to k-point densities of \SI{3.0}{\angstrom} and \SI{7.0}{\angstrom}.
For surface, ``low-k'' and ``high-k'' are k35 and k80 meshes.
For both systems, ``low-E'' and ``high-E'' correspond to $E_\mathrm{cutoff}$ values of 300 eV and 500 eV. 
For accurate surface energies, highly converged bulk energies are required, since the total energy of six bulk units is subtracted from that of the slab (i.e. three bulk units per surface). Therefore, the Alavi-Spencer scheme \cite{spencer2008singularity} was used to correct the Coulomb divergence ($\texttt{HFRCUT}$=-1 in VASP) during bulk calculations.

\clearpage

\subsection{\label{sec:}\srpbe Functional}

\input{ACS/figures/mol/RS.tex}

\begin{figure}[H]
    \centering
    \begin{tikzpicture}
        \node[inner sep=0pt, anchor=south west] (fig1) at (0, 0)
        {\includegraphics[width=0.47\textwidth]{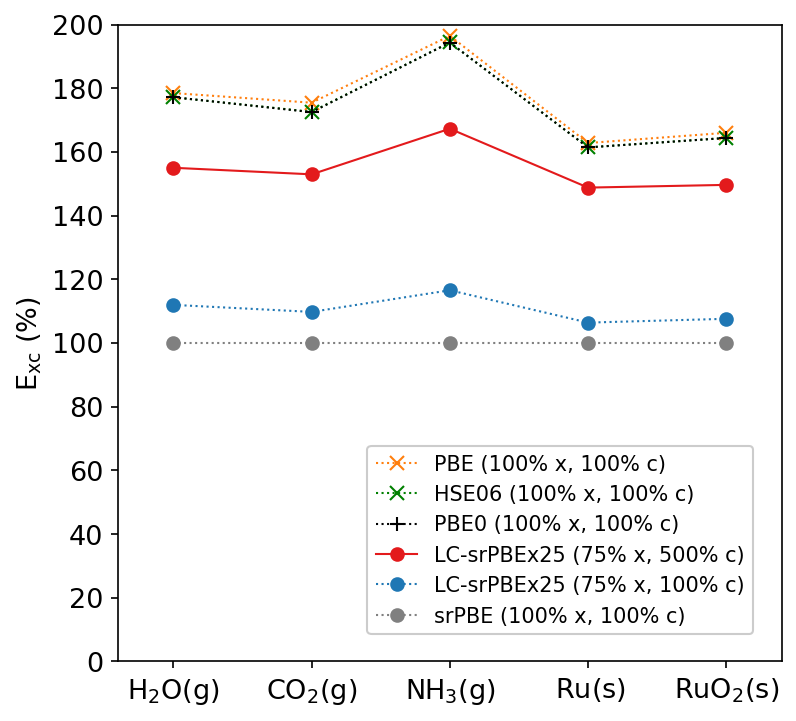}};
        \begin{scope}[x={(fig1.south east)},y={(fig1.north west)}]
            \node[anchor=north west] at (-0.04,1.03) {(a)};
        \end{scope}
        
        \hspace{1mm}
        \node[inner sep=0pt, anchor=south west] (fig2) at (8, 0) 
        {\includegraphics[width=0.47\textwidth]{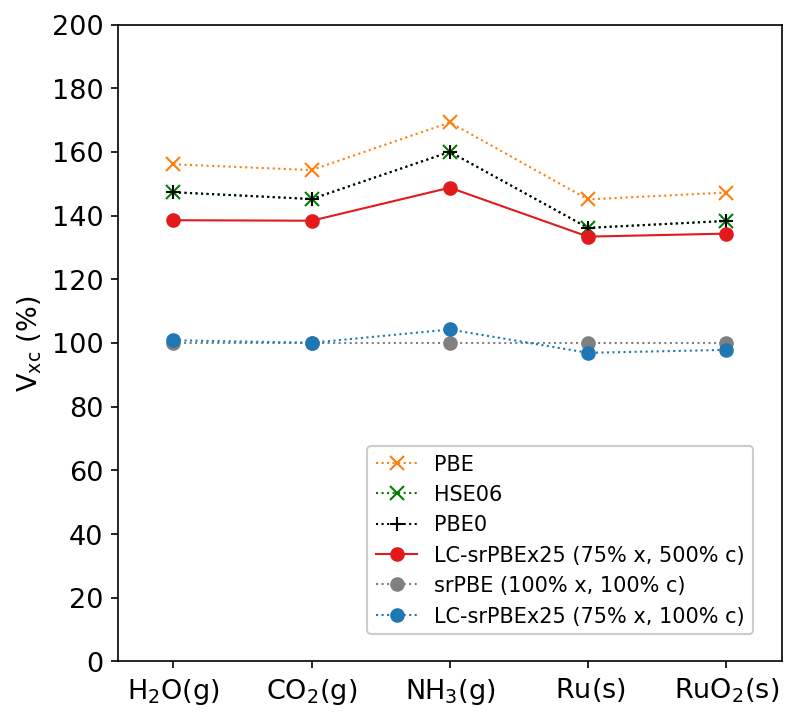}};
        \begin{scope}[x={(fig2.south east)},y={(fig2.north west)}]
            \node[anchor=north west] at (-0.04,1.03) {(b)};
        \end{scope}
    \end{tikzpicture}
    \caption{
    The xc energy (a) and the expectation value of xc potential (b), shown relative to srPBE with 100\% exchange and 100\% correlation, for different functionals. EXX contributions are included. Increasing the correlation contribution in \srpbe compensates for the underestimated exchange contributions.
    }
    \label{fig:Exc}
\end{figure}

\begin{figure} [H]
    \centering
    \includegraphics[width=0.9\textwidth]{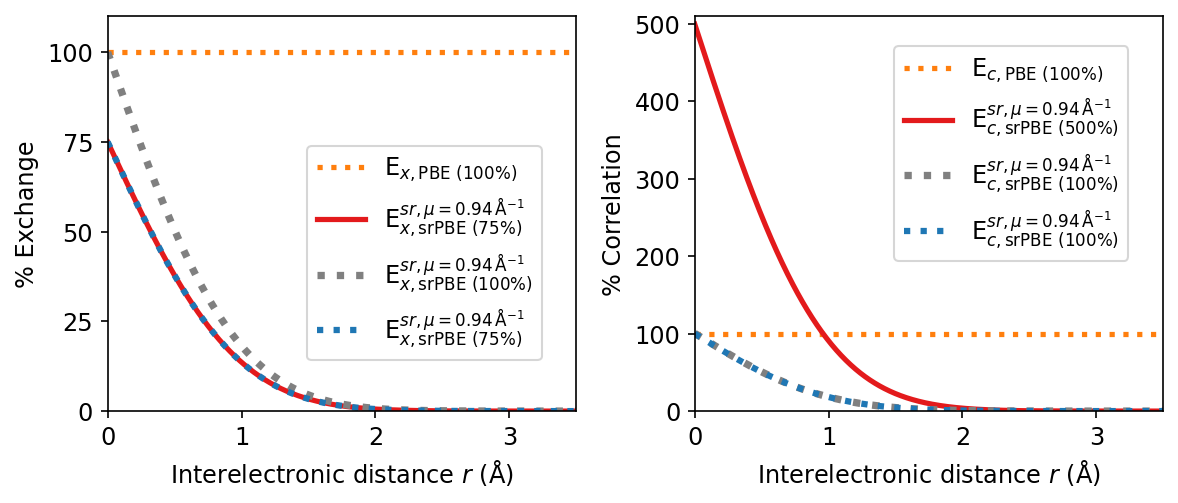}
    \caption{
    Range separation of exchange (left) and correlation (right) for PBE and \srpbe functionals with different amount of xc contributions.
    Colors correspond to \Cref{fig:Exc}.
    }
    \label{fig:RS_Exc}
\end{figure}


\input{ACS/tables/charge.tex}

\input{ACS/tables/dE_H2O.tex}

\begin{figure} [h!]
    \centering
    \includegraphics[width=0.6\textwidth]{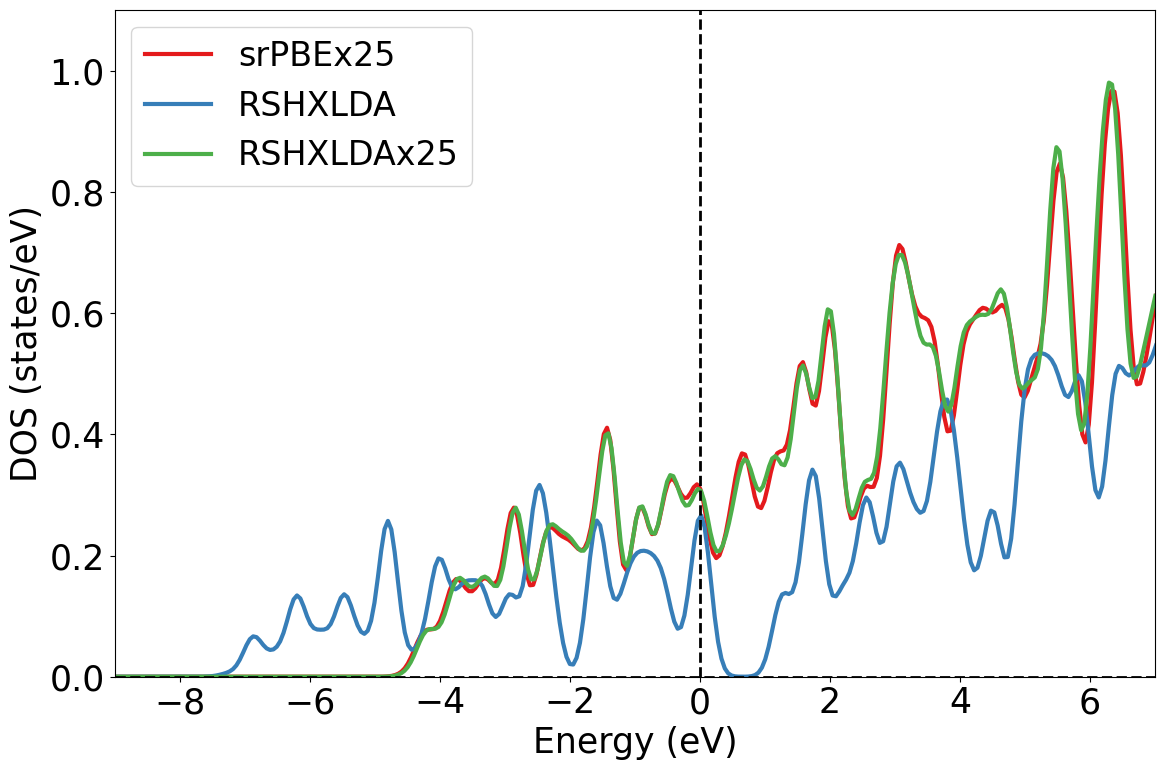}
    \caption{
    Density of states of bulk Na calculated using different functionals. \srpbe retains states near the Fermi level, unlike RSHXLDA with 100\% long-range EXX.
    }
    \label{fig:dos_Na}
\end{figure}

\subsection{\label{sec:SI_ads}Energy Convergence}

\subsubsection{\label{sec:SI_ads_metal}Metal bulk and surface}

\begin{figure} [H]
    \centering
    \includegraphics[width=0.9\textwidth]{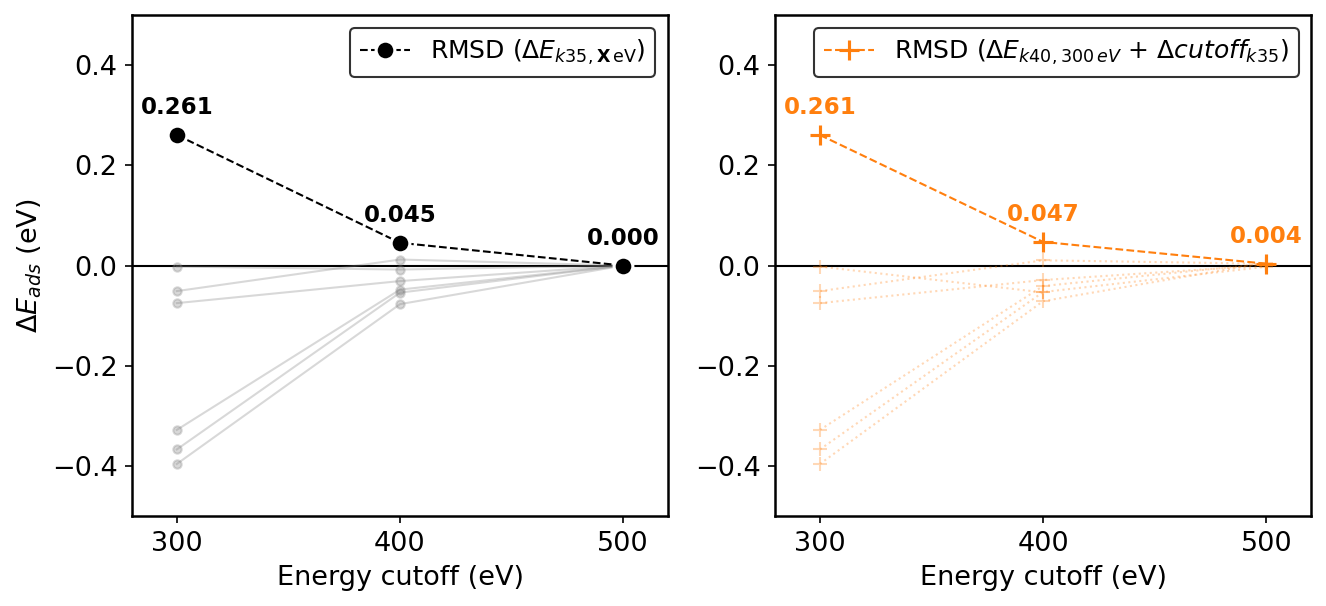}
    \caption{
    (Left) Convergence of adsorption energy on metal surfaces with respect to \Ecut at k35 mesh, referenced to $\Delta E_{\text{k35,\,500}}$. 
    (Right) Validation of the correction term \dEcut at different \Ecut, referenced to $\Delta E_{\text{k40, 500}}$. 
    The \dEcut-corrected $\Delta E_{\text{k40,\,300}}$ values closely match $\Delta E_{\text{k40, 500}}$, demonstrating the effectiveness of the correction.
    Semi-transparent symbols correspond to individual values, while opaque symbols indicate RMSDs relative to the references. 
    }
    \label{fig:conv_ecut}
\end{figure}

\begin{figure} [H]
    \centering
    \includegraphics[width=0.55\textwidth]{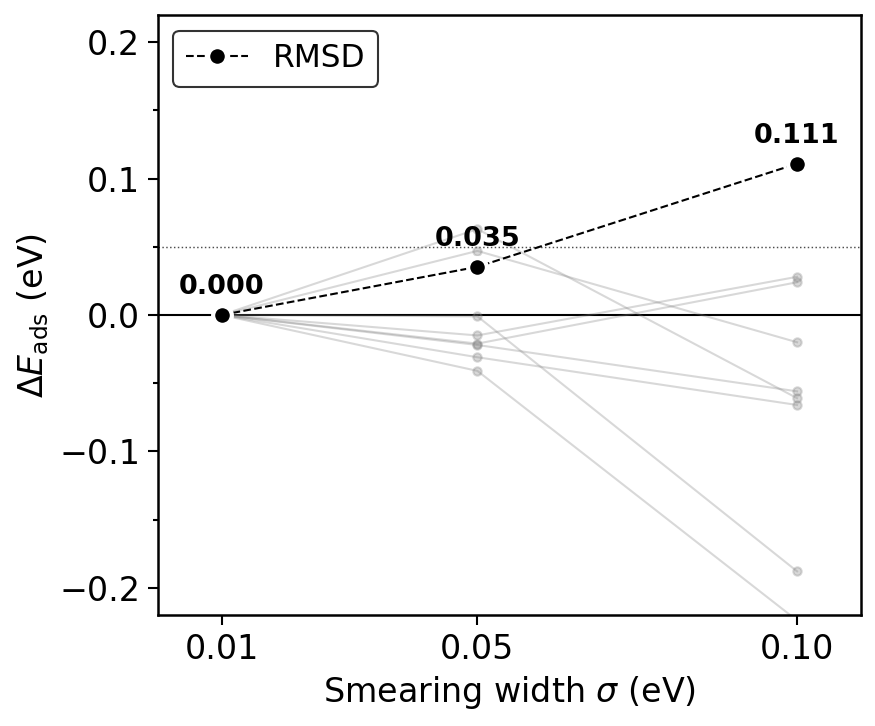}
    \caption{
    Convergence of adsorption energy on metal surfaces with respect to smearing width, referenced to values obtained with $\sigma$=\SI{0.01}{eV}. 
    Semi-transparent symbols correspond to individual values, while opaque symbols indicate RMSDs relative to the reference. 
    }
    \label{fig:conv_sigma}
\end{figure}

\begin{figure}[H]
    \centering
    \begin{tikzpicture}
        \node[inner sep=0pt, anchor=south west] (fig1) at (0, 0)
        {\includegraphics[width=0.55\textwidth]{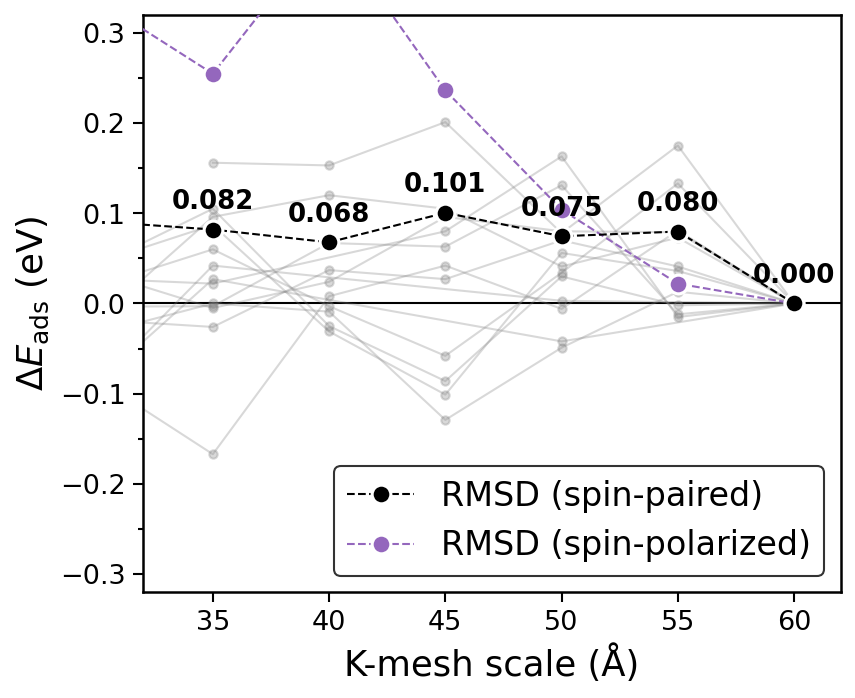}};
        \begin{scope}[x={(fig1.south east)},y={(fig1.north west)}]
        \end{scope}
        
    \end{tikzpicture}
    \caption{
    Convergence of adsorption energy on metal surfaces with respect to k-mesh scale, referenced to $\Delta E^{\text{k60}}$.
    Semi-transparent symbols correspond to individual values, while opaque symbols indicate RMSDs relative to the references. Black symbols correspond to spin-paired results of non-magnetic systems, and purple symbols correspond to spin-polarized results of Ni(111) systems.
    }
    \label{fig:conv_kpoint}
\end{figure}

\begin{figure} [H]
    \centering
    \includegraphics[width=1\textwidth]{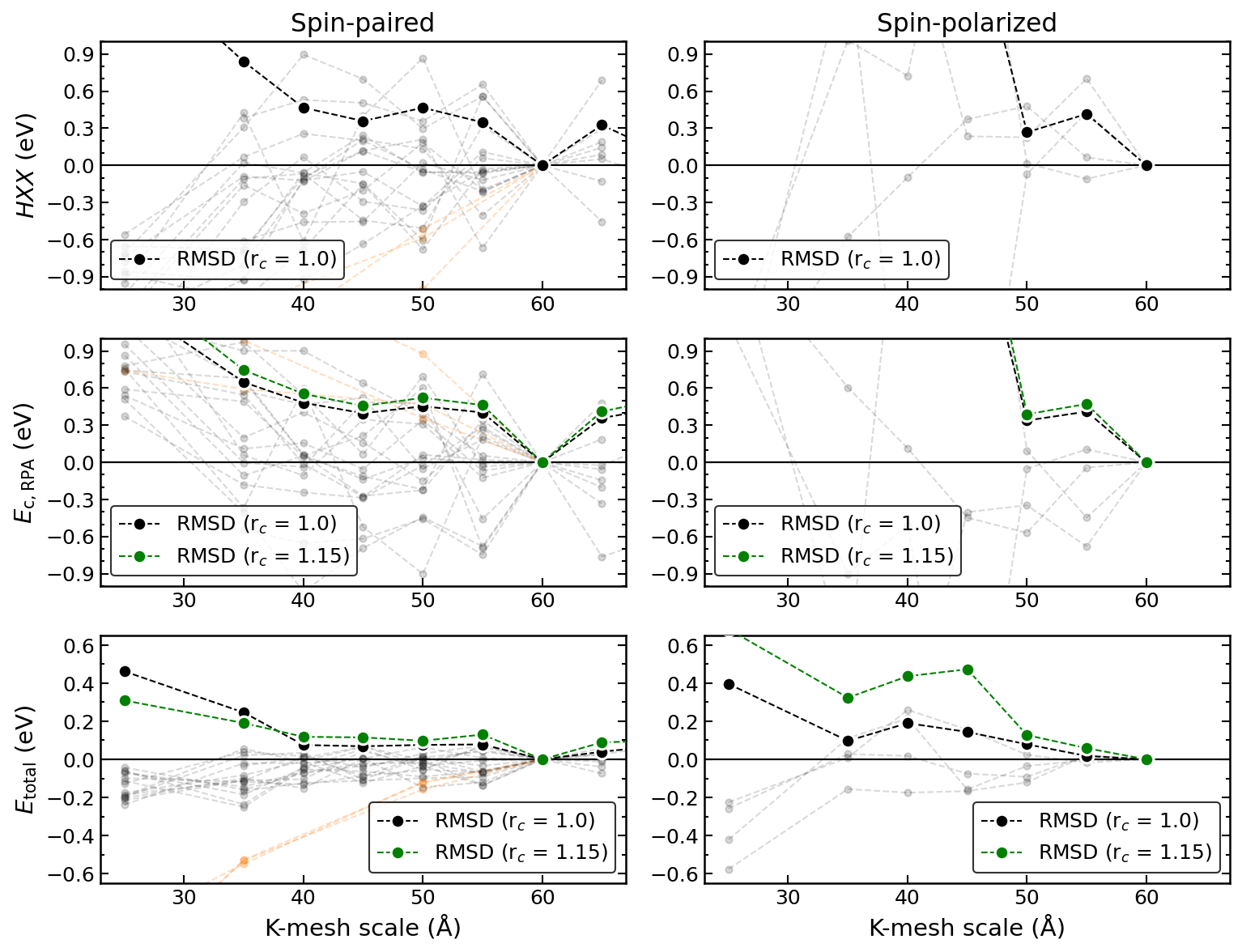}
    \caption{
    Convergence of HXX (top), RPA correlation energy (middle), and total energy (= HXX$+E_{\text{c,RPA}}$) (bottom) of metal surface systems with respect to k-mesh scale, referenced to k60 values ($r_c$=1.0 or $r_c$=1.15). The left panels correspond to spin-paired results of non-magnetic systems, and the right pannels correspond to spin-polarized results of Ni(111) systems.
    Semi-transparent symbols correspond to individual values ($r_c$=1.0), while opaque symbols indicate RMSDs relative to the references. Orange symbols correspond to 3$\times$3 slabs ($r_c$=1.0). Increasing the scaling factor from 1.0 to 1.15 worsens energy convergence.
    }
    \label{fig:Etot_conv_kpoint}
\end{figure}

\begin{figure} [H]
    \centering
    \includegraphics[width=0.8\textwidth]{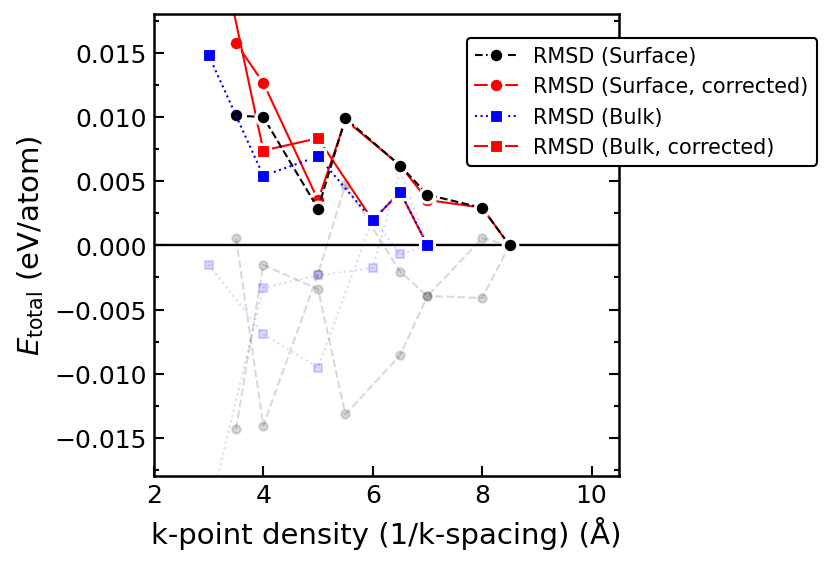}
    \caption{
    Convergence of total energy (= HXX$+E_{\text{c,RPA}}$) of bulk and surface systems of Pd and Pt with respect to k-mesh scale, referenced to k-point densities of 7 Å (bulk) and 8.5 Å (= k65) (surface) ($r_c$=1.02). 
    Semi-transparent symbols correspond to individual values, while opaque symbols indicate RMSDs relative to the references. Red symbols correspond to results with a correction to the non-RPA correlation part for partial occupancies (\texttt{ACFDT corr.} in VASP). The present work does not include this correction term.
    }
    \label{fig:conv_bulk_surf}
\end{figure}

\begin{figure} [H]
    \centering
    \includegraphics[width=0.55\textwidth]{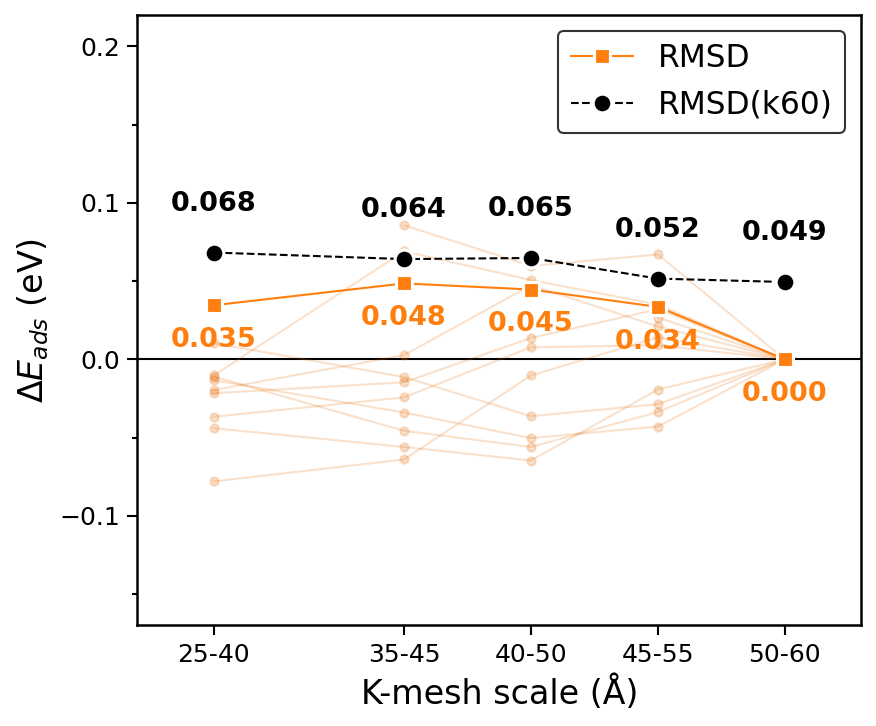}
    \caption{
    Convergence of averaged adsorption energy on metal surfaces with respect to k-mesh scales.
    Each y-axis value represents the averaged adsorption energy over three consecutive k-mesh scales (e.g., ``25--40'' values are averaged over k25, k35, and k40).
    The values are referenced to the averaged k50--k60 values (orange) or individual k60 values (black).
    Semi-transparent symbols correspond to individual values, while opaque symbols indicate RMSDs relative to the references.  
    }
    \label{fig:conv_kpoint_mean}
\end{figure}

\subsubsection{\label{sec:SI_ads_ox}Oxide surface}

For the MgO surface, a 3$\times$3 supercell with 2 layers was used for the baseline calculations. 
A 3$\times$3 supercell with four layers (4L) was used to obtain the correction for surface thickness ($\Delta$layer).
A 5$\times$5 supercell with two layers (2L) was used to obtain the correction for superell size ($\Delta$supercell). 
\begin{equation}
\Delta layer = \Delta E_{\text{k15, 400, 4L}} - \Delta E_{\text{k15, 400, 2L}} \nonumber
\end{equation}
\begin{equation}
\Delta supercell = \Delta E_{\text{k15, 400, 5$\times$5}} - \Delta E_{\text{k15, 400, 3$\times$3}} \nonumber
\end{equation}
\begin{equation}
    \Delta E_{\text{high-k, high-E, 4L, 5$\times$5}} \approx (\Delta E_{\text{high-k, low-E}} + \dEcut) + \Delta layer + \Delta supercell \nonumber
\end{equation}
The calculations for the \ce{Al2O3} surface used a 2$\times$2 supercell with 12 layers, without corrections for slab thickness or supercell size. 
For the \ce{TiO2} surface, a 1$\times$3 supercell with 6 layers (6L) was used for the baseline calculations, and a 1$\times$3 supercell with 7 layers (7L) was used to obtain the correction for surface thickness ($\Delta$layer).
\begin{equation}
\Delta layer = \Delta E_{\text{k1, 600, 7L}} - \Delta E_{\text{k1, 600, 6L}} \nonumber
\end{equation}
\begin{equation}
    \Delta E_{\text{high-k, high-E, 7L}} \approx (\Delta E_{\text{high-k, low-E}} + \dEcut) + \Delta layer \nonumber
\end{equation}

\begin{figure} [H]
    \centering
    \includegraphics[width=0.95\textwidth]{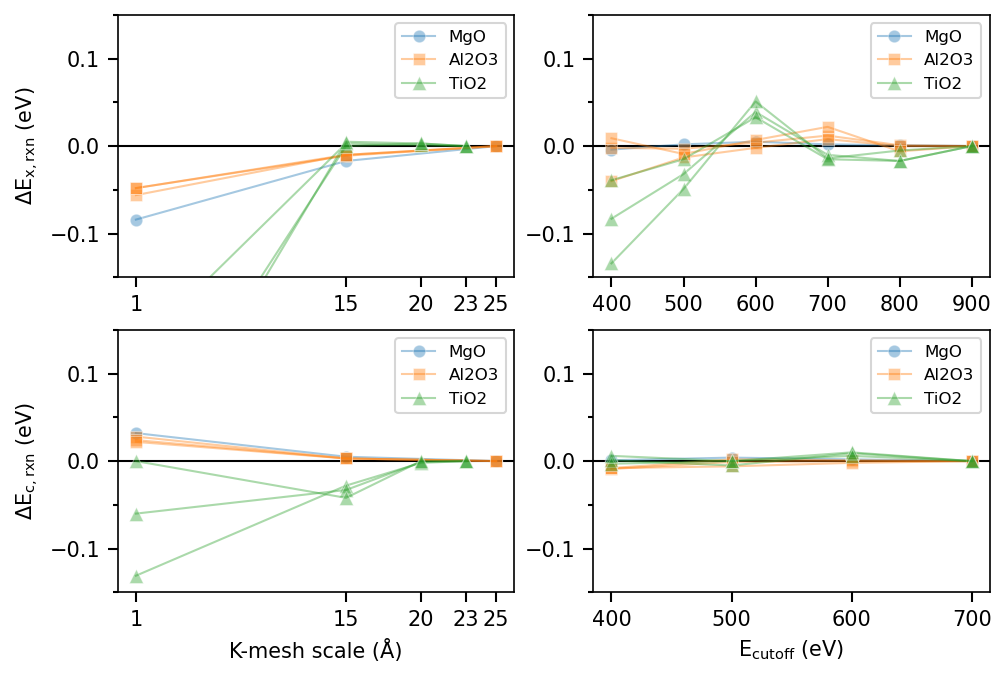}
    \caption{
    Convergence of surface reaction energy on the oxide surfaces. The top and bottom panels show the non-RPA correlation and RPA correlation contributions, respectively. Results are plotted with respect to k-mesh scale (left) and \Ecut (right). 
    }
    \label{fig:conv_ox}
\end{figure}

\begin{figure} [h!]
    \centering
    \includegraphics[width=0.95\textwidth]{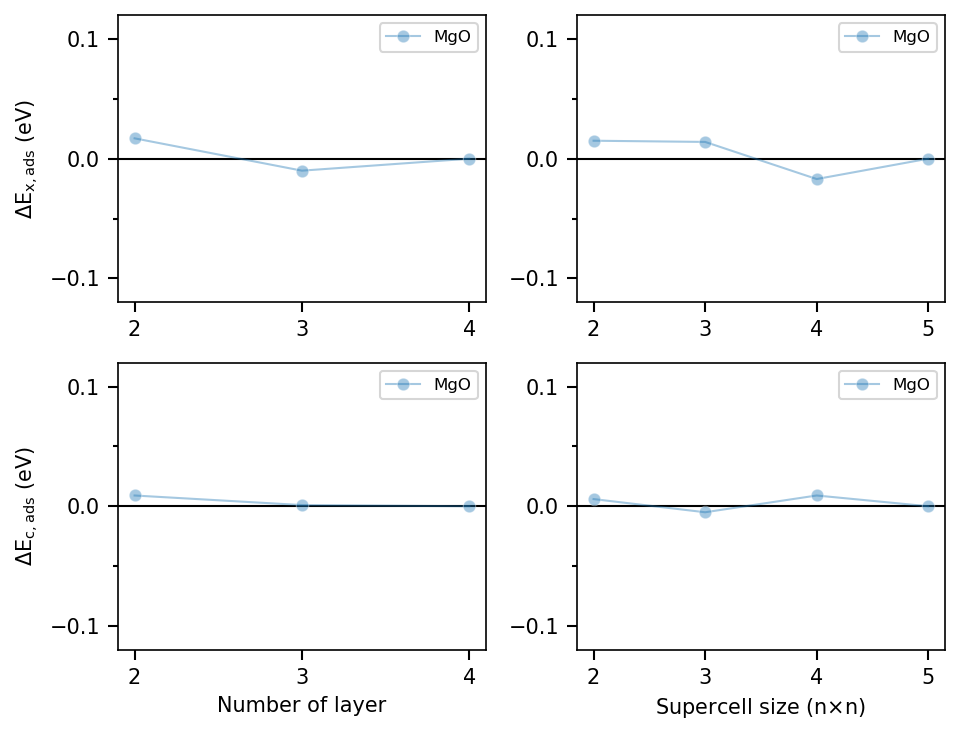}
    \caption{
    Convergence of adsorption energy on the MgO surface. The top and bottom panels show the non-RPA correlation and RPA correlation contributions, respectively. Results are plotted with respect to the number of surface layers (left) and the supercell size (right). A k-mesh scale of \SI{15}{\angstrom} was used. The surface structures have a supercell size of 3$\times$3 (left) and consist of two layers (right).  
    }
    \label{fig:conv_MgO}
\end{figure}

\clearpage

\subsubsection{\label{sec:}Energy of infinitely dilute gas  }

\begin{figure} [H]
    \centering
    \includegraphics[width=1.0\textwidth]{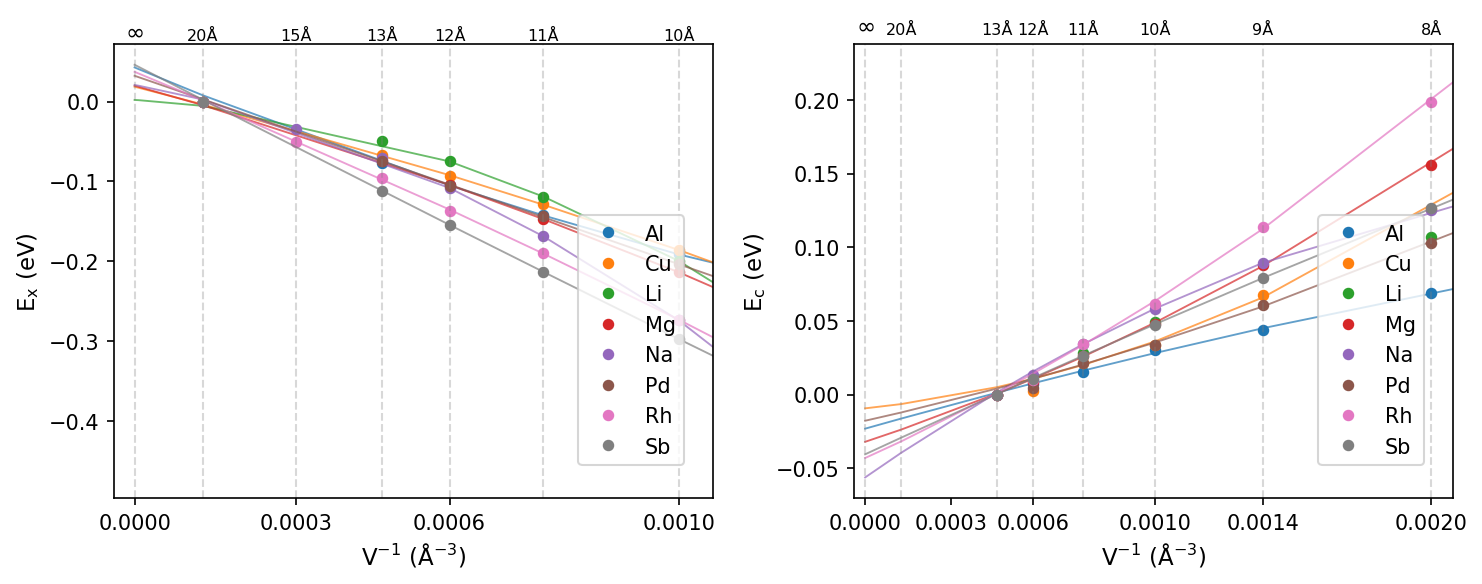}
    \caption{
    Volume dependence of \EHXX (a) and \Ecrpa (b) of gas species used for bulk properties, referenced to the energies at cell lengths of 20Å and 13Å, respectively. The solid lines represent the extrapolated energies using values at 10 -- 13Å (E$_\text{x}$) or 8 -- 11Å (E$_\text{c}$) cubic cells.
    }
    \label{fig:fs_atom}
\end{figure}

\begin{figure} [H]
    \centering
    \includegraphics[width=1.0\textwidth]{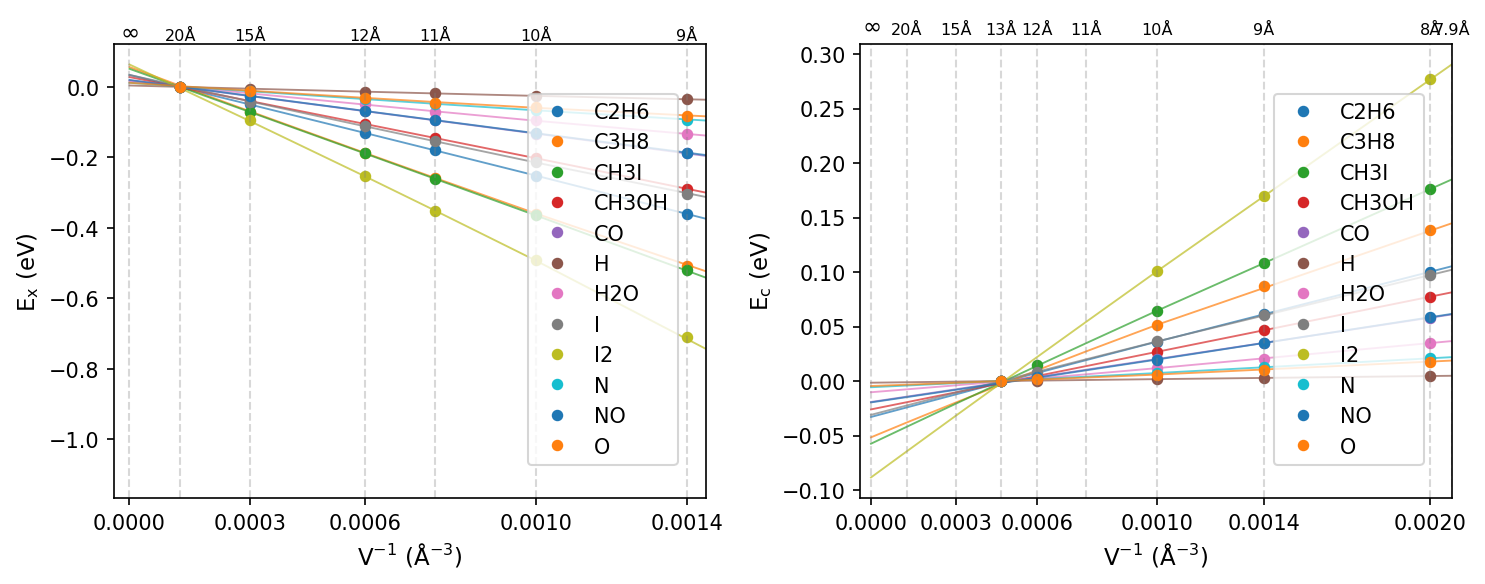}
    \caption{
    Volume dependence of \EHXX (a) and \Ecrpa (b) of adsorbates, referenced to the energies at cell lengths of 20Å and 13Å, respectively. The solid lines represent the extrapolated energies using values at 8 -- 11Å (E$_\text{x}$) or 8 -- 10Å (E$_\text{c}$) cubic cells.
    }
    \label{fig:fs_mol}
\end{figure}

\begin{figure} [H]
    \centering
    \includegraphics[width=1.0\textwidth]{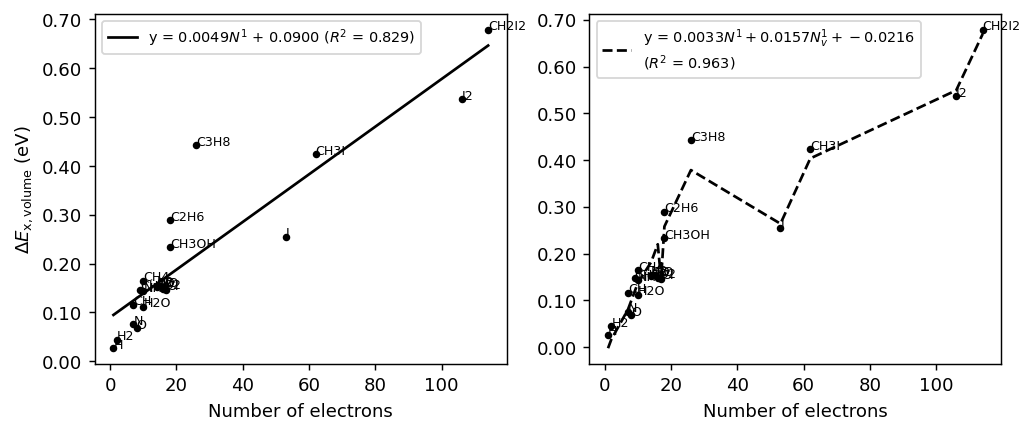}
    \includegraphics[width=1.0\textwidth]{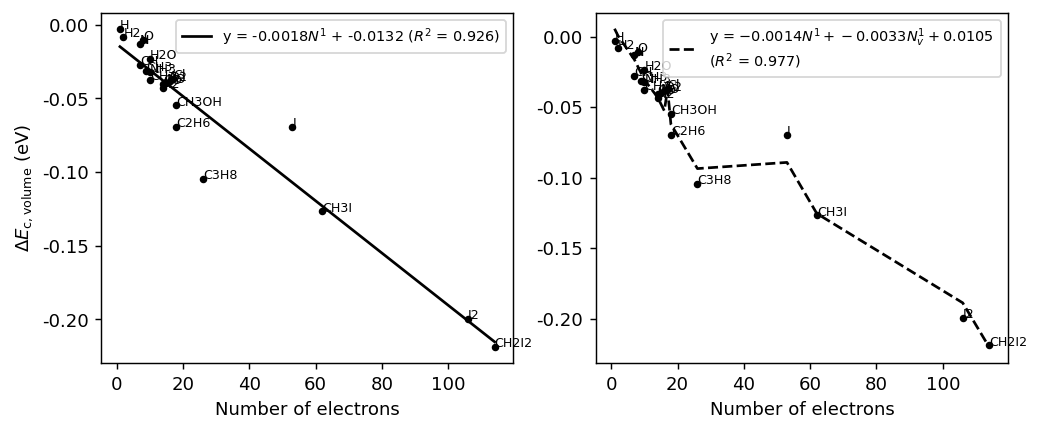}
    \caption{
    $\Delta E_{\mathrm{volume}}$ ($=E_{\infty}-E_{\text{10Å}}$) at \optVt level as a function of the total number of electrons ($N$) in gas-phase species. Top panels show the exchange contribution $\Delta E_{\mathrm{x,volume}}$, and bottom panels show the correlation contribution $\Delta E_{\mathrm{c,volume}}$. Left panels include linear fits in N, while right panels include fits as functions of $N$ and $N_\mathrm{v}$ (number of valence electrons). 
    }
    \label{fig:fs_mol_fit}
\end{figure}

\subsection{\label{sec:SI_note_RPA}Notes for RPA calculations}

To reduce the cost of SCF cycles, the q-point grid can be reduced (\texttt{NKRED}-related tags) to pre-converge the wavefunction.
In the next step, the calculation of unoccupied bands takes 500--2000 cpu hours, but only a single non-SCF step (one iteration) is required to determine the unoccupied bands.
In the low-scaling RPA routine in VASP, the calculation of unoccupied bands is done internally, but this can be avoided by manually setting the number of bands. 
Separating the calculation of unoccupied bands allows the use of parallelization settings that are not available during the RPA routine.
The actual evaluation of RPA correlation energy only takes $\sim$500 cpu hours, but this step requires memory of 4000--6000 GB (which can be reduced by $\sim$1.5 by lowering \texttt{NTAUPAR} tag to adjust parallelization of imaginary time calculation). 
The large memory requirement can be handled by adjusting the number of CPUs allocated per task (via \verb|--cpus-per-task| option in the Slurm workload manager), i.e., by using more nodes and less CPUs per node.

\subsection{\label{sec:SI_etc}Miscellaneous}

\begin{figure} [h!]
    \centering
    \includegraphics[width=0.95\textwidth]{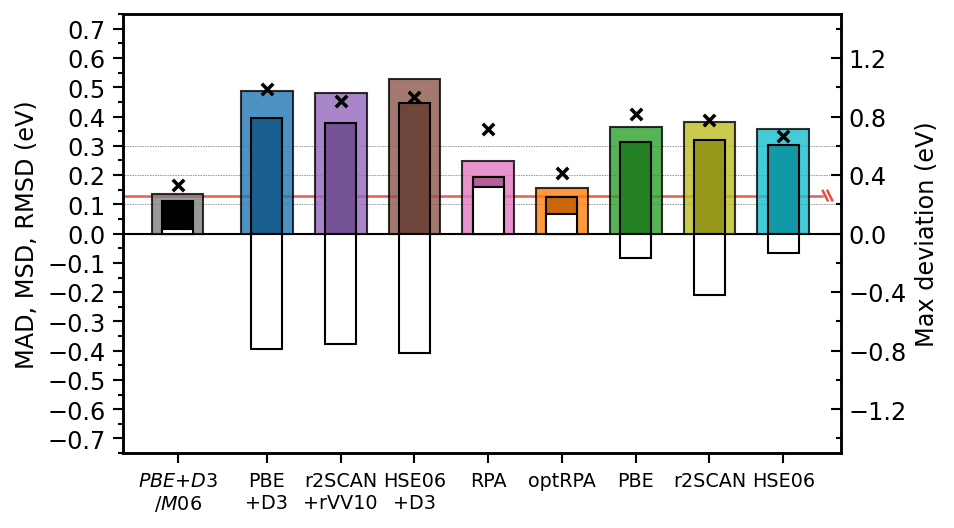}
    \caption{
    Deviations from ZPE-corrected experimental adsorption energies \cite{araujo2022ads38} on metals (Cu, Ru, Rh, Pd, Ir, Pt, and Ni) (n=\Nads). 
    \embed \cite{araujo2022ads38} refers to a hybrid scheme that combines cluster (M06 level) and periodic model (PBE$+$D3 level).
    Non-dissociative adsorption energies referenced to gas-phase adsorbates (e.g., \ce{O}) as in Ref.~\citenum{araujo2022ads38}.
    The adsorption energy values are defined on a per-adsorbate basis; for example, for O adsorption, the adsorption energy is $\Delta$E in (a, c) or $\Delta$E/2 in (b).
    Bars represent MAD (dark color), MSD (white color), and RMSD (light color), and \texttt{×} symbols indicate maximum deviations. Red horizontal lines denotes the transition metal chemical accuracy of \SI{3}{kcal\per\mol}.
    }
    \label{fig:ads_metal1_noD}
\end{figure}

\input{ACS/tables/LS_energies.tex}

\end{suppinfo}

\end{document}

%% file: ACS/tables/bulk_mag.tex
\begin{table}[H]
\centering
\fontsize{10}{12}\selectfont
  \setlength{\tabcolsep}{3pt}
  \caption{
  Spin magnetic moment ($\mu_\mathrm{B}$) of magnetic bulk solids calculated with different functionals. Values outside parentheses correspond to PBE+D3--optimized structures, while values in parentheses correspond to structures with experimental lattice constants.
  }
  \resizebox{0.55\textwidth}{!}{
    \begin{tabular}{p{2.6cm} C{1.8cm} C{1.8cm} C{1.8cm}  }
    \toprule
    \toprule
    \multicolumn{1}{c}{\textbf{Method}} & 
    \multicolumn{1}{c}{\textbf{Fe}} &
    \multicolumn{1}{c}{\textbf{Co}} &
    \multicolumn{1}{c}{\textbf{Ni}} \\
    \midrule
      PBE$+$D3       & 2.115 (2.191) & 1.586 (1.633) & 0.654 (0.649)    \\
      r2SCAN$+$rVV10   & 2.544 (2.657) & 1.734 (1.763) & 0.740 (0.775)    \\
      HSE06$+$D3     & 2.748 (2.823) & 1.841 (1.917) &  0.873 (0.888)   \\
      \srpbe      &  2.057 (2.144) & 1.586 (1.635) & 0.625 (0.652)  \\
      Experiment\cite{wijn1997bulk_mag_expt}   & 2.13 & 1.52 & 0.57   \\
    \bottomrule
    \bottomrule
    \end{tabular}
  }
  \vspace{0.2ex}
  
  \label{tab:bulk_mag}
\end{table}

%% file: ACS/tables/bulk_rel.tex
\begin{table}[H]
\centering
\fontsize{10}{12}\selectfont
  \setlength{\tabcolsep}{3pt}
  \caption{
  Relative energy (eV) of Co, \ce{MoO3}, and \ce{TiO2} bulks in different phases. The smallest value (bold) corresponds to the most stable phase within each formula. 
  Values outside parentheses correspond to PBE+D3--optimized structures, while values in parentheses correspond to structures with experimental lattice constants \cite{fox1999Co_lat}.
  }
  \resizebox{0.85\textwidth}{!}{
    \begin{tabular}{p{2.2cm} C{1.5cm} C{1.5cm} C{1.5cm} C{1.5cm} C{1.5cm} C{1.5cm} C{1.5cm} }
    \toprule
    \toprule
    \multicolumn{1}{c}{\textbf{}} & 
    \multicolumn{3}{c}{\textbf{Co}} &
    \multicolumn{2}{c}{\textbf{\ce{MoO3}}} &
    \multicolumn{2}{c}{\textbf{\ce{TiO2}}} \\
    \cmidrule(lr){2-4} \cmidrule(lr){5-6} \cmidrule(lr){7-8}  
    \multicolumn{1}{c}{\textbf{Method}} & \textbf{hcp Co} & \textbf{fcc Co} & \textbf{bcc Co} & 
    \textbf{$\alpha$-\ce{MoO3}} & \textbf{$\beta$-\ce{MoO3}} & 
    \textbf{rutile} & \textbf{anatase} \\
    
    \midrule
      PBE$+$D3      & \textbf{0} & 0.022 & 0.096 & 0 & \textbf{-0.075} & 0 & \textbf{-0.014}  \\
      r2SCAN$+$rVV10& 0 & -0.024 & \textbf{-0.037} & 0 & \textbf{-0.030} & 0 & \textbf{-0.002}     \\
      HSE06$+$D3    & 0 & -0.225 & \textbf{-0.267} & 0 & \textbf{-0.077} & 0 & \textbf{-0.009} \\
      DMC\cite{luo2016tio2_dmc} &  &  &  &  &      & 0  & \textbf{-0.041} \\
      \optVt        & \textbf{0} & 0.114 & 0.058 & \textbf{0} & 0.008 & \textbf{0} & 0.004  \\
                    & \textbf{(0)} & (0.106) & (0.059) &  &  &  &   \\
      Experiment     & stable & & & stable & & stable &    \\
    \bottomrule
    \bottomrule
    \end{tabular}
  }
  \vspace{0.2ex}
  
  \label{tab:bulk_rel}
\end{table}

%% file: ACS/figures/ads/ads_metal.tex
\begin{figure}[H]
    \centering
    \begin{tikzpicture}
        \node[inner sep=0pt, anchor=south west] (fig1) at (0, 0)
        {\includegraphics[width=0.5\textwidth]{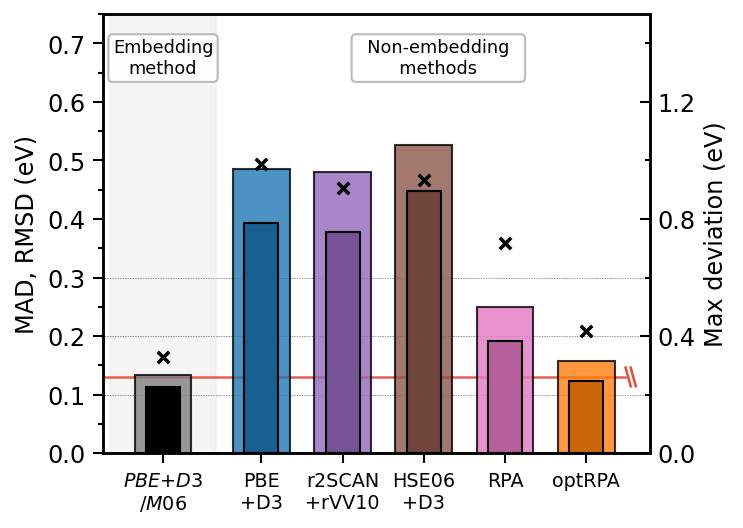}};
        \begin{scope}[x={(fig1.south east)},y={(fig1.north west)}]
            \node[anchor=north west] at (-0.04,1.03) {(a)};
        \end{scope}
        
        \hspace{1mm}
        \node[inner sep=0pt, anchor=south west] (fig2) at (0, -6) 
        {\includegraphics[width=0.5\textwidth]{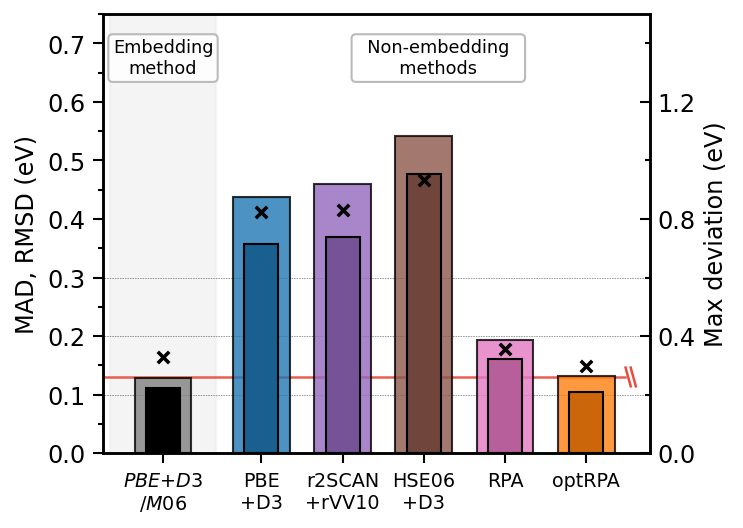}};
        \begin{scope}[x={(fig2.south east)},y={(fig2.north west)}]
            \node[anchor=north west] at (-0.04,1.03) {(b)};
        \end{scope}

        \node[inner sep=0pt, anchor=south west] (fig1) at (0, -11)
        {\includegraphics[width=0.5\textwidth]{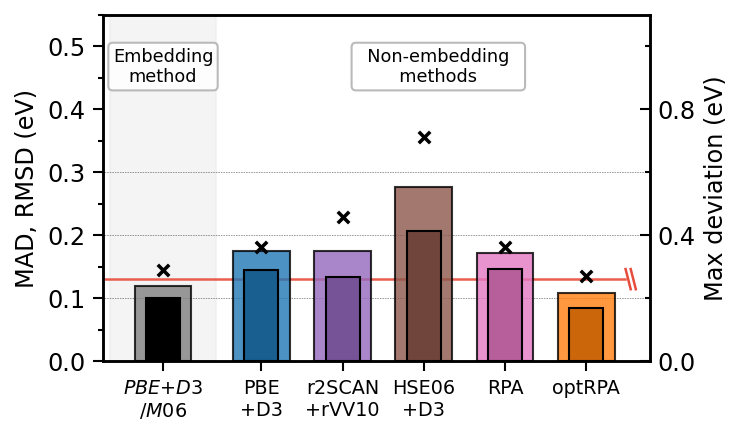}};
        \begin{scope}[x={(fig1.south east)},y={(fig1.north west)}]
            \node[anchor=north west] at (-0.04,1.03) {(c)};
        \end{scope}
    \end{tikzpicture}
    \caption{
    Absolute deviations from ZPE-corrected experimental adsorption energies \cite{araujo2022ads38} on metals (Cu, Ru, Rh, Pd, Ir, Pt, and Ni) (n=\Nads). 
    \embed \cite{araujo2022ads38} refers to a hybrid scheme that combines cluster (M06 level) and periodic model (PBE$+$D3 level).
    (a) Non-dissociative adsorption energies referenced to gas-phase adsorbates (e.g., \ce{O}) as in Ref.~\citenum{araujo2022ads38}.
    (b) Adsorption energies referenced to gas-phase molecules (e.g., \ce{O2}) as in ADS41\cite{mall2019ads41}. 
    (c) Non-dissociative adsorption energies obtained by referencing $E(\text{slab+adsorbate})-E(\text{slab})$ to least-squares elemental chemical potentials, compared against experimental values adjusted by the formation energies of adsorbates.
    The adsorption energy values are defined on a per-adsorbate basis; for example, for O adsorption, the adsorption energy is $\Delta$E in (a, c) or $\Delta$E/2 in (b).
    Bars represent MAD (dark color) and RMSD (light color), and \texttt{×} symbols indicate maximum deviations. Red horizontal lines denotes the transition metal chemical accuracy of \SI{3}{kcal\per\mol}.
    }
    \label{fig:ads_metal}
\end{figure}

%% file: ACS/figures/mol/RS.tex
\begin{figure}[h]
    \centering
    \begin{tikzpicture}
        \node[inner sep=0pt, anchor=south west] (fig1) at (0, 0)
        {\includegraphics[width=0.8\textwidth]{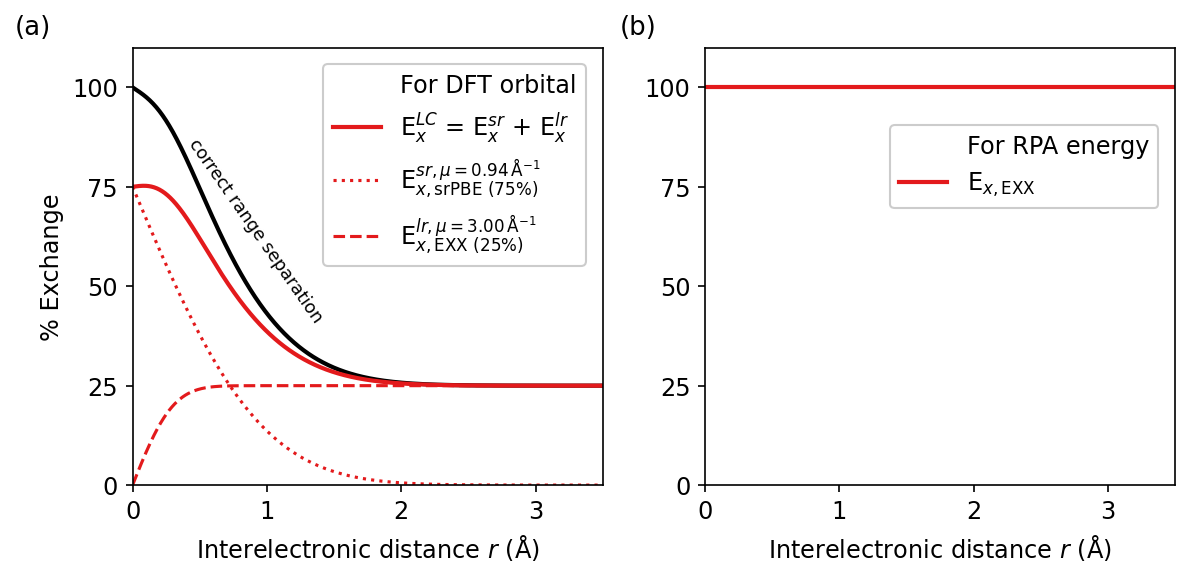}};
        \begin{scope}[x={(fig1.south east)},y={(fig1.north west)}]
        \end{scope}
        
        \hspace{1mm}
        \node[inner sep=0pt, anchor=south west] (fig2) at (0, -6.3) 
        {\includegraphics[width=0.8\textwidth]{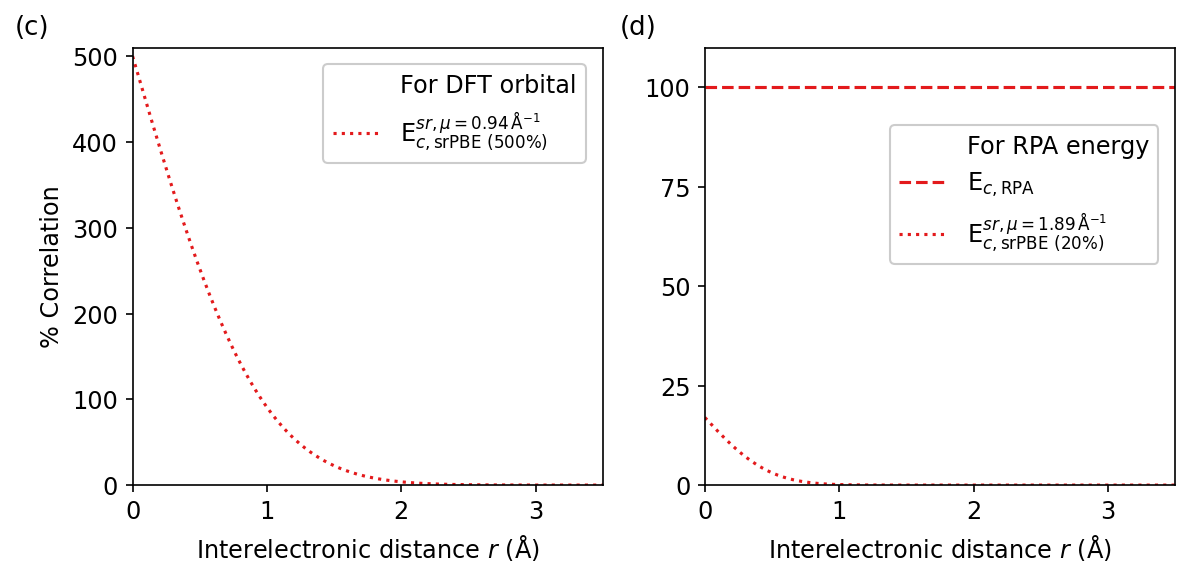}};
        \begin{scope}[x={(fig2.south east)},y={(fig2.north west)}]
        \end{scope}
    \end{tikzpicture}
    \caption{
    Range separation of exchange (a, b) and correlation (c, d) for \srpbe functional (a, c) and the RPA energy evaluation (b, d). The incorrect range separation of exchange in \srpbe (solid red line in (a)) is compensated by increasing the srPBE correlation to 500\% (c). In (d), a small fraction of srPBE correlation is applied when evaluating the exchange part of RPA energy to reduce the scaling constant closer to 1.
    }
    \label{fig:RS}
\end{figure}

%% file: ACS/tables/charge.tex
\begin{table}[h!]
\fontsize{10}{12}\selectfont
  \setlength{\tabcolsep}{3pt}
  \caption{MAD (RMSD) of charge densities relative to HSE06 values ($\times 10^{-4}\ \mathrm{Ry}^{-3}$).
  }
  \resizebox{0.7\textwidth}{!}{
    \begin{tabular}{p{2.7cm} C{2.3cm} C{2.4cm} C{2.3cm} C{2.4cm} C{1.0cm} }
    \toprule
    \toprule
    \multicolumn{1}{c}{\textbf{Method}} & 
    \multicolumn{1}{c}{\textbf{Si bulk}} &
    \multicolumn{1}{c}{\textbf{Si atom}} &
    \multicolumn{1}{c}{\textbf{\ce{H2O}}} \\
    
    \midrule
      HSE06$+$D3     & 0           & 0            & 0 \\
      PBE$+$D3       & 2.80 (4.55) & 0.05 (0.35)  & 0.12 (1.44) \\
      r2SCAN$+$rVV10 & 0.84 (1.43) & 0.05 (0.36)  & 0.03 (0.48) \\
      \srpbe    & 1.98 (3.41) & 0.13 (0.60)  & 0.24 (4.91) \\
      
    \bottomrule
    \bottomrule
    \end{tabular}
  }
  \vspace{0.2ex}
  
  \label{tab:chg}
\end{table}

%% file: ACS/tables/dE_H2O.tex
\begin{table}[h!]
\fontsize{10}{12}\selectfont
  \setlength{\tabcolsep}{3pt}
  \caption{\ce{H2O}(g) formation energy (eV) evaluated using various combinations of functionals. Columns correspond to functionals used to obtain charge densities and orbitals, and rows to those used to evaluate the energies in a non-self-consistent manner. Diagonal values represent self-consistent results.}
  \resizebox{0.8\textwidth}{!}{
    \begin{tabular}{p{2.7cm} C{2.3cm} C{2.4cm} C{2.3cm} C{2.4cm} C{1.0cm} }
    \toprule
    \toprule
    \multicolumn{1}{c}{\textbf{}} & 
    \multicolumn{1}{c}{\textbf{HSE06$+$D3}} &
    \multicolumn{1}{c}{\textbf{PBE$+$D3}} &
    \multicolumn{1}{c}{\textbf{r2SCAN$+$rVV10}} &
    \multicolumn{1}{c}{\textbf{\srpbe}} \\
    
    \midrule
      HSE06$+$D3     & -2.638 & -2.642 & -2.642 & -2.679  \\
      PBE$+$D3       & -2.516 & -2.512 & -2.518 & -2.531  \\
      r2SCAN$+$rVV10 & -2.495 & -2.497 & -2.491 & -2.551  \\
      \srpbe         & -3.761 & -3.735 & -3.782 & -3.718  \\
      
    \bottomrule
    \bottomrule
    \end{tabular}
  }
  \vspace{0.2ex}
  
  \label{tab:dE_H2O}
\end{table}

%% file: ACS/tables/LS_energies.tex
\begin{table}[h!]
\fontsize{12}{12}\selectfont
  \setlength{\tabcolsep}{3pt}
  \caption{RMSDs of least-squares energies obtained from slab-based adsorption energies ($E_{\mathrm{slab+ads}} - E_{\mathrm{slab}}$) and gas-phase adsorbate energies ($E_{\mathrm{ads(g)}}$) to separate errors originating from slabs and gas-phase adsorbates. 
  For $E_{\mathrm{ads(g)}}$, each adsorbate is included multiple times based on its occurrence in the adsorption dataset (e.g., O adsorbate is used three times). } 
  \resizebox{0.95\textwidth}{!}{
    \begin{tabular}{C{3.5cm} C{3cm} C{2cm} C{2cm} C{2.0cm} C{2.0cm} C{2.0cm}}
    \toprule
    \toprule
    \multicolumn{1}{c}{\boldmath{$E_i$} } &
    \multicolumn{1}{c}{\textbf{\embed}} &
    \multicolumn{1}{c}{\textbf{PBE}} &
    \multicolumn{1}{c}{\textbf{r2SCAN}} &
    \multicolumn{1}{c}{\textbf{HSE06}} &
    \multicolumn{1}{c}{\textbf{RPA}} &
    \multicolumn{1}{c}{\textbf{\optVt}} \\

      \text{(in Eq. \eqref{eqn:mu_ls}) } & & \textbf{$+$D3} & \textbf{$+$rVV10} & \textbf{$+$D3} & & \\
      
    \midrule
    $E_{\mathrm{slab+ads}} - E_{\mathrm{slab}}$          
    &0.119	&0.175	&0.175	&0.277	&0.172	&0.108 \\
    $E_{\mathrm{ads(g)}}$   
    & 0.051 & 0.172 & 0.139 & 0.135 & 0.115 & 0.050 \\
    
    \bottomrule
    \bottomrule
    \end{tabular}
  }
  \par\vspace{1.0ex}
  \resizebox{0.95\textwidth}{!}{
    \begin{tabular}{C{3.5cm} C{3cm} C{2cm} C{2cm} C{2.0cm} C{2.0cm} C{2.0cm}}
    \toprule
    \toprule
    \multicolumn{1}{c}{\boldmath{$E_i$} } &
    \multicolumn{1}{c}{} &
    \multicolumn{1}{c}{\textbf{PBE}} &
    \multicolumn{1}{c}{\textbf{r2SCAN}} &
    \multicolumn{1}{c}{\textbf{HSE06}} &
    \multicolumn{1}{c}{} &
    \multicolumn{1}{c}{} \\

      \text{(in Eq. \eqref{eqn:mu_ls}) } & & & & & & \\
      
    \midrule
    $E_{\mathrm{slab+ads}} - E_{\mathrm{slab}}$      
    & &0.175	&0.170	&0.267 & & \\
    $E_{\mathrm{ads(g)}}$   
    & &0.164	&0.126	&0.127 & & \\

    \bottomrule
    \bottomrule
    \end{tabular}
  }

  \label{tab:ls_rmsd}
\end{table}

%% file: ACS/acs.bib
@inbook{ashcroft1976physics,
   author = {Neil W Ashcroft and N D Mermin},
   pages = {337},
   publisher = {Saunders College Publishing},
   title = {Solid State Physics},
   year = {1976}
}

@article{gygi1986hfrcut0,
   author = {F. Gygi and A. Baldereschi},
   doi = {10.1103/PhysRevB.34.4405},
   issn = {0163-1829},
   issue = {6},
   journal = {Physical Review B},
   month = {9},
   pages = {4405-4408},
   title = {Self-consistent Hartree-Fock and screened-exchange calculations in solids: Application to silicon},
   volume = {34},
   url = {https://link.aps.org/doi/10.1103/PhysRevB.34.4405},
   year = {1986}
}

@article{zaera1991CH3I_ads,
   author = {Francisco Zaera and Helmuth Hoffmann},
   doi = {10.1021/j100169a042},
   issn = {0022-3654},
   issue = {16},
   journal = {The Journal of Physical Chemistry},
   month = {8},
   pages = {6297-6303},
   title = {Detection of chemisorbed methyl and methylene groups: surface chemistry of methyl iodide on platinum(111)},
   volume = {95},
   url = {https://pubs.acs.org/doi/abs/10.1021/j100169a042},
   year = {1991}
}

@article{kresse1993vasp,
   author = {G. Kresse and J. Hafner},
   doi = {10.1103/PhysRevB.47.558},
   issn = {0163-1829},
   issue = {1},
   journal = {Physical Review B},
   month = {1},
   pages = {558-561},
   title = {Ab initio molecular dynamics for liquid metals},
   volume = {47},
   url = {https://link.aps.org/doi/10.1103/PhysRevB.47.558},
   year = {1993}
}

@article{massidda1993hfrcut0,
   author = {S. Massidda and M. Posternak and A. Baldereschi},
   doi = {10.1103/PhysRevB.48.5058},
   issn = {0163-1829},
   issue = {8},
   journal = {Physical Review B},
   month = {8},
   pages = {5058-5068},
   title = {Hartree-Fock LAPW approach to the electronic properties of periodic systems},
   volume = {48},
   url = {https://link.aps.org/doi/10.1103/PhysRevB.48.5058},
   year = {1993}
}

@article{fan1994CH3I_ads,
   author = {Jingfu Fan and Michael Trenary},
   doi = {10.1021/la00022a044},
   issn = {0743-7463},
   issue = {10},
   journal = {Langmuir},
   month = {10},
   pages = {3649-3657},
   title = {Symmetry and the Surface Infrared Selection Rule for the determination of the Structure of Molecules on Metal Surfaces},
   volume = {10},
   url = {https://pubs.acs.org/doi/abs/10.1021/la00022a044},
   year = {1994}
}

@article{french1995CH3I_ads,
   author = {C. French and I. Harrison},
   doi = {10.1016/0039-6028(95)00778-4},
   issn = {00396028},
   issue = {1-3},
   journal = {Surface Science},
   month = {11},
   pages = {85-100},
   title = {Orientation and decomposition kinetics of methyl iodide on Pt(111)},
   volume = {342},
   url = {https://linkinghub.elsevier.com/retrieve/pii/0039602895007784},
   year = {1995}
}

@article{kresse1996vasp1,
   author = {G. Kresse and J. Furthmüller},
   doi = {10.1103/PhysRevB.54.11169},
   issn = {0163-1829},
   issue = {16},
   journal = {Physical Review B},
   month = {10},
   pages = {11169-11186},
   title = {Efficient iterative schemes for ab initio total-energy calculations using a plane-wave basis set},
   volume = {54},
   url = {https://link.aps.org/doi/10.1103/PhysRevB.54.11169},
   year = {1996}
}

@article{kresse1996vasp2,
   author = {G. Kresse and J. Furthmüller},
   doi = {10.1016/0927-0256(96)00008-0},
   issn = {09270256},
   issue = {1},
   journal = {Computational Materials Science},
   month = {7},
   pages = {15-50},
   title = {Efficiency of ab-initio total energy calculations for metals and semiconductors using a plane-wave basis set},
   volume = {6},
   url = {https://linkinghub.elsevier.com/retrieve/pii/0927025696000080},
   year = {1996}
}

@article{perdew1996pbe,
   author = {John P. Perdew and Kieron Burke and Matthias Ernzerhof},
   doi = {10.1103/PhysRevLett.77.3865},
   issn = {0031-9007},
   issue = {18},
   journal = {Physical Review Letters},
   month = {10},
   pages = {3865-3868},
   title = {Generalized Gradient Approximation Made Simple},
   volume = {77},
   url = {https://link.aps.org/doi/10.1103/PhysRevLett.77.3865},
   year = {1996}
}

@book{wijn1997bulk_mag_expt,
   author = {HPJ Wijn},
   city = {Berlin/Heidelberg},
   doi = {10.1007/b52851},
   editor = {H. P. J. Wijn},
   isbn = {3-540-60334-4},
   pages = {41},
   publisher = {Springer-Verlag},
   title = {3d, 4d and 5d Elements, Alloys and Compounds},
   volume = {32A},
   url = {http://materials.springer.com/bp/docs/978-3-540-44932-4},
   year = {1997}
}

@article{vitos1998Esurf_expt,
   author = {L. Vitos and A.V. Ruban and H.L. Skriver and J. Kollár},
   doi = {10.1016/S0039-6028(98)00363-X},
   issn = {00396028},
   issue = {1-2},
   journal = {Surface Science},
   month = {8},
   pages = {186-202},
   title = {The surface energy of metals},
   volume = {411},
   url = {https://linkinghub.elsevier.com/retrieve/pii/S003960289800363X},
   year = {1998}
}

@article{kresse1999vasp_paw,
   author = {G. Kresse and D. Joubert},
   doi = {10.1103/PhysRevB.59.1758},
   issn = {0163-1829},
   issue = {3},
   journal = {Physical Review B},
   month = {1},
   pages = {1758-1775},
   title = {From ultrasoft pseudopotentials to the projector augmented-wave method},
   volume = {59},
   url = {https://link.aps.org/doi/10.1103/PhysRevB.59.1758},
   year = {1999}
}

@article{fox1999Co_lat,
   author = {S. Fox and H. J. F. Jansen},
   doi = {10.1103/PhysRevB.60.4397},
   issn = {0163-1829},
   issue = {7},
   journal = {Physical Review B},
   month = {8},
   pages = {4397-4400},
   title = {Total energy of trigonal and tetragonal cobalt},
   volume = {60},
   year = {1999}
}

@article{yan2000rpa_poorSR,
   author = {Zidan Yan and John P. Perdew and Stefan Kurth},
   doi = {10.1103/PhysRevB.61.16430},
   issn = {0163-1829},
   issue = {24},
   journal = {Physical Review B},
   month = {6},
   pages = {16430-16439},
   title = {Density functional for short-range correlation: Accuracy of the random-phase approximation for isoelectronic energy changes},
   volume = {61},
   url = {https://link.aps.org/doi/10.1103/PhysRevB.61.16430},
   year = {2000}
}

@article{lin2001twist_average,
   author = {C. Lin and F. H. Zong and D. M. Ceperley},
   doi = {10.1103/PhysRevE.64.016702},
   issn = {1063651X},
   issue = {1},
   journal = {Physical Review E - Statistical Physics, Plasmas, Fluids, and Related Interdisciplinary Topics},
   pages = {12},
   title = {Twist-averaged boundary conditions in continuum quantum Monte Carlo algorithms},
   volume = {64},
   year = {2001}
}

@article{goll2005srpbe,
   author = {Erich Goll and Hans Joachim Werner and Hermann Stoll},
   doi = {10.1039/b509242f},
   issn = {14639076},
   issue = {23},
   journal = {Physical Chemistry Chemical Physics},
   month = {12},
   pages = {3917-3923},
   pmid = {19810319},
   title = {A short-range gradient-corrected density functional in long-range coupled-cluster calculations for rare gas dimers},
   volume = {7},
   year = {2005}
}

@article{zhao2005BH76b,
   author = {Yan Zhao and Núria González-Garda and Donald G. Truhlar},
   doi = {10.1021/jp045141s},
   issn = {10895639},
   issue = {9},
   journal = {Journal of Physical Chemistry A},
   month = {3},
   pages = {2012-2018},
   title = {Benchmark database of barrier heights for heavy atom transfer, nucleophilic substitution, association, and unimolecular reactions and its use to test theoretical methods},
   volume = {109},
   year = {2005}
}

@article{zhao2005BH76a,
   author = {Yan Zhao and Benjamin J. Lynch and Donald G. Truhlar},
   doi = {10.1039/b416937a},
   issn = {14639076},
   issue = {1},
   journal = {Physical Chemistry Chemical Physics},
   month = {1},
   pages = {43-52},
   title = {Multi-coefficient extrapolated density functional theory for thermochemistry and thermochemical kinetics},
   volume = {7},
   year = {2005}
}

@article{furche2005rpa_atom,
   author = {Filipp Furche and Troy Van Voorhis},
   doi = {10.1063/1.1884112},
   issn = {00219606},
   issue = {16},
   journal = {Journal of Chemical Physics},
   month = {4},
   title = {Fluctuation-dissipation theorem density-functional theory},
   volume = {122},
   year = {2005}
}

@article{furche2005rpa-hybrid,
   author = {Filipp Furche and Troy Van Voorhis},
   doi = {10.1063/1.1884112},
   issn = {0021-9606},
   issue = {16},
   journal = {The Journal of Chemical Physics},
   month = {4},
   title = {Fluctuation-dissipation theorem density-functional theory},
   volume = {122},
   url = {https://pubs.aip.org/jcp/article/122/16/164106/348352/Fluctuation-dissipation-theorem-density-functional},
   year = {2005}
}

@article{toulouse2005srpbe,
   author = {Julien Toulouse and François Colonna and Andreas Savin},
   doi = {10.1063/1.1824896},
   issn = {00219606},
   issue = {1},
   journal = {Journal of Chemical Physics},
   title = {Short-range exchange and correlation energy density functionals: Beyond the local-density approximation},
   volume = {122},
   year = {2005}
}

@article{perdew2005jacob,
   author = {John P. Perdew and Adrienn Ruzsinszky and Jianmin Tao and Viktor N. Staroverov and Gustavo E. Scuseria and Gábor I. Csonka},
   doi = {10.1063/1.1904565},
   issn = {00219606},
   issue = {6},
   journal = {Journal of Chemical Physics},
   month = {8},
   title = {Prescription for the design and selection of density functional approximations: More constraint satisfaction with fewer fits},
   volume = {123},
   year = {2005}
}

@article{weigend2005qzv,
   author = {Florian Weigend and Reinhart Ahlrichs},
   doi = {10.1039/b508541a},
   issn = {14639076},
   journal = {Physical Chemistry Chemical Physics},
   pages = {3297-3305},
   pmid = {16240044},
   title = {Balanced basis sets of split valence, triple zeta valence and quadruple zeta valence quality for H to Rn: Design and assessment of accuracy},
   year = {2005}
}

@article{goll2006srpbe,
   author = {Erich Goll and Hans Joachim Werner and Hermann Stoll and Thierry Leininger and Paola Gori-Giorgi and Andreas Savin},
   doi = {10.1016/j.chemphys.2006.05.020},
   issn = {03010104},
   issue = {1-3},
   journal = {Chemical Physics},
   month = {10},
   pages = {276-282},
   title = {A short-range gradient-corrected spin density functional in combination with long-range coupled-cluster methods: Application to alkali-metal rare-gas dimers},
   volume = {329},
   year = {2006}
}

@article{karton2006w4,
   author = {Amir Karton and Elena Rabinovich and Jan M.L. Martin and Branko Ruscic},
   doi = {10.1063/1.2348881},
   issn = {00219606},
   issue = {14},
   journal = {Journal of Chemical Physics},
   title = {W4 theory for computational thermochemistry: In pursuit of confident sub-kJ/mol predictions},
   volume = {125},
   year = {2006}
}

@article{krukau2006hse06,
   author = {Aliaksandr V. Krukau and Oleg A. Vydrov and Artur F. Izmaylov and Gustavo E. Scuseria},
   doi = {10.1063/1.2404663},
   issn = {00219606},
   issue = {22},
   journal = {Journal of Chemical Physics},
   title = {Influence of the exchange screening parameter on the performance of screened hybrid functionals},
   volume = {125},
   year = {2006}
}

@article{huang2006embed_pc_metal,
   author = {Patrick Huang and Emily A. Carter},
   doi = {10.1063/1.2336428},
   issn = {00219606},
   issue = {8},
   journal = {Journal of Chemical Physics},
   pmid = {16964996},
   title = {Self-consistent embedding theory for locally correlated configuration interaction wave functions in condensed matter},
   volume = {125},
   year = {2006}
}

@article{feller2007ccsdt,
   author = {David Feller and Kirk A. Peterson},
   doi = {10.1063/1.2464112},
   issn = {00219606},
   issue = {11},
   journal = {Journal of Chemical Physics},
   title = {Probing the limits of accuracy in electronic structure calculations: Is theory capable of results uniformly better than "chemical accuracy"?},
   volume = {126},
   year = {2007}
}

@article{labat2007tio2_dft,
   author = {Frédéric Labat and Philippe Baranek and Christophe Domain and Christian Minot and Carlo Adamo},
   doi = {10.1063/1.2717168},
   issn = {00219606},
   issue = {15},
   journal = {Journal of Chemical Physics},
   title = {Density functional theory analysis of the structural and electronic properties of TiO2 rutile and anatase polytypes: Performances of different exchange-correlation functionals},
   volume = {126},
   year = {2007}
}

@article{deyonker2007TMaccuracy,
   author = {Nathan J. DeYonker and Kirk A. Peterson and Gideon Steyl and Angela K. Wilson and Thomas R. Cundari},
   doi = {10.1021/jp0715023},
   issn = {10895639},
   issue = {44},
   journal = {Journal of Physical Chemistry A},
   pages = {11269-11277},
   pmid = {17500547},
   title = {Quantitative computational thermochemistry of transition metal species},
   volume = {111},
   year = {2007}
}

@article{gerber2007rshxlda,
   author = {Iann C. Gerber and János G. Ángyán and Martijn Marsman and Georg Kresse},
   doi = {10.1063/1.2759209},
   issn = {00219606},
   issue = {5},
   journal = {Journal of Chemical Physics},
   title = {Range separated hybrid density functional with long-range Hartree-Fock exchange applied to solids},
   volume = {127},
   year = {2007}
}

@article{irikura2007zpe_diatom,
   author = {Karl K. Irikura},
   doi = {10.1063/1.2436891},
   issn = {00472689},
   issue = {2},
   journal = {Journal of Physical and Chemical Reference Data},
   pages = {389-397},
   publisher = {American Institute of Physics Inc.},
   title = {Experimental vibrational zero-point energies: Diatomic molecules},
   volume = {36},
   year = {2007}
}

@article{harl2008rpa_30y,
   author = {Judith Harl and Georg Kresse},
   doi = {10.1103/PhysRevB.77.045136},
   issn = {10980121},
   issue = {4},
   journal = {Physical Review B - Condensed Matter and Materials Physics},
   month = {1},
   title = {Cohesive energy curves for noble gas solids calculated by adiabatic connection fluctuation-dissipation theory},
   volume = {77},
   year = {2008}
}

@article{chai2008hf_incompatible,
   author = {Jeng Da Chai and Martin Head-Gordon},
   doi = {10.1063/1.2834918},
   issn = {00219606},
   issue = {8},
   journal = {Journal of Chemical Physics},
   title = {Systematic optimization of long-range corrected hybrid density functionals},
   volume = {128},
   year = {2008}
}

@article{spencer2008singularity,
   author = {James Spencer and Ali Alavi},
   doi = {10.1103/PhysRevB.77.193110},
   issn = {10980121},
   issue = {19},
   journal = {Physical Review B - Condensed Matter and Materials Physics},
   month = {5},
   title = {Efficient calculation of the exact exchange energy in periodic systems using a truncated Coulomb potential},
   volume = {77},
   year = {2008}
}

@article{gruneis2009sosex,
   author = {Andreas Grüneis and Martijn Marsman and Judith Harl and Laurids Schimka and Georg Kresse},
   doi = {10.1063/1.3250347},
   issn = {00219606},
   issue = {15},
   journal = {Journal of Chemical Physics},
   title = {Making the random phase approximation to electronic correlation accurate},
   volume = {131},
   year = {2009}
}

@article{janesko2009LC-RPA,
   author = {Benjamin G. Janesko and Thomas M. Henderson and Gustavo E. Scuseria},
   doi = {10.1063/1.3090814},
   issn = {00219606},
   issue = {8},
   journal = {Journal of Chemical Physics},
   title = {Long-range-corrected hybrids including random phase approximation correlation},
   volume = {130},
   year = {2009}
}

@article{ren2009rpa-hybrid,
   author = {Xinguo Ren and Patrick Rinke and Matthias Scheffler},
   doi = {10.1103/PhysRevB.80.045402},
   issn = {10980121},
   issue = {4},
   journal = {Physical Review B - Condensed Matter and Materials Physics},
   month = {8},
   title = {Exploring the random phase approximation: Application to CO adsorbed on Cu(111)},
   volume = {80},
   year = {2009}
}

@article{neese2009ccsdt_PNO,
   author = {Frank Neese and Andreas Hansen and Dimitrios G. Liakos},
   doi = {10.1063/1.3173827},
   issn = {00219606},
   issue = {6},
   journal = {Journal of Chemical Physics},
   title = {Efficient and accurate approximations to the local coupled cluster singles doubles method using a truncated pair natural orbital basis},
   volume = {131},
   year = {2009}
}

@article{harl2010Ecorr,
   author = {Judith Harl and Laurids Schimka and Georg Kresse},
   doi = {10.1103/PhysRevB.81.115126},
   issn = {10980121},
   issue = {11},
   journal = {Physical Review B - Condensed Matter and Materials Physics},
   month = {3},
   title = {Assessing the quality of the random phase approximation for lattice constants and atomization energies of solids},
   volume = {81},
   year = {2010}
}

@article{lars2010gmtkn24,
   author = {Lars Goerigk and Stefan Grimme},
   doi = {10.1021/ct900489g},
   issn = {15499618},
   issue = {1},
   journal = {Journal of Chemical Theory and Computation},
   month = {1},
   pages = {107-126},
   title = {A general database for main group thermochemistry, kinetics, and noncovalent interactions - Assessment of common and reparameterized (meta-)GGA density functionals},
   volume = {6},
   year = {2010}
}

@article{needs2010qmc_casino,
   author = {R. J. Needs and M. D. Towler and N. D. Drummond and P. López Ríos},
   doi = {10.1088/0953-8984/22/2/023201},
   issn = {09538984},
   issue = {2},
   journal = {Journal of Physics Condensed Matter},
   title = {Continuum variational and diffusion quantum Monte Carlo calculations},
   volume = {22},
   year = {2010}
}

@article{schimka2010rpa_surf,
   author = {L. Schimka and J. Harl and A. Stroppa and A. Grüneis and M. Marsman and F. Mittendorfer and G. Kresse},
   doi = {10.1038/nmat2806},
   issn = {14764660},
   issue = {9},
   journal = {Nature Materials},
   pages = {741-744},
   publisher = {Nature Publishing Group},
   title = {Accurate surface and adsorption energies from many-body perturbation theory},
   volume = {9},
   year = {2010}
}

@article{ruzsinszky2010dRPA,
   author = {Adrienn Ruzsinszky and John P. Perdew and Gábor I. Csonka},
   doi = {10.1021/ct900518k},
   issn = {15499618},
   issue = {1},
   journal = {Journal of Chemical Theory and Computation},
   month = {1},
   pages = {127-134},
   title = {The RPA atomization energy puzzle},
   volume = {6},
   year = {2010}
}

@article{grimme2011d3bj,
   author = {Stefan Grimme and Stephan Ehrlich and Lars Goerigk},
   doi = {10.1002/jcc.21759},
   issn = {01928651},
   issue = {7},
   journal = {Journal of Computational Chemistry},
   month = {5},
   pages = {1456-1465},
   pmid = {21370243},
   title = {Effect of the damping function in dispersion corrected density functional theory},
   volume = {32},
   year = {2011}
}

@article{karton2011w411,
   author = {Amir Karton and Shauli Daon and Jan M.L. Martin},
   doi = {10.1016/j.cplett.2011.05.007},
   issn = {00092614},
   issue = {4-6},
   journal = {Chemical Physics Letters},
   month = {7},
   pages = {165-178},
   title = {W4-11: A high-confidence benchmark dataset for computational thermochemistry derived from first-principles W4 data},
   volume = {510},
   year = {2011}
}

@article{irelan2011LC-RPA,
   author = {Robert M. Irelan and Thomas M. Henderson and Gustavo E. Scuseria},
   doi = {10.1063/1.3630951},
   issn = {00219606},
   issue = {9},
   journal = {Journal of Chemical Physics},
   month = {9},
   title = {Long-range-corrected hybrids using a range-separated Perdew-Burke-Ernzerhof functional and random phase approximation correlation},
   volume = {135},
   year = {2011}
}

@article{rezac2011s66,
   author = {Jan Řezáč and Kevin E. Riley and Pavel Hobza},
   doi = {10.1021/ct2002946},
   issn = {15499626},
   issue = {8},
   journal = {Journal of Chemical Theory and Computation},
   month = {8},
   pages = {2427-2438},
   publisher = {American Chemical Society},
   title = {S66: A well-balanced database of benchmark interaction energies relevant to biomolecular structures},
   volume = {7},
   year = {2011}
}

@article{neese2012orca,
   author = {Frank Neese},
   doi = {10.1002/wcms.81},
   issn = {17590876},
   issue = {1},
   journal = {Wiley Interdisciplinary Reviews: Computational Molecular Science},
   month = {1},
   pages = {73-78},
   title = {The ORCA program system},
   volume = {2},
   year = {2012}
}

@article{paier2012rSE,
   author = {Joachim Paier and Xinguo Ren and Patrick Rinke and Gustavo E. Scuseria and Andreas Grüneis and Georg Kresse and Matthias Scheffler},
   doi = {10.1088/1367-2630/14/4/043002},
   issn = {13672630},
   journal = {New Journal of Physics},
   month = {4},
   title = {Assessment of correlation energies based on the random-phase approximation},
   volume = {14},
   year = {2012}
}

@article{peterson2012ccsdt,
   author = {Kirk A. Peterson and David Feller and David A. Dixon},
   doi = {10.1007/s00214-011-1079-5},
   issn = {1432881X},
   issue = {1},
   journal = {Theoretical Chemistry Accounts},
   month = {1},
   publisher = {Springer New York LLC},
   title = {Chemical accuracy in ab initio thermochemistry and spectroscopy: Current strategies and future challenges},
   volume = {131},
   year = {2012}
}

@article{cohen2012DFTbroad,
   author = {Aron J. Cohen and Paula Mori-Sánchez and Weitao Yang},
   doi = {10.1021/cr200107z},
   issn = {0009-2665},
   issue = {1},
   journal = {Chemical Reviews},
   month = {1},
   pages = {289-320},
   title = {Challenges for Density Functional Theory},
   volume = {112},
   url = {https://pubs.acs.org/doi/10.1021/cr200107z},
   year = {2012}
}

@article{jiang2012TMaccuracy,
   author = {Wanyi Jiang and Nathan J. Deyonker and John J. Determan and Angela K. Wilson},
   doi = {10.1021/jp205710e},
   issn = {10895639},
   issue = {2},
   journal = {Journal of Physical Chemistry A},
   month = {1},
   pages = {870-885},
   pmid = {22107449},
   title = {Toward accurate theoretical thermochemistry of first row transition metal complexes},
   volume = {116},
   year = {2012}
}

@article{jang2012Co_dft,
   author = {Young Rok Jang and Byung Deok Yu},
   doi = {10.1143/JPSJ.81.114715},
   issn = {00319015},
   issue = {11},
   journal = {Journal of the Physical Society of Japan},
   month = {11},
   title = {Hybrid functional study of the structural and electronic properties of Co and Ni},
   volume = {81},
   year = {2012}
}

@article{yan2013rpa_oxde,
   author = {Jun Yan and Jens S. Hummelshøj and Jens K. Nørskov},
   doi = {10.1103/PhysRevB.87.075207},
   issn = {10980121},
   issue = {7},
   journal = {Physical Review B - Condensed Matter and Materials Physics},
   month = {2},
   title = {Formation energies of group i and II metal oxides using random phase approximation},
   volume = {87},
   year = {2013}
}

@article{olsen2013rpa_gpaw,
   author = {Thomas Olsen and Kristian S. Thygesen},
   doi = {10.1103/PhysRevB.87.075111},
   issn = {10980121},
   issue = {7},
   journal = {Physical Review B - Condensed Matter and Materials Physics},
   month = {2},
   title = {Random phase approximation applied to solids, molecules, and graphene-metal interfaces: From van der Waals to covalent bonding},
   volume = {87},
   year = {2013}
}

@article{ren2013rSE,
   author = {Xinguo Ren and Patrick Rinke and Gustavo E. Scuseria and Matthias Scheffler},
   doi = {10.1103/PhysRevB.88.035120},
   issn = {10980121},
   issue = {3},
   journal = {Physical Review B - Condensed Matter and Materials Physics},
   month = {7},
   title = {Renormalized second-order perturbation theory for the electron correlation energy: Concept, implementation, and benchmarks},
   volume = {88},
   year = {2013}
}

@article{olsen2013ralda_bulk,
   author = {Thomas Olsen and Kristian S. Thygesen},
   doi = {10.1103/PhysRevB.88.115131},
   issn = {10980121},
   issue = {11},
   journal = {Physical Review B - Condensed Matter and Materials Physics},
   month = {9},
   title = {Beyond the random phase approximation: Improved description of short-range correlation by a renormalized adiabatic local density approximation},
   volume = {88},
   year = {2013}
}

@article{sund2013singularity,
   author = {Ravishankar Sundararaman and T. A. Arias},
   doi = {10.1103/PhysRevB.87.165122},
   issn = {10980121},
   issue = {16},
   journal = {Physical Review B - Condensed Matter and Materials Physics},
   month = {4},
   title = {Regularization of the Coulomb singularity in exact exchange by Wigner-Seitz truncated interactions: Towards chemical accuracy in nontrivial systems},
   volume = {87},
   year = {2013}
}

@article{rolik2013ccsdt_LNO,
   author = {Zoltán Rolik and Lóránt Szegedy and István Ladjánszki and Bence Ladóczki and Mihály Kállay},
   doi = {10.1063/1.4819401},
   issn = {00219606},
   issue = {9},
   journal = {Journal of Chemical Physics},
   month = {9},
   title = {An efficient linear-scaling CCSD(T) method based on local natural orbitals},
   volume = {139},
   year = {2013}
}

@article{riplinger2013ccsdt_DLPNO,
   author = {Christoph Riplinger and Frank Neese},
   doi = {10.1063/1.4773581},
   issn = {00219606},
   issue = {3},
   journal = {Journal of Chemical Physics},
   month = {1},
   title = {An efficient and near linear scaling pair natural orbital based local coupled cluster method},
   volume = {138},
   year = {2013}
}

@article{jain2013MP,
   author = {Anubhav Jain and Shyue Ping Ong and Geoffroy Hautier and Wei Chen and William Davidson Richards and Stephen Dacek and Shreyas Cholia and Dan Gunter and David Skinner and Gerbrand Ceder and Kristin A. Persson},
   doi = {10.1063/1.4812323},
   issn = {2166-532X},
   issue = {1},
   journal = {APL Materials},
   month = {7},
   publisher = {American Institute of Physics Inc.},
   title = {Commentary: The Materials Project: A materials genome approach to accelerating materials innovation},
   volume = {1},
   url = {https://pubs.aip.org/apm/article/1/1/011002/119685/Commentary-The-Materials-Project-A-materials},
   year = {2013}
}

@article{karp2013CH3ads,
   author = {Eric M. Karp and Trent L. Silbaugh and Charles T. Campbell},
   doi = {10.1021/ja400899p},
   issn = {00027863},
   issue = {13},
   journal = {Journal of the American Chemical Society},
   month = {4},
   pages = {5208-5211},
   title = {Energetics of adsorbed CH3 on Pt(111) by calorimetry},
   volume = {135},
   year = {2013}
}

@article{sabatini2013rvv10,
   author = {Riccardo Sabatini and Tommaso Gorni and Stefano De Gironcoli},
   doi = {10.1103/PhysRevB.87.041108},
   issn = {10980121},
   issue = {4},
   journal = {Physical Review B - Condensed Matter and Materials Physics},
   month = {1},
   title = {Nonlocal van der Waals density functional made simple and efficient},
   volume = {87},
   year = {2013}
}

@article{boese2013embed_ccsdt_ox,
   author = {A. Daniel Boese and Joachim Sauer},
   doi = {10.1039/c3cp52321g},
   issn = {14639076},
   issue = {39},
   journal = {Physical Chemistry Chemical Physics},
   month = {10},
   pages = {16481-16493},
   title = {Accurate adsorption energies of small molecules on oxide surfaces: CO-MgO(001)},
   volume = {15},
   year = {2013}
}

@article{kaltak2014rpa_cubic,
   author = {Merzuk Kaltak and Jiří Klimeš and Georg Kresse},
   doi = {10.1103/PhysRevB.90.054115},
   issn = {1550235X},
   issue = {5},
   journal = {Physical Review B - Condensed Matter and Materials Physics},
   month = {8},
   title = {Cubic scaling algorithm for the random phase approximation: Self-interstitials and vacancies in Si},
   volume = {90},
   year = {2014}
}

@article{olsen2014rA,
   author = {Thomas Olsen and Kristian S. Thygesen},
   doi = {10.1103/PhysRevLett.112.203001},
   issn = {10797114},
   issue = {20},
   journal = {Physical Review Letters},
   month = {5},
   publisher = {American Physical Society},
   title = {Accurate ground-state energies of solids and molecules from time-dependent density-functional theory},
   volume = {112},
   year = {2014}
}

@article{peverati2014DFTbroad,
   author = {Roberto Peverati and Donald G. Truhlar},
   doi = {10.1098/rsta.2012.0476},
   issn = {1364-503X},
   issue = {2011},
   journal = {Philosophical Transactions of the Royal Society A: Mathematical, Physical and Engineering Sciences},
   month = {3},
   pages = {20120476},
   publisher = {Royal Society},
   title = {Quest for a universal density functional: the accuracy of density functionals across a broad spectrum of databases in chemistry and physics},
   volume = {372},
   url = {https://royalsocietypublishing.org/doi/10.1098/rsta.2012.0476},
   year = {2014}
}

@article{wolcott2014CHads,
   author = {Christopher A. Wolcott and Isabel X. Green and Trent L. Silbaugh and Ye Xu and Charles T. Campbell},
   doi = {10.1021/jp505494a},
   issn = {19327455},
   issue = {50},
   journal = {Journal of Physical Chemistry C},
   month = {12},
   pages = {29310-29321},
   publisher = {American Chemical Society},
   title = {Energetics of adsorbed CH2 and CH on Pt(111) by calorimetry: The dissociative adsorption of diiodomethane},
   volume = {118},
   year = {2014}
}

@article{shin201zpe_C,
   author = {Hyeondeok Shin and Sinabro Kang and Jahyun Koo and Hoonkyung Lee and Jeongnim Kim and Yongkyung Kwon},
   doi = {10.1063/1.4867544},
   issn = {00219606},
   issue = {11},
   journal = {Journal of Chemical Physics},
   month = {3},
   title = {Cohesion energetics of carbon allotropes: Quantum Monte Carlo study},
   volume = {140},
   year = {2014}
}

@article{jauho2015rapbe_oxide,
   author = {Thomas S. Jauho and Thomas Olsen and Thomas Bligaard and Kristian S. Thygesen},
   doi = {10.1103/PhysRevB.92.115140},
   issn = {1550235X},
   issue = {11},
   journal = {Physical Review B - Condensed Matter and Materials Physics},
   month = {9},
   publisher = {American Physical Society},
   title = {Improved description of metal oxide stability: Beyond the random phase approximation with renormalized kernels},
   volume = {92},
   year = {2015}
}

@article{patrick2015ralda_bulk,
   author = {Christopher E. Patrick and Kristian S. Thygesen},
   doi = {10.1063/1.4919236},
   issn = {10897690},
   issue = {10},
   journal = {Journal of Chemical Physics},
   month = {9},
   publisher = {American Institute of Physics Inc.},
   title = {Adiabatic-connection fluctuation-dissipation DFT for the structural properties of solids - The renormalized ALDA and electron gas kernels},
   volume = {143},
   year = {2015}
}

@article{ruzsinszky2015sosex_rse,
   author = {Adrienn Ruzsinszky and Igor Ying Zhang and Matthias Scheffler},
   doi = {10.1063/1.4932306},
   issn = {00219606},
   issue = {14},
   journal = {Journal of Chemical Physics},
   month = {10},
   publisher = {American Institute of Physics Inc.},
   title = {Insight into organic reactions from the direct random phase approximation and its corrections},
   volume = {143},
   year = {2015}
}

@article{wellendorff2015ce39,
   author = {Jess Wellendorff and Trent L. Silbaugh and Delfina Garcia-Pintos and Jens K. Nørskov and Thomas Bligaard and Felix Studt and Charles T. Campbell},
   doi = {10.1016/j.susc.2015.03.023},
   issn = {00396028},
   journal = {Surface Science},
   pages = {36-44},
   publisher = {Elsevier},
   title = {A benchmark database for adsorption bond energies to transition metal surfaces and comparison to selected DFT functionals},
   volume = {640},
   year = {2015}
}

@article{gautier2015Oads,
   author = {Sarah Gautier and Stephan N. Steinmann and Carine Michel and Paul Fleurat-Lessard and Philippe Sautet},
   doi = {10.1039/c5cp04534g},
   issn = {14639076},
   issue = {43},
   journal = {Physical Chemistry Chemical Physics},
   month = {10},
   pages = {28921-28930},
   publisher = {Royal Society of Chemistry},
   title = {Molecular adsorption at Pt(111). How accurate are DFT functionals?},
   volume = {17},
   year = {2015}
}

@article{riplinger2016ccsdt_scaling,
   author = {Christoph Riplinger and Peter Pinski and Ute Becker and Edward F. Valeev and Frank Neese},
   doi = {10.1063/1.4939030},
   issn = {00219606},
   issue = {2},
   journal = {Journal of Chemical Physics},
   month = {1},
   publisher = {American Institute of Physics Inc.},
   title = {Sparse maps - A systematic infrastructure for reduced-scaling electronic structure methods. II. Linear scaling domain based pair natural orbital coupled cluster theory},
   volume = {144},
   year = {2016}
}

@article{kubas2016ccsdt_oxide,
   author = {Adam Kubas and Daniel Berger and Harald Oberhofer and Dimitrios Maganas and Karsten Reuter and Frank Neese},
   doi = {10.1021/acs.jpclett.6b01845},
   issn = {19487185},
   issue = {20},
   journal = {Journal of Physical Chemistry Letters},
   month = {10},
   pages = {4207-4212},
   publisher = {American Chemical Society},
   title = {Surface Adsorption Energetics Studied with "gold Standard" Wave-Function-Based Ab Initio Methods: Small-Molecule Binding to TiO2(110)},
   volume = {7},
   year = {2016}
}

@article{luo2016tio2_dmc,
   author = {Ye Luo and Anouar Benali and Luke Shulenburger and Jaron T. Krogel and Olle Heinonen and Paul R.C. Kent},
   doi = {10.1088/1367-2630/18/11/113049},
   issn = {13672630},
   issue = {11},
   journal = {New Journal of Physics},
   month = {11},
   publisher = {Institute of Physics Publishing},
   title = {Phase stability of TiO2 polymorphs from diffusion Quantum Monte Carlo},
   volume = {18},
   year = {2016}
}

@article{lej2016psp,
   author = {Kurt Lejaeghere and Gustav Bihlmayer and Torbjörn Björkman and Peter Blaha and Stefan Blügel and Volker Blum and Damien Caliste and Ivano E. Castelli and Stewart J. Clark and Andrea Dal Corso and Stefano De Gironcoli and Thierry Deutsch and John Kay Dewhurst and Igor Di Marco and Claudia Draxl and Marcin Dułak and Olle Eriksson and José A. Flores-Livas and Kevin F. Garrity and Luigi Genovese and Paolo Giannozzi and Matteo Giantomassi and Stefan Goedecker and Xavier Gonze and Oscar Grånäs and E. K.U. Gross and Andris Gulans and François Gygi and D. R. Hamann and Phil J. Hasnip and N. A.W. Holzwarth and Diana Iuşan and Dominik B. Jochym and François Jollet and Daniel Jones and Georg Kresse and Klaus Koepernik and Emine Küçükbenli and Yaroslav O. Kvashnin and Inka L.M. Locht and Sven Lubeck and Martijn Marsman and Nicola Marzari and Ulrike Nitzsche and Lars Nordström and Taisuke Ozaki and Lorenzo Paulatto and Chris J. Pickard and Ward Poelmans and Matt I.J. Probert and Keith Refson and Manuel Richter and Gian Marco Rignanese and Santanu Saha and Matthias Scheffler and Martin Schlipf and Karlheinz Schwarz and Sangeeta Sharma and Francesca Tavazza and Patrik Thunström and Alexandre Tkatchenko and Marc Torrent and David Vanderbilt and Michiel J. Van Setten and Veronique Van Speybroeck and John M. Wills and Jonathan R. Yates and Guo Xu Zhang and Stefaan Cottenier},
   doi = {10.1126/science.aad3000},
   issn = {10959203},
   issue = {6280},
   journal = {Science},
   month = {3},
   pmid = {27013736},
   publisher = {American Association for the Advancement of Science},
   title = {Reproducibility in density functional theory calculations of solids},
   volume = {351},
   year = {2016}
}

@article{margraf2017w411re,
   author = {Johannes T. Margraf and Duminda S. Ranasinghe and Rodney J. Bartlett},
   doi = {10.1039/c7cp00757d},
   issn = {14639076},
   issue = {15},
   journal = {Physical Chemistry Chemical Physics},
   pages = {9798-9805},
   pmid = {28361143},
   publisher = {Royal Society of Chemistry},
   title = {Automatic generation of reaction energy databases from highly accurate atomization energy benchmark sets},
   volume = {19},
   year = {2017}
}

@article{goerigk2017gmtkn55,
   author = {Lars Goerigk and Andreas Hansen and Christoph Bauer and Stephan Ehrlich and Asim Najibi and Stefan Grimme},
   doi = {10.1039/c7cp04913g},
   issn = {14639076},
   issue = {48},
   journal = {Physical Chemistry Chemical Physics},
   pages = {32184-32215},
   pmid = {29110012},
   publisher = {Royal Society of Chemistry},
   title = {A look at the density functional theory zoo with the advanced GMTKN55 database for general main group thermochemistry, kinetics and noncovalent interactions},
   volume = {19},
   year = {2017}
}

@article{medvedev2017DFTaccuracy,
   author = {Michael G. Medvedev and Ivan S. Bushmarinov and Jianwei Sun and John P. Perdew and Konstantin A. Lyssenko},
   doi = {10.1126/science.aah5975},
   issn = {0036-8075},
   issue = {6320},
   journal = {Science},
   month = {1},
   pages = {49-52},
   title = {Density functional theory is straying from the path toward the exact functional},
   volume = {355},
   url = {https://www.science.org/doi/10.1126/science.aah5975},
   year = {2017}
}

@article{nagy2017ccsdt_LNO,
   author = {Péter R. Nagy and Mihály Kállay},
   doi = {10.1063/1.4984322},
   issn = {00219606},
   issue = {21},
   journal = {Journal of Chemical Physics},
   month = {6},
   pmid = {28576082},
   publisher = {American Institute of Physics Inc.},
   title = {Optimization of the linear-scaling local natural orbital CCSD(T) method: Redundancy-free triples correction using Laplace transform},
   volume = {146},
   year = {2017}
}

@article{krogel2017qmc_psp,
   author = {Jaron T. Krogel and P. R.C. Kent},
   doi = {10.1063/1.4986951},
   issn = {00219606},
   issue = {24},
   journal = {Journal of Chemical Physics},
   month = {6},
   pmid = {28668053},
   publisher = {American Institute of Physics Inc.},
   title = {Magnitude of pseudopotential localization errors in fixed node diffusion quantum Monte Carlo},
   volume = {146},
   year = {2017}
}

@article{dd2017qmc_ads,
   author = {Katharina Doblhoff-Dier and Jörg Meyer and Philip E. Hoggan and Geert Jan Kroes},
   doi = {10.1021/acs.jctc.7b00344},
   issn = {15499626},
   issue = {7},
   journal = {Journal of Chemical Theory and Computation},
   month = {7},
   pages = {3208-3219},
   pmid = {28514594},
   publisher = {American Chemical Society},
   title = {Quantum Monte Carlo Calculations on a Benchmark Molecule-Metal Surface Reaction: H2 + Cu(111)},
   volume = {13},
   year = {2017}
}

@article{saritas2017dmc_Eform,
   author = {Kayahan Saritas and Tim Mueller and Lucas Wagner and Jeffrey C. Grossman},
   doi = {10.1021/acs.jctc.6b01179},
   issn = {15499626},
   issue = {5},
   journal = {Journal of Chemical Theory and Computation},
   month = {5},
   pages = {1943-1951},
   pmid = {28358499},
   publisher = {American Chemical Society},
   title = {Investigation of a Quantum Monte Carlo Protocol To Achieve High Accuracy and High-Throughput Materials Formation Energies},
   volume = {13},
   year = {2017}
}

@article{duanmu2017ce39reref,
   author = {Kaining Duanmu and Donald G. Truhlar},
   doi = {10.1021/acs.jctc.6b01156},
   issn = {15499626},
   issue = {2},
   journal = {Journal of Chemical Theory and Computation},
   month = {2},
   pages = {835-842},
   pmid = {27983852},
   publisher = {American Chemical Society},
   title = {Validation of Density Functionals for Adsorption Energies on Transition Metal Surfaces},
   volume = {13},
   year = {2017}
}

@article{schmidt2018benchmarkRPA,
   author = {Per S. Schmidt and Kristian S. Thygesen},
   doi = {10.1021/acs.jpcc.7b12258},
   issn = {19327455},
   issue = {8},
   journal = {Journal of Physical Chemistry C},
   month = {3},
   pages = {4381-4390},
   publisher = {American Chemical Society},
   title = {Benchmark Database of Transition Metal Surface and Adsorption Energies from Many-Body Perturbation Theory},
   volume = {122},
   year = {2018}
}

@article{lehtola2018libxc,
   author = {Susi Lehtola and Conrad Steigemann and Micael J.T. Oliveira and Miguel A.L. Marques},
   doi = {10.1016/j.softx.2017.11.002},
   issn = {23527110},
   journal = {SoftwareX},
   month = {1},
   pages = {1-5},
   publisher = {Elsevier B.V.},
   title = {Recent developments in LIBXC — A comprehensive library of functionals for density functional theory},
   volume = {7},
   year = {2018}
}

@article{zhang2018bulk_prop,
   author = {Guo Xu Zhang and Anthony M. Reilly and Alexandre Tkatchenko and Matthias Scheffler},
   doi = {10.1088/1367-2630/aac7f0},
   issn = {13672630},
   issue = {6},
   journal = {New Journal of Physics},
   month = {6},
   publisher = {Institute of Physics Publishing},
   title = {Performance of various density-functional approximations for cohesive properties of 64 bulk solids},
   volume = {20},
   year = {2018}
}

@article{kim2018qmcpack,
   author = {Jeongnim Kim and Andrew D Baczewski and Todd D Beaudet and Anouar Benali and M Chandler Bennett and Mark A Berrill and Nick S Blunt and Edgar Josué Landinez Borda and Michele Casula and David M Ceperley and Simone Chiesa and Bryan K Clark and Raymond C Clay and Kris T Delaney and Mark Dewing and Kenneth P Esler and Hongxia Hao and Olle Heinonen and Paul R C Kent and Jaron T Krogel and Ilkka Kylänpää and Ying Wai Li and M Graham Lopez and Ye Luo and Fionn D Malone and Richard M Martin and Amrita Mathuriya and Jeremy McMinis and Cody A Melton and Lubos Mitas and Miguel A Morales and Eric Neuscamman and William D Parker and Sergio D Pineda Flores and Nichols A Romero and Brenda M Rubenstein and Jacqueline A R Shea and Hyeondeok Shin and Luke Shulenburger and Andreas F Tillack and Joshua P Townsend and Norm M Tubman and Brett Van Der Goetz and Jordan E Vincent and D ChangMo Yang and Yubo Yang and Shuai Zhang and Luning Zhao},
   doi = {10.1088/1361-648X/aab9c3},
   issn = {0953-8984},
   issue = {19},
   journal = {Journal of Physics: Condensed Matter},
   month = {5},
   pages = {195901},
   pmid = {29582782},
   publisher = {Institute of Physics Publishing},
   title = {QMCPACK: an open source ab initio quantum Monte Carlo package for the electronic structure of atoms, molecules and solids},
   volume = {30},
   url = {https://iopscience.iop.org/article/10.1088/1361-648X/aab9c3},
   year = {2018}
}

@article{khiri2018soc_zpe_I,
   author = {Dorra Khiri and Ivan Černušák and Florent Louis},
   doi = {10.1021/acs.jpca.8b04748},
   issn = {15205215},
   issue = {32},
   journal = {Journal of Physical Chemistry A},
   month = {8},
   pages = {6546-6557},
   pmid = {30016100},
   publisher = {American Chemical Society},
   title = {Theoretical Study of the Reactions of H Atoms with CH3I and CH2I2},
   volume = {122},
   year = {2018}
}

@article{iron2019mobh35,
   author = {Mark A. Iron and Trevor Janes},
   doi = {10.1021/acs.jpca.9b01546},
   issn = {15205215},
   issue = {17},
   journal = {Journal of Physical Chemistry A},
   month = {5},
   pages = {3761-3781},
   pmid = {30973722},
   publisher = {American Chemical Society},
   title = {Evaluating Transition Metal Barrier Heights with the Latest Density Functional Theory Exchange-Correlation Functionals: The MOBH35 Benchmark Database},
   volume = {123},
   year = {2019}
}

@article{olsen2019rA_benchmark,
   author = {Thomas Olsen and Christopher E. Patrick and Jefferson E. Bates and Adrienn Ruzsinszky and Kristian S. Thygesen},
   doi = {10.1038/s41524-019-0242-8},
   issn = {2057-3960},
   issue = {1},
   journal = {npj Computational Materials},
   month = {11},
   pages = {106},
   publisher = {Nature Publishing Group},
   title = {Beyond the RPA and GW methods with adiabatic xc-kernels for accurate ground state and quasiparticle energies},
   volume = {5},
   url = {https://www.nature.com/articles/s41524-019-0242-8},
   year = {2019}
}

@article{mall2019ads41,
   author = {Shaama Mallikarjun Sharada and Rasmus K.B. Karlsson and Yasheng Maimaiti and Johannes Voss and Thomas Bligaard},
   doi = {10.1103/PhysRevB.100.035439},
   issn = {24699969},
   issue = {3},
   journal = {Physical Review B},
   month = {7},
   publisher = {American Physical Society},
   title = {Adsorption on transition metal surfaces: Transferability and accuracy of DFT using the ADS41 dataset},
   volume = {100},
   year = {2019}
}

@article{nagy2019ccsdt_LNO,
   author = {Péter R. Nagy and Mihály Kállay},
   doi = {10.1021/acs.jctc.9b00511},
   issn = {15499626},
   issue = {10},
   journal = {Journal of Chemical Theory and Computation},
   month = {10},
   pages = {5275-5298},
   pmid = {31465219},
   publisher = {American Chemical Society},
   title = {Approaching the Basis Set Limit of CCSD(T) Energies for Large Molecules with Local Natural Orbital Coupled-Cluster Methods},
   volume = {15},
   year = {2019}
}

@article{hsing2019qmc_ads,
   author = {Cheng Rong Hsing and Chun Ming Chang and Ching Cheng and Ching Ming Wei},
   doi = {10.1021/acs.jpcc.9b03780},
   issn = {19327455},
   issue = {25},
   journal = {Journal of Physical Chemistry C},
   month = {6},
   pages = {15659-15664},
   publisher = {American Chemical Society},
   title = {Quantum Monte Carlo Studies of CO Adsorption on Transition Metal Surfaces},
   volume = {123},
   year = {2019}
}

@article{zhang2019MoO3_dft,
   author = {Tianyu Zhang and Xiaofeng Yang and Qingfeng Ge},
   doi = {10.1063/1.5113787},
   issn = {00219606},
   issue = {4},
   journal = {Journal of Chemical Physics},
   month = {7},
   pmid = {31370548},
   publisher = {American Institute of Physics Inc.},
   title = {Surface chemistry and reactivity of \{$\alpha$\}-MoO3 toward methane: A SCAN-functional based DFT study},
   volume = {151},
   year = {2019}
}

@article{gould2019gRPAp,
   author = {Tim Gould and Adrienn Ruzsinszky and John P. Perdew},
   doi = {10.1103/PhysRevA.100.022515},
   issn = {24699934},
   issue = {2},
   journal = {Physical Review A},
   month = {8},
   publisher = {American Physical Society},
   title = {Simple self-interaction correction to random-phase-approximation-like correlation energies},
   volume = {100},
   year = {2019}
}

@article{shee2019TMaccuracy,
   author = {James Shee and Benjamin Rudshteyn and Evan J. Arthur and Shiwei Zhang and David R. Reichman and Richard A. Friesner},
   doi = {10.1021/acs.jctc.9b00083},
   issn = {15499626},
   issue = {4},
   journal = {Journal of Chemical Theory and Computation},
   month = {4},
   pages = {2346-2358},
   pmid = {30883110},
   publisher = {American Chemical Society},
   title = {On Achieving High Accuracy in Quantum Chemical Calculations of 3 d Transition Metal-Containing Systems: A Comparison of Auxiliary-Field Quantum Monte Carlo with Coupled Cluster, Density Functional Theory, and Experiment for Diatomic Molecules},
   volume = {15},
   year = {2019}
}

@article{sauer2019embed_ccsdt_ox,
   author = {Joachim Sauer},
   doi = {10.1021/acs.accounts.9b00506},
   issn = {15204898},
   issue = {12},
   journal = {Accounts of Chemical Research},
   month = {12},
   pages = {3502-3510},
   pmid = {31765121},
   publisher = {American Chemical Society},
   title = {Ab Initio Calculations for Molecule-Surface Interactions with Chemical Accuracy},
   volume = {52},
   year = {2019}
}

@article{Maristella2019embed_mp2-ccsdt_ox,
   author = {Maristella Alessio and Denis Usvyat and Joachim Sauer},
   doi = {10.1021/acs.jctc.8b01122},
   issn = {15499626},
   issue = {2},
   journal = {Journal of Chemical Theory and Computation},
   month = {2},
   pages = {1329-1344},
   pmid = {30596490},
   publisher = {American Chemical Society},
   title = {Chemically Accurate Adsorption Energies: CO and H 2 O on the MgO(001) Surface},
   volume = {15},
   year = {2019}
}

@article{dohm2020mobh29,
   author = {Sebastian Dohm and Markus Bursch and Andreas Hansen and Stefan Grimme},
   doi = {10.1021/acs.jctc.9b01266},
   issn = {15499626},
   issue = {3},
   journal = {Journal of Chemical Theory and Computation},
   month = {3},
   pages = {2002-2012},
   pmid = {32074450},
   publisher = {American Chemical Society},
   title = {Semiautomated Transition State Localization for Organometallic Complexes with Semiempirical Quantum Chemical Methods},
   volume = {16},
   year = {2020}
}

@article{tran2020mag_bulk,
   author = {Fabien Tran and Guillaume Baudesson and Jesús Carrete and Georg K.H. Madsen and Peter Blaha and Karlheinz Schwarz and David J. Singh},
   doi = {10.1103/PhysRevB.102.024407},
   issn = {24699969},
   issue = {2},
   journal = {Physical Review B},
   month = {7},
   publisher = {American Physical Society},
   title = {Shortcomings of meta-GGA functionals when describing magnetism},
   volume = {102},
   year = {2020}
}

@article{jana2020mag_bulk,
   author = {Subrata Jana and Abhilash Patra and Lucian A. Constantin and Prasanjit Samal},
   doi = {10.1063/1.5131530},
   issn = {00219606},
   issue = {4},
   journal = {Journal of Chemical Physics},
   month = {1},
   pmid = {32007058},
   publisher = {American Institute of Physics Inc.},
   title = {Screened range-separated hybrid by balancing the compact and slowly varying density regimes: Satisfaction of local density linear response},
   volume = {152},
   year = {2020}
}

@article{kent2020qmcpack,
   author = {P. R.C. Kent and Abdulgani Annaberdiyev and Anouar Benali and M. Chandler Bennett and Edgar Josué Landinez Borda and Peter Doak and Hongxia Hao and Kenneth D. Jordan and Jaron T. Krogel and Ilkka Kylänpaä and Joonho Lee and Ye Luo and Fionn D. Malone and Cody A. Melton and Lubos Mitas and Miguel A. Morales and Eric Neuscamman and Fernando A. Reboredo and Brenda Rubenstein and Kayahan Saritas and Shiv Upadhyay and Guangming Wang and Shuai Zhang and Luning Zhao},
   doi = {10.1063/5.0004860},
   issn = {10897690},
   issue = {17},
   journal = {Journal of Chemical Physics},
   month = {5},
   pmid = {32384844},
   publisher = {American Institute of Physics Inc.},
   title = {QMCPACK: Advances in the development, efficiency, and application of auxiliary field and real-space variational and diffusion quantum Monte Carlo},
   volume = {152},
   year = {2020}
}

@article{borlido2020psp,
   author = {Pedro Borlido and Jan Doumont and Fabien Tran and Miguel A.L. Marques and Silvana Botti},
   doi = {10.1021/acs.jctc.0c00214},
   issn = {15499626},
   issue = {6},
   journal = {Journal of Chemical Theory and Computation},
   month = {6},
   pages = {3620-3627},
   pmid = {32407117},
   publisher = {American Chemical Society},
   title = {Validation of Pseudopotential Calculations for the Electronic Band Gap of Solids},
   volume = {16},
   year = {2020}
}

@article{sheldon2021rpa_ads,
   author = {Christopher Sheldon and Joachim Paier and Joachim Sauer},
   doi = {10.1063/5.0071995},
   issn = {10897690},
   issue = {17},
   journal = {Journal of Chemical Physics},
   month = {11},
   pmid = {34742209},
   publisher = {American Institute of Physics Inc.},
   title = {Adsorption of CH4on the Pt(111) surface: Random phase approximation compared to density functional theory},
   volume = {155},
   year = {2021}
}

@article{dumaz2021citeDFT,
   author = {Marie Dumaz and Reese Boucher and Miguel A.L. Marques and Aldo H. Romero},
   doi = {10.1007/s11192-021-04057-z},
   issn = {15882861},
   issue = {8},
   journal = {Scientometrics},
   month = {8},
   pages = {6681-6695},
   publisher = {Springer Science and Business Media B.V.},
   title = {Authorship and citation cultural nature in Density Functional Theory from solid state computational packages},
   volume = {126},
   year = {2021}
}

@article{mihm2021mb_conv,
   author = {Tina N. Mihm and Tobias Schäfer and Sai Kumar Ramadugu and Laura Weiler and Andreas Grüneis and James J. Shepherd},
   doi = {10.1038/s43588-021-00165-1},
   issn = {26628457},
   issue = {12},
   journal = {Nature Computational Science},
   month = {12},
   pages = {801-808},
   publisher = {Springer Nature},
   title = {A shortcut to the thermodynamic limit for quantum many-body calculations of metals},
   volume = {1},
   year = {2021}
}

@article{schafer2021fs_mol,
   author = {Tobias Schäfer and Nathan Daelman and Núria López},
   doi = {10.1021/acs.jpclett.1c01589},
   issn = {19487185},
   issue = {27},
   journal = {Journal of Physical Chemistry Letters},
   month = {7},
   pages = {6277-6283},
   pmid = {34212726},
   publisher = {American Chemical Society},
   title = {Cerium Oxides without U: The Role of Many-Electron Correlation},
   volume = {12},
   year = {2021}
}

@article{galimberti2021r2scan,
   author = {Daria Ruth Galimberti and Joachim Sauer},
   doi = {10.1021/acs.jctc.1c00519},
   issn = {15499626},
   issue = {9},
   journal = {Journal of Chemical Theory and Computation},
   month = {9},
   pages = {5849-5862},
   pmid = {34459582},
   publisher = {American Chemical Society},
   title = {Chemically Accurate Vibrational Free Energies of Adsorption from Density Functional Theory Molecular Dynamics: Alkanes in Zeolites},
   volume = {17},
   year = {2021}
}

@article{semidalas2022mobh35rev,
   author = {Emmanouil Semidalas and Jan M.L. Martin},
   doi = {10.1021/acs.jctc.1c01126},
   issn = {15499626},
   issue = {2},
   journal = {Journal of Chemical Theory and Computation},
   month = {2},
   pages = {883-898},
   pmid = {35045709},
   publisher = {American Chemical Society},
   title = {The MOBH35 Metal-Organic Barrier Heights Reconsidered: Performance of Local-Orbital Coupled Cluster Approaches in Different Static Correlation Regimes},
   volume = {18},
   year = {2022}
}

@article{voss2022oxide,
   author = {J Voss},
   doi = {10.1088/2399-6528/ac6069},
   issn = {2399-6528},
   issue = {3},
   journal = {Journal of Physics Communications},
   month = {3},
   pages = {035009},
   publisher = {IOP Publishing Ltd},
   title = {Hubbard-corrected oxide formation enthalpies without adjustable parameters},
   volume = {6},
   url = {https://iopscience.iop.org/article/10.1088/2399-6528/ac6069},
   year = {2022}
}

@article{araujo2022ads38,
   author = {Rafael B. Araujo and Gabriel L.S. Rodrigues and Egon Campos dos Santos and Lars G.M. Pettersson},
   doi = {10.1038/s41467-022-34507-y},
   issn = {20411723},
   issue = {1},
   journal = {Nature Communications},
   month = {12},
   pmid = {36369277},
   publisher = {Nature Research},
   title = {Adsorption energies on transition metal surfaces: towards an accurate and balanced description},
   volume = {13},
   year = {2022}
}

@article{isaacs2022soc_I2,
   author = {Eric B. Isaacs and Hyeondeok Shin and Abdulgani Annaberdiyev and Chris Wolverton and Lubos Mitas and Anouar Benali and Olle Heinonen},
   doi = {10.1103/PhysRevB.105.224110},
   issn = {24699969},
   issue = {22},
   journal = {Physical Review B},
   month = {6},
   publisher = {American Physical Society},
   title = {Assessing the accuracy of compound formation energies with quantum Monte Carlo},
   volume = {105},
   year = {2022}
}

@article{ning2022r2scan_rvv10,
   author = {Jinliang Ning and Manish Kothakonda and James W. Furness and Aaron D. Kaplan and Sebastian Ehlert and Jan Gerit Brandenburg and John P. Perdew and Jianwei Sun},
   doi = {10.1103/PhysRevB.106.075422},
   issn = {24699969},
   issue = {7},
   journal = {Physical Review B},
   month = {8},
   publisher = {American Physical Society},
   title = {Workhorse minimally empirical dispersion-corrected density functional with tests for weakly bound systems: R2SCAN+rVV10},
   volume = {106},
   year = {2022}
}

@article{grotjahn2023mobh28,
   author = {Robin Grotjahn and Martin Kaupp},
   doi = {10.1002/ijch.202200021},
   issn = {18695868},
   issue = {7-8},
   journal = {Israel Journal of Chemistry},
   month = {8},
   publisher = {John Wiley and Sons Inc},
   title = {A Look at Real-World Transition-Metal Thermochemistry and Kinetics with Local Hybrid Functionals},
   volume = {63},
   year = {2023}
}

@article{shi2023ccsdt_oxide,
   author = {Benjamin X. Shi and Andrea Zen and Venkat Kapil and Péter R. Nagy and Andreas Grüneis and Angelos Michaelides},
   doi = {10.1021/jacs.3c09616},
   issn = {15205126},
   issue = {46},
   journal = {Journal of the American Chemical Society},
   month = {11},
   pages = {25372-25381},
   pmid = {37948071},
   publisher = {American Chemical Society},
   title = {Many-Body Methods for Surface Chemistry Come of Age: Achieving Consensus with Experiments},
   volume = {145},
   year = {2023}
}

@article{masios2023CCSDcT,
   author = {Nikolaos Masios and Andreas Irmler and Tobias Schäfer and Andreas Grüneis},
   doi = {10.1103/PhysRevLett.131.186401},
   issn = {10797114},
   issue = {18},
   journal = {Physical Review Letters},
   month = {11},
   pmid = {37977639},
   publisher = {American Physical Society},
   title = {Averting the Infrared Catastrophe in the Gold Standard of Quantum Chemistry},
   volume = {131},
   year = {2023}
}

@article{wexler2023Oads,
   author = {Robert B. Wexler and Emily A. Carter},
   doi = {10.1002/adts.202200592},
   issn = {25130390},
   issue = {10},
   journal = {Advanced Theory and Simulations},
   month = {10},
   publisher = {John Wiley and Sons Inc},
   title = {Oxygen-Chlorine Chemisorption Scaling for Seawater Electrolysis on Transition Metals: The Role of Redox},
   volume = {6},
   year = {2023}
}

@article{kreitz2023error_cancel,
   author = {Bjarne Kreitz and Kento Abeywardane and C. Franklin Goldsmith},
   doi = {10.1021/acs.jctc.3c00112},
   issn = {1549-9618},
   issue = {13},
   journal = {Journal of Chemical Theory and Computation},
   month = {7},
   pages = {4149-4162},
   title = {Linking Experimental and Ab Initio Thermochemistry of Adsorbates with a Generalized Thermochemical Hierarchy},
   volume = {19},
   year = {2023}
}

@article{szaro2023rpa_ads_c2,
   author = {Nicholas A. Szaro and Mubarak Bello and Charles H. Fricke and Olajide H. Bamidele and Andreas Heyden},
   doi = {10.1021/acs.jpclett.3c02723},
   issn = {19487185},
   issue = {48},
   journal = {Journal of Physical Chemistry Letters},
   month = {12},
   pages = {10769-10778},
   pmid = {38011289},
   publisher = {American Chemical Society},
   title = {Benchmarking the Accuracy of Density Functional Theory against the Random Phase Approximation for the Ethane Dehydrogenation Network on Pt(111)},
   volume = {14},
   year = {2023}
}

@article{wei2023embed_rpa_metal,
   author = {Ziyang Wei and John Mark P. Martirez and Emily A. Carter},
   doi = {10.1063/5.0181229},
   issn = {10897690},
   issue = {19},
   journal = {Journal of Chemical Physics},
   month = {11},
   pmid = {37971031},
   publisher = {American Institute of Physics Inc.},
   title = {Introducing the embedded random phase approximation: H2 dissociative adsorption on Cu(111) as an exemplar},
   volume = {159},
   year = {2023}
}

@article{kim2024nitrogen,
   author = {Honghui Kim and Neung-Kyung Yu and Nianhan Tian and Andrew J. Medford},
   doi = {10.1021/acs.jpcc.4c01497},
   issn = {1932-7447},
   issue = {27},
   journal = {The Journal of Physical Chemistry C},
   month = {7},
   pages = {11159-11175},
   title = {Assessing Exchange-Correlation Functionals for Heterogeneous Catalysis of Nitrogen Species},
   volume = {128},
   url = {https://pubs.acs.org/doi/10.1021/acs.jpcc.4c01497},
   year = {2024}
}

@article{voss2024Fe_mag,
   author = {Johannes Voss},
   doi = {10.1002/jcc.27366},
   issn = {0192-8651},
   issue = {21},
   journal = {Journal of Computational Chemistry},
   month = {8},
   pages = {1829-1845},
   pmid = {38668453},
   publisher = {John Wiley and Sons Inc},
   title = {Machine learning for accuracy in density functional approximations},
   volume = {45},
   url = {https://onlinelibrary.wiley.com/doi/10.1002/jcc.27366},
   year = {2024}
}

@article{oudot2024rpa_ads,
   author = {B. Oudot and K. Doblhoff-Dier},
   doi = {10.1063/5.0220465},
   issn = {10897690},
   issue = {5},
   journal = {Journal of Chemical Physics},
   month = {8},
   pmid = {39092949},
   publisher = {American Institute of Physics},
   title = {Reaction barriers at metal surfaces computed using the random phase approximation: Can we beat DFT in the generalized gradient approximation?},
   volume = {161},
   year = {2024}
}

@article{ye2024ccsdt_oxide,
   author = {Hong Zhou Ye and Timothy C. Berkelbach},
   doi = {10.1039/d4fd00041b},
   issn = {13645498},
   journal = {Faraday Discussions},
   month = {3},
   publisher = {Royal Society of Chemistry},
   title = {Adsorption and vibrational spectroscopy of CO on the surface of MgO from periodic local coupled-cluster theory},
   year = {2024}
}

@article{ye2024ccsdt_oxide2,
   author = {Hong-Zhou Ye and Timothy C. Berkelbach},
   month = {2},
   pages = {arXiv:2309.14640},
   title = {Ab initio surface chemistry with chemical accuracy},
   url = {http://arxiv.org/abs/2309.14640},
   year = {2024}
}

@article{alavi2024ccsdt_periodic,
   author = {Ali Alavi and Kemal Atalar and Timothy C. Berkelbach and George H. Booth and Garnet Kin-Lic Chan and Francesco A. Evangelista and Tamar Goldzak and Andreas Grüneis and Gaurav Harsha and Venkat Kapil and Peter Knowles and Marie-Bernadette Lepetit and Julia Liebert and Arman Nejad and Verena A. Neufeld and Trinidad Novoa and Katarzyna Pernal and Felix Plasser and Umatur Rehman and Benjamin X. Shi and David P. Tew and Zikuan Wang and Carlos Mejuto-Zaera and Dominika Zgid and Andrew Zhu and Tianyu Zhu and Martijn A. Zwijnenburg},
   doi = {10.1039/D4FD90044H},
   issn = {1359-6640},
   journal = {Faraday Discussions},
   month = {10},
   pages = {682-707},
   publisher = {Royal Society of Chemistry},
   title = {Correlation in extended systems: general discussion},
   volume = {254},
   url = {https://xlink.rsc.org/?DOI=D4FD90044H},
   year = {2024}
}

@article{ye2024ccsdt_periodic,
   author = {Hong Zhou Ye and Timothy C. Berkelbach},
   doi = {10.1021/acs.jctc.4c00936},
   issn = {15499626},
   journal = {Journal of Chemical Theory and Computation},
   pmid = {39376105},
   publisher = {American Chemical Society},
   title = {Periodic Local Coupled-Cluster Theory for Insulators and Metals},
   year = {2024}
}

@article{carbone2024CCSDcT,
   author = {Johanna P. Carbone and Andreas Irmler and Alejandro Gallo and Tobias Schäfer and William Z. Van Benschoten and James J. Shepherd and Andreas Grüneis},
   doi = {10.1039/d4fd00085d},
   issn = {13645498},
   journal = {Faraday Discussions},
   month = {8},
   pages = {586-597},
   pmid = {39169819},
   publisher = {Royal Society of Chemistry},
   title = {CO adsorption on Pt(111) studied by periodic coupled cluster theory},
   volume = {254},
   year = {2024}
}

@article{sauer2024rpa_review,
   author = {Joachim Sauer},
   doi = {10.1016/j.jcat.2024.115482},
   issn = {10902694},
   journal = {Journal of Catalysis},
   month = {5},
   publisher = {Academic Press Inc.},
   title = {The future of computational catalysis},
   volume = {433},
   year = {2024}
}

@misc{kreitz2025thermo,
   author = {Bjarne Kreitz and Gabriel S. Gusmão and Dingqi Nai and Sushree Jagriti Sahoo and Andrew A. Peterson and David H. Bross and C. Franklin Goldsmith and Andrew J. Medford},
   doi = {10.1039/d4cs00768a},
   issn = {14604744},
   issue = {2},
   journal = {Chemical Society Reviews},
   month = {11},
   pages = {560-589},
   pmid = {39611700},
   publisher = {Royal Society of Chemistry},
   title = {Unifying thermochemistry concepts in computational heterogeneous catalysis},
   volume = {54},
   year = {2024}
}

@article{sheldon2024embed_rpa_metal,
   author = {Christopher Sheldon and Joachim Paier and Denis Usvyat and Joachim Sauer},
   doi = {10.1021/acs.jctc.3c01308},
   issn = {15499626},
   issue = {5},
   journal = {Journal of Chemical Theory and Computation},
   month = {3},
   pages = {2219-2227},
   pmid = {38330551},
   publisher = {American Chemical Society},
   title = {Hybrid RPA:DFT Approach for Adsorption on Transition Metal Surfaces: Methane and Ethane on Platinum (111)},
   volume = {20},
   year = {2024}
}

@article{shi2025autoSKZCAM,
   author = {Benjamin X. Shi and Andrew S. Rosen and Tobias Schäfer and Andreas Grüneis and Venkat Kapil and Andrea Zen and Angelos Michaelides},
   doi = {10.1038/s41557-025-01884-y},
   issn = {1755-4330},
   journal = {Nature Chemistry},
   month = {8},
   title = {An accurate and efficient framework for modelling the surface chemistry of ionic materials},
   url = {https://www.nature.com/articles/s41557-025-01884-y},
   year = {2025}
}

@article{schafer2025CCSDcT,
   author = {Tobias Schäfer},
   doi = {10.1021/acs.jpclett.4c03134},
   issn = {1948-7185},
   issue = {1},
   journal = {The Journal of Physical Chemistry Letters},
   month = {1},
   pages = {17-23},
   pmid = {39690878},
   publisher = {American Chemical Society},
   title = {Ground States for Metals from Converged Coupled Cluster Calculations},
   volume = {16},
   url = {https://pubs.acs.org/doi/10.1021/acs.jpclett.4c03134},
   year = {2025}
}

@article{porter2026error_cancel,
   author = {Seth G. Porter and Bjarne Kreitz},
   doi = {10.1039/d5fd00144g},
   issn = {1359-6640},
   journal = {Faraday Discussions},
   publisher = {Royal Society of Chemistry (RSC)},
   title = {Thermophysical properties of adsorbates with beyond-DFT accuracy from DFT data through error cancellation},
   year = {2026}
}

@misc{paw_vasp,
   title = {Available PAW pseudopotentials in VASP},
   url = {https://www.vasp.at/wiki/index.php/Available_pseudopotentials}
}

@misc{W411_nonMR,
   title = {cis/trans-HO3, Be2, B2, C2, BN, OF, F2O, FOO, FOOF, Cl2O, ClOO, OClO, O3, S3, and S4}
}

@misc{atct2025,
   author = {Branko Ruscic and David Bross},
   title = {Active Thermochemical Tables (ATcT) values based on ver. 1.220 of the Thermochemical Network}
}

@misc{cccbdb,
   title = {NIST Computational Chemistry Comparison and Benchmark Database, NIST Standard Reference Database Number 101 Release 22, May 2022, Editor: Russell D. Johnson III .http://cccbdb.nist.gov/}
}
